\documentclass[10pt]{article}
\usepackage{graphicx}
\usepackage{float}
\usepackage{amsmath}
\allowdisplaybreaks[1]
\usepackage{amsfonts}
\usepackage{jheppub}
\usepackage{multirow}
\usepackage{colortbl}
\definecolor{kugray5}{RGB}{224,224,224}
\usepackage{color}
\usepackage{physics} 			
\usepackage{bm}					

\usepackage[labelformat=simple]{caption,subcaption} 

\usepackage{units}

\newcommand{\mw}{m_W}
\def\Re{{\rm Re}}

\def\rG{r_\Gamma}
\def\rZ{r_Z}
\def\rt{r_t}
\def\al{\alpha}
\def\be{\beta}
\def\de{\delta}
\def\ga{\gamma}
\def\ro{\rho}
\def\si{\sigma}
\def\cA{{\cal A}}
\def\cB{{\cal B}}
\def\cN{{\cal N}}
\def\cM{{\cal M}}
\def\cO{{\cal O}}
\def\cF{{\cal F}}

\newcommand{\Zop}{\hat{\bm{Z}}}		
\newcommand{\TW}[1]{\widetilde{#1}} 
\def\mz{m_Z}
\def\mw{m_W}
\def\mh{m_H}

\def\gs{g_s}
\def\gW{g_W}
\def\cosW{\cos \theta_W}

\newcommand{\Eqno}[1]{Eq.~\eqref{#1}}
\newcommand{\beq}{\begin{equation}}
\newcommand{\eeq}{\end{equation}}
\newcommand{\beqn}{\begin{eqnarray}}
\newcommand{\eeqn}{\end{eqnarray}}
\newcommand{\slsh}{\rlap{$\;\!\!\not$}}     
\newcommand{\as}{\alpha_S}
\newcommand{\ep}{\epsilon}
\newcommand{\e}{\epsilon}
\newcommand{\nn}{\nonumber}
\newcommand\numberthis{\addtocounter{equation}{1}\tag{\theequation}}

\title{Two loop correction to interference in $gg \to ZZ$}

\author{
    John M. Campbell \\
    Fermilab, Batavia, IL 60510, USA \\
    R. Keith Ellis \\
    IPPP, University of Durham, South Road, Durham DH1 3LE, UK\\
    Michal Czakon, Sebastian Kirchner \\
    Institut f\"ur Theoretische Teilchenphysik und Kosmologie, RWTH Aachen University, \\ D-52056 Aachen, Germany\\
    E-mails:
    {\tt johnmc@fnal.gov}, {\tt keith.ellis@durham.ac.uk}, {\tt mczakon@physik.rwth-aachen.de},
    {\tt kirchner@physik.rwth-aachen.de}.
	}

\preprint{FERMILAB-PUB-16-113-T,IPPP/16/28,TTK-16-12}

\abstract{
We present results for the production of a pair of on-shell $Z$ bosons 
via gluon fusion. This process occurs both through the production and decay 
of the Higgs boson, and through continuum production where the $Z$ boson couples 
to a loop of massless quarks or to a massive quark.
We calculate the interference of the two processes and 
its contribution to the cross section up to and including order $O(\alpha_s^3)$.
The two-loop contributions to the amplitude are all known analytically, except for
the continuum production through loops of top quarks of mass $m$. The latter contribution 
is important for the invariant mass of the two $Z$ bosons, (as measured by the mass of their leptonic decay products, $m_{4l}$), 
in a regime where $m_{4l} \geq 2 m$ because of the contributions of longitudinal bosons.
We examine all the contributions
to the virtual amplitude involving top quarks, as expansions about the heavy top 
quark limit combined with a conformal mapping and \emph{Pad\'e 
approximants}.
Comparison with the analytic results, where known, allows us to assess the validity
of the heavy quark expansion, and it extensions.  
We give results for the NLO corrections to this interference, including both real and 
virtual radiation. }
\keywords{QCD, Hadron colliders, LHC}

\begin{document}

\maketitle

\section{Introduction} 
\label{sec:intro}

The production of four charged leptons is a process of great
importance at the LHC. It was one of the discovery channels of the
Higgs boson at the LHC. It also provides fundamental tests of the
gauge structure of the electroweak theory through the high energy
behaviour.  Four charged leptons are predominantly produced by quark
anti-quark annihilation; the mediation is by photons or $Z$ bosons
dependent on the mass of the four leptons, $m_{4l}$.

A smaller contribution, which however grows with energy is provided by
gluon-gluon fusion.  The Higgs boson is of course produced in this
channel; in the Standard Model (SM) this occurs predominantly through
the mediation of a loop of top quarks.  As pointed out by Kauer and
Passarino~\cite{Kauer:2012hd}, despite the narrow width of the Higgs
boson, the Higgs-mediated diagram gives a significant contribution for
$m_{4l}> \mh$.  If we examine the tail of the Higgs-mediated diagrams
there are three phenomena occurring:
\begin{itemize} 
\item The opening of the threshold for the production of real on-shell $Z$ bosons, $m_{4l}> 2 \mz$. 
\item The region $m_{4l}=2 m$, ($m$ is the top quark mass) where the 
loop diagrams develop an imaginary part.
\item The large $m_{4l}$ region, $ m_{4l}>2 m$, where the destructive interference between the 
Higgs-mediated diagrams leading to $Z$ bosons and the continuum production 
of on-shell $Z$ bosons is most important.
\end{itemize}

\begin{figure}[ht]
\begin{center}
\includegraphics[angle=270,scale=0.7]{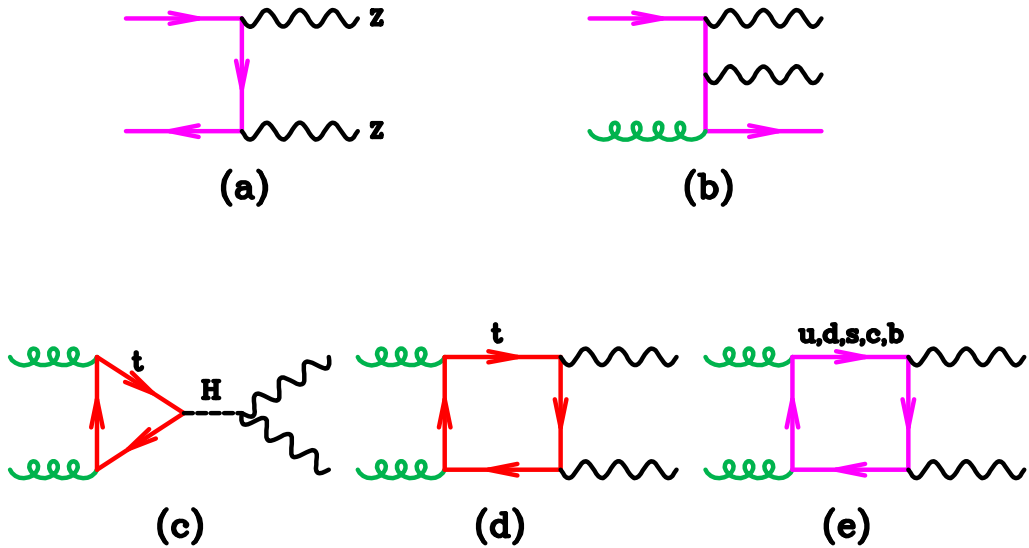}
\caption{Representative diagrams for the $ZZ$ production. In the following we will
suppress the $Z$-decays to leptons.}
\label{ZZprod}
\end{center}
\end{figure}
\renewcommand{\baselinestretch}{1.5} 

A feature of this tail is that it depends on the couplings of
the Higgs boson to the initial and final state particles but not on
the width of the Higgs boson. Assuming the couplings of the on- and
off-peak Higgs-mediated amplitudes are the same, it has been proposed
to use this property to derive upper bounds on the width of the Higgs
boson~\cite{Caola:2013yja}.  Note that models with different
on- and off-peak couplings can be constructed~\cite{Englert:2014aca}.

In the following we shall refer to the production of the bosons
$V_1,V_2$.  Gluon-gluon fusion first contributes to the cross section
for electroweak gauge boson production $pp\to V_1V_2$ as shown in
Fig.~\ref{ZZprod}(c)-(e) at $\cO(\as^2)$, which is the
next-to-next-to-leading-order (NNLO) with respect to the leading-order
(LO) QCD process shown in Fig.~\ref{ZZprod}(a); no two-loop $gg\to V_1
V_2$ amplitudes participate in this order in perturbation theory.

In the context of the Higgs boson width, however, the interference
between the Higgs-mediated $Z$ boson pair-production and the Standard
Model continuum at next-to-leading-order (NLO) QCD already requires
knowledge of the one- and two-loop $gg\to (H\to) V_1 V_2$
amplitudes. The requirement for more precise estimates to the Higgs
boson width were emphasised
in~\cite{ATLAS-CONF-2014-042,Khachatryan:2014iha,Melnikov:2015laa}.
Signal-background interference effects beyond the leading order 
have been considered in ref.~\cite{Bonvini:2013jha} for the process 
$gg\to H \to W^+ W^-$ for the case of a heavy Higgs boson.

In this work we will limit ourselves to the $Z$ boson pair final state,
due to its importance at the LHC. At LO~\cite{Dawson:1993qf} and
NLO~\cite{Spira:1995rr,Harlander:2005rq,Anastasiou:2006hc,Aglietti:2006tp}
the amplitudes for single Higgs boson production have been known for
quite some time.  At LO, the amplitude for the SM continuum $gg\to ZZ$
process occurs via massless and massive fermion loops and 
results are available in each case~~\cite{Glover:1988rg,Kauer:2012ma,Campbell:2013una,Campbell:2014gua}.

The situation, however, is different for the NLO continuum process,
although vast progress in terms of two-loop amplitudes has been
made~\cite{Gehrmann:2014bfa,Cascioli:2014yka,Caola:2014iua,Gehrmann:2015ora,Caola:2015ila,vonManteuffel:2015msa}.
Recently two-loop $gg\to ZZ$ amplitudes\footnote{Actually, the results
in~\cite{Caola:2015ila} and~\cite{vonManteuffel:2015msa} allow for
arbitrary off-shell electroweak gauge bosons in the final state.}
via \emph{massless} quarks became
available~\cite{Caola:2015ila,vonManteuffel:2015msa}. The complete
computation of two-loop amplitudes with \emph{massive} internal quark
loops, on the other hand, is commonly assumed to be just beyond
present technical capabilities.
Although the contribution of the top quark loops to these diagrams is smaller
than the contribution of the light quarks in the region just above the $Z$-pair
threshold, in the high $m_{4l}$ region the amplitude is dominated by the
contributions of longitudinal $Z$ bosons that couple to the top quark loops.
Recently a first heavy top quark
approximation for the two-loop $gg\to ZZ$ amplitude with internal top
quarks was published~\cite{Melnikov:2015laa}. In that work only the
leading term in the $s/m^2$ expansion was considered. In that
approximation, the vector-coupling of the $Z$ boson to the top quark
does not contribute. In addition an approximate treatment of this process at higher orders, based on soft gluon resummation,
was presented in Ref.~\cite{Li:2015jva}.

In the present work we will push this analysis further. We start by
presenting our results for the LO and NLO Higgs-mediated $ZZ$
production in terms of the $s/m^2$ expansion in
Sec.~\ref{sec:higgs_production_via_gluon_fusion}, despite the fact 
that the full result is known. This part
is required for the later interference with the SM
continuum. Furthermore, it is well suited to introduce our notation in
Sec.~\ref{sub:ggH_preliminaries} and to assess the validity of the
approximation methods with respect to the exact known (N)LO amplitudes
in Sec.~\ref{sub:mass_expansion_and_its_improvements}.

The results for the LO and virtual NLO contributions to the SM
continuum with massive quark loops will be given in
Sec.~\ref{sec:virtual_corrections_to_sm_zz_production_via_massive_quark_loops}
as a large-mass expansion (LME) with terms up to $(s/m^2)^6$.
We will limit our discussion to the interference between the
Higgs-mediated term and the continuum term. Similar to~\cite{Melnikov:2015laa} we will consider
\emph{on-shell} $Z$ bosons in the final state.  A theoretical
predictions for off-shell $Z$ bosons would be optimal, but in order to
reduce the number of scales in the problem, we restrict ourselves to
on-shell $Z$ bosons. Since we are primarily interested in the high-mass 
behaviour this is an appropriate approximation. 
A limited number of scales is beneficial when we
consider the extension of our approach to a full calculation.
In Sec.~\ref{sec:real_corrections_to_sm_zz_production} we summarize our treatment of the real
radiation contribution, which makes use of results already presented in Ref.~\cite{Campbell:2014gua}.

The results of our calculation, including loops of both massless and massive quarks,
will be presented in Sec.~\ref{sec:results}.  We will compare the effects of the NLO
corrections to the interference contribution with the corresponding corrections to the
Higgs diagrams alone.  In addition, we will discuss
the impact of our results on analyses of the off-shell region that
aim to bound the Higgs boson width.

All expansion results from Sec.~\ref{sec:higgs_production_via_gluon_fusion} and Sec.~\ref{ssub:non_anomalous_diagrams} are provided via ancillary files on \texttt{arXiv} as \texttt{FORM} and \texttt{Mathematica} readable code.

\section{Higgs Production in Gluon Fusion and Decay to $ZZ$} 
\label{sec:higgs_production_via_gluon_fusion}

In this section we give a detailed discussion of single Higgs boson
production at LO and NLO QCD and its subsequent decay to a pair of
on-shell $Z$ bosons.  As mentioned earlier the LO and NLO amplitudes
for single Higgs boson production have been known for a long time;
either approximate results in terms of Taylor expansions in the
inverse of the top quark mass
$s/m^2$~\cite{Dawson:1990zj,Dawson:1993qf,Harlander:2002wh,Anastasiou:2002yz,Aglietti:2006tp,Harlander:2009bw,Pak:2009bx}
or results keeping the exact top mass
dependence~\cite{Djouadi:1991tka,Aglietti:2006tp}.

It is understood that, whenever feasible and available, the exact
results for LO and NLO amplitudes are used. However, we are mainly
interested in approximations to the interference contributions
$\Re\bra{\cA_\text{LO}}\ket{\cB_\text{(N)LO}}$, where $\cA$ denotes
the Higgs-mediated and $\cB$ the SM continuum amplitude. Since no
exact results are available for $\cB_\text{NLO}$ we will use the,
so-called, large-mass expansion~\cite{Smirnov:2002pj} as an
approximation of the SM continuum. Hence, for consistency, we also
perform the expansion of the Higgs-mediated amplitude $\cA$ to high
powers in $s/m^2$.  Expansion of the two-loop Higgs-mediated amplitude
$\cA_\text{NLO}$ and its comparison to available results from the
literature provides moreover a helpful check of our expansion routines
due to the general structure of the LME.

Furthermore, the large-mass expansion in powers of $s/m^2$ is formally only
valid below the threshold of top quark pair-production, as $m$ is
assumed to be much larger than any other scale in the problem,
e.g. $s\ll m^2$. As extensively discussed in literature the naive LME
can be drastically improved at (and even far above) threshold by
taking the next mass threshold into account, see Ref~\cite{Smirnov:2002pj} and
references within, or by rescaling
the approximated NLO result by the exact LO result, see e.g. 
Refs~\cite{Pak:2009dg,Grigo:2013rya}. We will address
this issue in Sec.~\ref{ssub:ggH_comparison_lme_with_full_result} and try to 
draw conclusions for the SM continuum.

\subsection{Preliminaries} 
\label{sub:ggH_preliminaries}

\begin{figure}[t]
	\centering
	\begin{subfigure}{.4\textwidth}
		\includegraphics[angle=270,scale=.6]{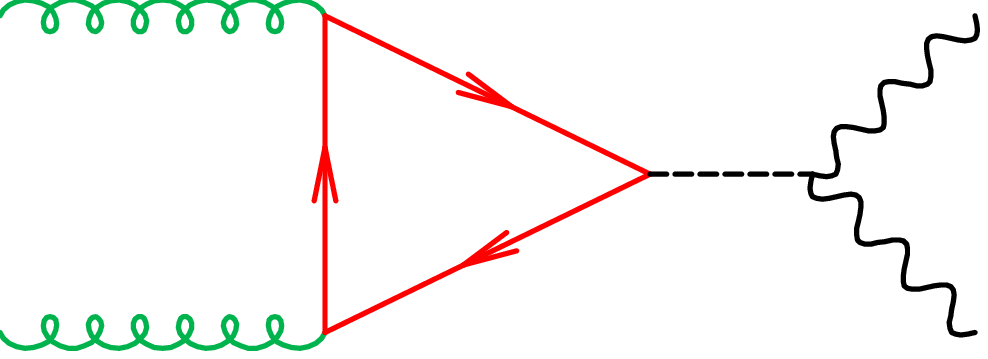}
		\caption{}
		\label{sfig:ggH_LO_amp}
	\end{subfigure}
	\hspace{.5cm}
	\begin{subfigure}{.4\textwidth}
		\includegraphics[angle=270,scale=.6]{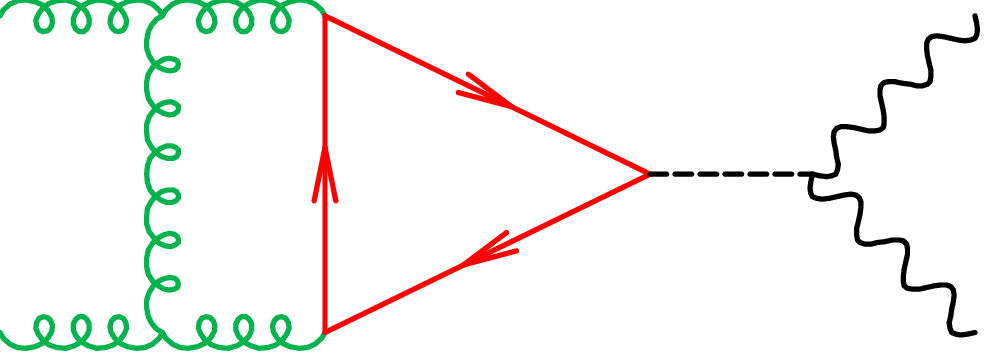}
		\caption{}
	\end{subfigure}
	
	\vspace{.5cm}
	\begin{subfigure}{.4\textwidth}
		\includegraphics[angle=270,scale=.6]{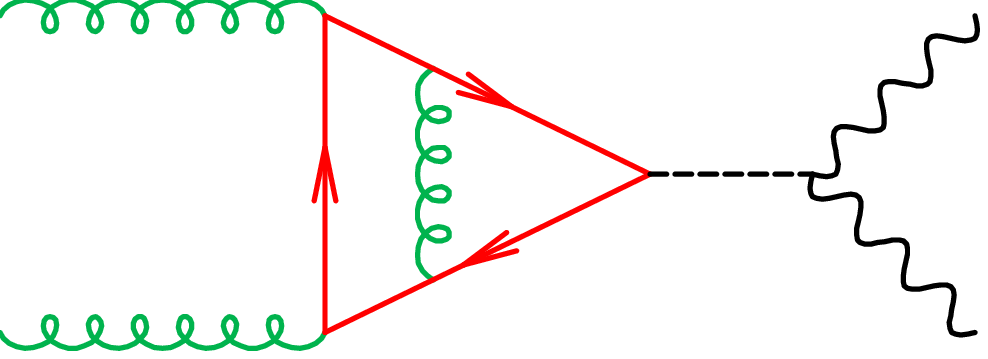}
		\caption{}
	\end{subfigure}
	\hspace{.5cm}
	\begin{subfigure}{.4\textwidth}
		\includegraphics[angle=270,scale=.6]{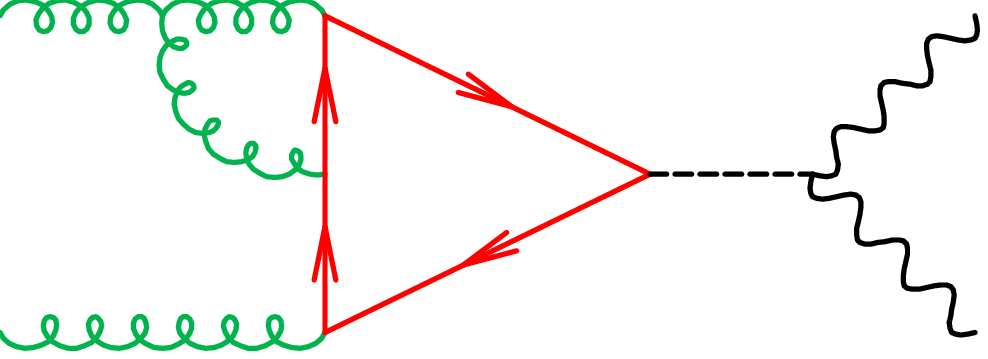}
		\caption{}
	\end{subfigure}
	\caption{Representative diagrams for the LO+NLO virtual $gg\to H\to ZZ$ amplitude.}
	\label{fig:ggH_NLO_amps}
\end{figure}

The amplitudes for single Higgs boson production
\begin{equation}
	g(p_1,\al,A) + g(p_2,\be,B) \to H(p_1+p_2),\;\;s= (p_1+p_2)^2 \, ,
\end{equation}
are illustrated in Fig.~\ref{fig:ggH_NLO_amps} for the one-loop and two-loop case.
The largest contribution is due to the internal
massive top quark loop; in the following we will ignore the
contribution of other quarks for the Higgs production process.

The $gg\to H$ amplitude, with color (Lorentz) indices $A,B (\alpha,\beta)$ for the initial state gluons, 
can be written as
\begin{equation}
	\label{eq:ggH_amp_decomposition}
	\ket{\cA^{0,AB}_{\alpha\beta}(\as^0,m^0,\mu,\ep)} =- i
        \delta^{AB}\, \frac{\gW}{2 \mw} \frac{4}{3}\;\left(g_{\alpha\beta}\, p_1 \cdot p_2 - p_{1,\beta}p_{2,\alpha} \right)\; \ket{\cA^0(\as^0,m^0,\mu,\ep)}\,,
\end{equation}
such that the reduced matrix element  
$\ket{\cA^0(\as^0,m^0,\mu,\ep)}$ is dimensionless
and can be expressed as a function of $\mu^2/s$ and $r_t=m^2/s$.
The bare on-shell amplitudes admit the perturbative expansion
\begin{equation}
	\label{eq:ggH_amp_bare_expansion}
	\ket{\cA^0(\as^0,m^0,\mu,\ep)} = \frac{\as^0}{4\pi} \ket{\cA^{0,(1)}(m^0,\mu,\ep)} + \left(\frac{\as^0}{4\pi}\right)^2 \ket{\cA^{0,(2)}(m^0,\mu,\ep)} + \cO{\left((\as^0)^3\right)}\,,
\end{equation}
where we introduced the parameter $\ep$ from dimensional regularisation 
in $d=4-2\ep$ space-time dimensions 
and $\mu$ to keep the amplitudes dimensionless. 
The calculation is performed in Conventional Dimensional Regularisation (CDR) and the following definition of the $d$-dimensional loop integral
measure 
\begin{equation}
	\label{eq:loopnorm_ggH}
	\int{\frac{d^4p}{(2\pi)^4}} \longrightarrow \mu^{2\ep} \underbrace{\frac{e^{\ep \ga_E}}{(4\pi)^\ep}}_{\equiv S_\ep} \cdot \int{\frac{d^dp}{(2\pi)^d}} 
\end{equation}
is used in accordance with the $\overline{MS}$-scheme, to avoid the proliferation of 
unnecessary $\ga_E-\log(4\pi)$ terms.

The ultraviolet (UV) renormalised amplitudes are given by
\begin{equation}
	\label{eq:ggH_renormalisation_definition}
	\ket{\cA^r(\as^{(n_f)}(\mu),m,\mu,\ep)} = Z_{m} Z_g \ket{\cA^0(\as^0,m^0,\mu,\ep)}\,,
\end{equation}
where $Z_g$ denotes the on-shell gluon renormalisation
constant. The $Ht\bar{t}$ vertex is renormalised, according
to~\cite{Braaten:1980yq}, by $g_H^0 = Z_{m} \,g_H$ with $g_H$ being
the Yukawa coupling for the top quark. The bare top quark mass is related to the renormalised mass, $m$, by $m^0 = Z_m m$. The necessary on-shell 
renormalisation constants are given by
\begin{align}
	Z_g &= 1- \frac{\as^{(n_f)}}{4\pi} T_F \left(\frac{\mu^2}{m^2}\right)^\ep \cdot \frac{4}{3\ep} + \cO\left((\as^{(n_f)})^2,\ep \right) \quad \text{and} \\
	Z_m &= 1- \frac{\as^{(n_f)}}{4\pi} C_F \left(\frac{\mu^2}{m^2}\right)^\ep \left[ \frac{3}{\ep} +4 \right] + \cO\left((\as^{(n_f)})^2,\ep \right)\,,
\end{align}
with $T_F=1/2$. See appendix A of~\cite{Baernreuther:2013caa} and
references therein for more information. The mass renormalisation
enters as an overall factor in \Eqno{eq:ggH_renormalisation_definition} because
of the renormalisation of the Yukawa coupling, and also implicitly in 
the relationship between the bare and renormalised mass.
We will always present mass-renormalised results
in the following.

The strong coupling constant is renormalised in the $\overline{MS}$-scheme according to
\begin{equation}
	\as^0  = Z_{\as}^{(n_f)} \as^{(n_f)}(\mu)\,,
\end{equation}
with~\cite{Baernreuther:2013caa}
\begin{equation}
	\label{eq:Zas_as}
	Z_{\as}^{(n_f)} = 1- \frac{\as^{(n_f)}}{4\pi} \frac{\beta_0^{(n_f)}}{\ep} + \cO\left((\as^{(n_f)})^2\right) \quad \text{and}\quad \beta_0^{(n_f)} = \frac{11}{3}C_A-\frac{4}{3}T_F n_f\,,
\end{equation}
where $n_f=6$ denotes the number of fermions and $\be_0^{(n_f)}$ the coefficient of the beta function. The explicit scale dependence of the renormalised strong coupling constant $\as^{(n_f)}(\mu)$ is dropped in the following to simplify our notation.
All of our quantities are computed in five-flavour $(n_l=5)$ QCD. Hence, we decouple the top quark from the QCD running via
\begin{equation}
	\as^{(n_f)} = \xi_{\as} \as^{(n_l)} \quad \text{and}\quad \xi_{\as} = 1+ \frac{\as^{(n_l)}}{4\pi} T_F \left[ \frac{4}{3}\log\left(\frac{\mu^2}{m^2}\right)\right] + \cO\left((\as^{(n_l)})^2,\ep \right) \,,
\end{equation}
with $n_l$ the number of light quarks.

After UV renormalisation the two-loop amplitude still contains divergences of infrared origin. The structure of these divergences is, however, completely understood at two-loop level. The finite remainder is defined by infrared (IR) renormalisation
\begin{equation}
	\label{eq:FinRem_def}
	\ket{\cF_{\cA,\cB}\left( \as^{(n_l)},m,\mu \right)} = \left(\Zop_{gg}^{(n_l)}\right)^{-1} \ket{\cM_{\cA,\cB}^r \left( \as^{(n_l)},m,\mu,\ep \right)} \,.
\end{equation}
Expanding \Eqno{eq:FinRem_def} in $\as^{(n_l)}/(4\pi)$ yields the explicit expressions for the LO and NLO finite remainders
\begin{align}
	\label{eq:FinRem_1_def}
	\ket{\cF_{\cA,\cB}^{(1)}(m,\mu)} &= \ket{\cM_{\cA,\cB}^{r,(1)} \left( m,\mu\right)} \quad \text{and} \\
	\ket{\cF_{\cA,\cB}^{(2)}(m,\mu)} &= \ket{\cM_{\cA,\cB}^{r,(2)}(m,\mu,\ep)} - \Zop_{gg}^{(n_l,1)}\, \ket{\cM_{\cA,\cB}^{r,(1)}(m,\mu,\ep)} \, .
	\label{eq:FinRem_2_def}
\end{align}
The infrared renormalisation matrix $\Zop_{gg}^{(n_l)}$ is taken from~\cite{Ferroglia:2009ii,Baernreuther:2013caa,Czakon:2014oma} and reads for the gluon-gluon initial state with colourless final state in terms of the renormalised strong coupling constant
\begin{equation}
	\label{eq:Zop_as}
	\Zop_{gg}^{(n_l)} = 1+\frac{\as^{(n_l)}}{4\pi} \, \Zop_{gg}^{(n_l,1)}  = 1+\frac{\as^{(n_l)}}{4\pi}\left(\frac{-2C_A}{\ep^2}-\frac{2C_A \log(-\mu^2/s)+\beta_0^{(n_l)}}{\ep}\right) +\cO\left( (\as^{(n_l)})^2\right) \,.
\end{equation}

In the end we are interested in the amplitude for the process
\beq
g(p_1)+g(p_2) \to H \to Z(p_3)+Z(p_4) \, ,
\eeq
and we set up momentum conservation as $p_1+p_2 = p_3+p_4$. For the calculation at hand we also need the decay amplitude $H \to ZZ$, see Fig.~\ref{sfig:ggH_LO_amp}, which is given by
\beqn
\ket{\cM^{\rho\si}}_{H\to ZZ} &=&i \gW \frac{\mw}{\cos^2 \theta_W} g^{\rho \sigma} \; .
\label{decayamplitude}
\eeqn
Combining Eqs.~(\ref{eq:ggH_amp_decomposition},\ref{decayamplitude}) the full amplitude for production and decay is
\beq
\label{eq:ggHZZ_amplitude_structure}
\ket{\cA^{\al\be\rho\si,AB}_\text{ggHZZ}(\as^{(n_l)},m,\mu,\ep)} =
 \cN\, \de^{AB} \, \frac{4}{3}\; \frac{s}{s-\mh^2} \,\ket{\cA(\as^{(n_l)},m,\mu,\ep)} \cdot
\Bigg( g^{\alpha \beta} -\frac{p_2^{\alpha} p_1^{\beta}}{p_1\cdot p_2}\Bigg) g^{\rho \sigma} ,
\eeq
where we have defined an overall normalisation factor,
\beq
\cN= i \left( \frac{\gW}{2 \cosW} \right)^2 \,.
\label{eq:Ndef}
\eeq
From this it is straightforward to square the amplitude to obtain the result for the
Higgs-mediated diagrams alone.  The sum over the polarisations of the gluons and the $Z$ bosons of momentum $p$ can be
performed as usual with the projection operators,
\beq
P_g^{\mu \nu}=-g^{\mu \nu}, \hspace{1cm} P_Z^{\rho \beta}(p) = -g^{\rho \beta}+\frac{p^\rho p^\beta}{\mz^2} \, .
\label{Zpolsum}
\eeq
Using these projectors we get the subsidiary result
\beqn
P_{Z}^{\rho\sigma}(p_3)P_{Z \, \rho \sigma}(p_4)&=&2
\Bigg[\frac{(d-2)}{2}+\frac{1}{8}\frac{(s-2 \mz^2)^2}{\mz^4}\Bigg] \, .
\eeqn
Including also the sum over colors yields the matrix element squared for the signal in this channel,
(The statistical factor for identical $Z$ bosons is not included).
\begin{align*}
	\label{eq:MEsq_ggH}
	\mathcal{S}_{gg} &\equiv \bra{\cA^{\al\be\rho\si,AB}_\text{ggHZZ}(\as^{(n_l)},m,\mu,\ep)}\ket{\cA^{AB}_{ \text{ggHZZ},\al\be\rho'\si'}(\as^{(n_l)},m,\mu,\ep)}  P_{Z \rho'}^{\rho}(p_3) P_{Z\sigma'}^{\sigma}(p_4) \numberthis \\
	&= |\cN|^2 \frac{64 N_A}{9} \left(\frac{s}{s-\mh^2}\right)^2 \bra{\cA(\as^{(n_l)},m,\mu,\ep)}\ket{\cA(\as^{(n_l)},m,\mu,\ep)}\cdot (1-\ep) \bigg[ 1-\ep + \frac{1}{8}\bigg(\frac{1}{\rZ}-2\bigg)^2\bigg] \,,
\end{align*}
where we use the notation $\rZ=\mz^2/s$ and $N_A= N_c^2-1 = 8$.

\subsection{Large-Mass Expansion and Improvements} 
\label{sub:mass_expansion_and_its_improvements}

Using the aforementioned conventions we can compute the leading- and
next-to-leading-order amplitude $\ket{\cA^{(1,2)}(m,\mu,\ep)}$ for
single Higgs boson production. Although we always work with the loop
measure $S_\ep = \exp(\ep \ga_E)(4\pi)^{-\ep}$ we factor out
\begin{equation}
	\label{eq:loop_measure_factors}
	S_\ep c_\Gamma = \frac{e^{\ep \ga_E}}{(4\pi^\ep)}\cdot \Gamma(1+\ep)(4\pi)^\ep = 1+\ep^2\frac{\pi^2}{12}
+\cO(\ep^3) \, ,
\end{equation}
in the results presented below to keep factors of $\pi^2$ implicit. The dimensional dependent factor $c_\Gamma$ denotes the somewhat more natural loop measure, because it cancels exactly the $\Gamma(1+\ep)$ factor obtained by the loop integration.

The exactly known leading-order result in $d$-dimensions ($d=4-2\ep$) yields~\cite{Resnick:1973vg,Georgi:1977gs,Dawson:1993qf,Anastasiou:2006hc,Harlander:2009bw}
\begin{align*}
	\label{eq:ME_ggH_LO_exact}
	\ket{\cA^{(1)}(m,\mu,\ep)} &= S_\ep c_\Gamma \cdot 3 r_t  \numberthis\\
	&\times \Bigg( \frac{2\ep}{1-\ep} B_0\left(p_1+p_2;m,m\right) - \left(1-\frac{4}{1-\ep} r_t\right)s\, C_0\left(p_1,p_2;m,m,m\right) \Bigg)\,,
\end{align*}
where $s=(p_1+p_2)^2$. The
definitions of the integrals $B_0$ and $C_0$ are given in
appendix~\ref{Intdef}.

The essential idea of the large-mass expansion based on the method of
\emph{expansion by regions}~\cite{Smirnov:2002pj} is that the
integration domain is divided into different regions where the loop
momenta are soft, $k_i \sim p_i \ll m$ or hard, $p_i \ll k_i\sim m$.
The external momenta $p_i \ll m$ are always assumed to be small. In
the expansion of one-loop integrals only the region of a hard loop
momentum $k_1\sim m$ exists, because all propagators are associated
with the large mass $m$. As a result the one-loop expansion consists
only of a naive Taylor expansion and its result is given in terms of
simple massive one-loop vacuum integrals.

The two-loop integral expansion is more involved since the hard as
well as the soft region must be considered. The first region results,
with the help of~\cite{Davydychev:1995nq}, in scalar massive two-loop
vacuum integrals. The soft region produces a product of massive
one-loop vacuum integrals and massless one-loop bubble and triangle
integrals. All occurring integrals are well known and, although, the
intermediate expressions become huge, the final results are remarkably
simple, as can be seen below. We use our own fully automatic
\emph{in-house} software to perform the large-mass expansion, relying
extensively on the features of \texttt{FORM}~\cite{Vermaseren:2000nd}
and \texttt{Mathematica}.  For a similar approach to Higgs boson
pair-production, see e.g.~\cite{Dawson:1998py}.

Using the large-mass expansion for the $B_0$ and $C_0$ integral, given in 
Sec.~\ref{Intdef}, the corresponding expansion of the full result for 
$\ket{\cA^{(1)}(m,\mu,\ep)}$ in $d$ dimensions is 
\begin{align*}
	\label{eq:ggH_LME1L}
	\ket{\cA^{(1)}(m,\mu,\ep)} &=S_\ep c_\Gamma \left(\frac{\mu^2}{m^2}\right)^\ep \Bigg\{
	1
	+\frac{1}{r_t} \left[ \frac{7 (1+\ep)}{120} \right] 
	+\frac{1}{r_t^2} \left[ \frac{1}{336} \left(2+3 \ep+\ep^2\right) \right] \numberthis \\
	&\hspace{-2cm}+ \frac{1}{r_t^3} \left[ \frac{13 \left(6+11 \ep+6 \ep^2\right)}{100800} \right] 
	+ \frac{1}{r_t^4} \left[ \frac{24+50 \ep+35 \ep^2}{207900} \right]
	+ \frac{1}{r_t^5} \left[ \frac{19 \left(120+274 \ep+225 \ep^2\right)}{121080960} \right] \\
	&\hspace{-2cm}+ \frac{1}{r_t^6} \left[ \frac{180+441 \ep+406 \ep^2}{55036800} \right]
	+ \frac{1}{r_t^7} \left[ \frac{1260+3267 \ep+3283 \ep^2}{2117187072} \right] 
	+ \frac{1}{r_t^8} \left[ \frac{10080+27396 \ep+29531 \ep^2}{89791416000} \right] \\
	&\hspace{-2cm}+ \frac{1}{r_t^9} \left[ \frac{31 \left(10080+28516 \ep+32575 \ep^2\right)}{14340021696000} \right] 
	+ \frac{1}{r_t^{10}} \left[ \frac{50400+147620 \ep+177133 \ep^2}{11640723494400} \right] 
	+ \cO\left(1/r_t^{11}, \ep^3\right) \Bigg\}\,.
\end{align*}
Similarly the two-loop result can be expressed in terms of the leading-order amplitude $\ket{\bar{\cA}^{(1)}(m,\mu,\ep)} = (S_\ep c_\Gamma (\mu^2/m^2)^\ep)^{-1} \ket{\cA^{(1)}(m,\mu,\ep)}$ and with only mass renormalisation included
\begin{align*}
	\label{eq:ggH_LME2L}
	\ket{\cA^{0,(2)}(m,\mu,\ep)} &= \left(S_\ep c_\Gamma \left(\frac{\mu^2}{m^2}\right)^\ep \right)^2 \Bigg\{C_A \Bigg[ \left(-\frac{2}{\ep^2}\left(\frac{m^2}{-s-i\ep}\right)^\ep + \frac{\pi^2}{3}\right) \ket{\bar{\cA}^{(1)}(m,\mu,\ep)}  \\
	&\hspace{-2cm}+ 5 
	+ \frac{1}{r_t}\frac{29}{360} 
	+ \frac{1}{r_t^2}\frac{1}{2520} 
	- \frac{1}{r_t^3}\frac{29}{56000} 
	- \frac{1}{r_t^4}\frac{3329}{24948000} 
	- \frac{1}{r_t^5}\frac{1804897}{63567504000}
	- \frac{1}{r_t^6}\frac{41051}{7063056000} \\
	&\hspace{-2cm}- \frac{1}{r_t^7}\frac{156811}{132324192000} 
	- \frac{1}{r_t^8}\frac{74906179}{307984556880000}
	- \frac{1}{r_t^9}\frac{834852479}{16562725058880000} 
	- \frac{1}{r_t^{10}}\frac{2412657613}{228565605812544000} \Bigg]\\
	&\hspace{-2cm}- 3 C_F \Bigg[ 
	1 
	-\frac{1}{r_t}\frac{61}{270} 
	-\frac{1}{r_t^2}\frac{554}{14175} 
	-\frac{1}{r_t^3}\frac{104593}{15876000}
	-\frac{1}{r_t^4}\frac{87077}{74844000}
	- \frac{1}{r_t^5}\frac{13518232199}{62931828960000} \numberthis \\
	&\hspace{-2cm}- \frac{1}{r_t^6}\frac{673024379}{16362275529600} 
	-\frac{1}{r_t^7}\frac{225626468867}{27815868400320000}
	- \frac{1}{r_t^8} \frac{51518310883673}{31445839226561760000} \\
	&\hspace{-2cm}- \frac{1}{r_t^9}\frac{24341081985219}{72122692986023680000} 
	- \frac{1}{r_t^{10}} \frac{2035074335031827}{28792409364206167680000} \Bigg] 
	+ \cO\left(1/{r_t^{11}},\ep\right) \Bigg\} \,.
\end{align*}
The first terms of \Eqno{eq:ggH_LME1L} and
\Eqno{eq:ggH_LME2L} fully agree with available results in the
literature~\cite{Harlander:2009bw,Pak:2009bx}. Especially the NLO
corrections presented in~\cite{Harlander:2009bw} cover terms in the
expansion up to $\cO\left(1/r_t^2,\ep^2\right)$ and we find full
agreement with our results for the amplitudes as well as the cross
sections.  The analytic results for the exact LO and NLO amplitude
$\cA$, keeping the full top mass dependence, can be taken
from~\cite{Anastasiou:2006hc,Beerli:2008zz}\footnote{The overall sign
of the NLO term differs between the published
paper~\cite{Anastasiou:2006hc} and the thesis of
Beerli~\cite{Beerli:2008zz}. We believe that the sign in the latter
is correct, which is also supported by the comparison with the
NLO results using the large-mass expansion~\cite{Harlander:2009bw,Pak:2009bx}.}. 
The NLO results for the virtual amplitude have also
been checked by our own independent program, using
\texttt{GiNaC}~\cite{Bauer:2000cp} to evaluate the harmonic
polylogarithms.  This serves as a further independent check of the
mass expansion results in \Eqno{eq:ggH_LME1L} and
\Eqno{eq:ggH_LME2L}. This agreement will be illustrated in
Sec.~\ref{ssub:ggH_comparison_lme_with_full_result}.

The radius of convergence of the large-mass expansion is
given by $s/(4m^2)\lesssim 1$. The polynomial
growth leads to an extremely good convergence below and close to
threshold of top quark pair-production, as shown later.

\subsubsection{Rescaling with Exact Leading-Order Result} 
\label{ssub:rescaling_with_exact_leading_order_result}

Above threshold, however, naively no convergence with respect to the exact result can be
expected. At least two procedures exist which
lead to major improvements in terms of convergence of the expanded
result even above threshold\footnote{The region above threshold could
  also be approximated by fitting a suitable ansatz to the high-energy
  limit~\cite{Marzani:2008az,Harlander:2009mq,Pak:2009dg}. This,
  however, would require additional knowledge of the high-energy
  behaviour and is beyond the scope of this work.}. 
We recall these procedures in this subsection and the next.

A well known method of extending the naive large-mass expansion of the NLO cross section beyond its range of validity relies on factoring out the LO cross section with exact top mass dependence,
\begin{equation}
	\label{eq:LME_rescaled}
	\sigma^\text{NLO}_{\text{imp},N} \equiv \sigma^\text{LO}_\text{exact} \cdot \frac{\sigma^\text{NLO}_\text{exp}}{\sigma^\text{LO}_\text{exp}} = \sigma^\text{LO}_\text{exact} \cdot \frac{\sum\limits_{n=0}^N \; c_n^\text{NLO} (1/r_t)^n}{\sum\limits_{n=0}^N \; c_n^\text{LO} (1/r_t)^n} \,.
\end{equation} 
The numerator and denominator are expanded to the same order in
$1/r_t$. It was argued for single Higgs boson production
in~\cite{Pak:2009dg} and for Higgs boson pair-production
in~\cite{Grigo:2013rya} that varying $N$ in the above formula allows to
check for additional power corrections. Including sufficient orders in
the expansion should lead to stable approximations
$\sigma^\text{NLO}_{\text{imp},N}$.

The method relies on the expansion of numerator and denominator in \Eqno{eq:LME_rescaled}
and evidently, requires the knowledge of all of the ingredients in
terms of series expansions. Although this requirement usually does not
pose any problem \emph{per se} it might turn out to be disadvantageous
in certain cases. In our particular case at hand, we require the SM
continuum as well as the Higgs-mediated amplitude as large-mass
expansions. Certainly the Higgs-mediated amplitude is well known at LO
and NLO including its full top mass dependence. Any approximation of
this amplitude poses a potential threat of introducing unnecessary
uncertainties. We will discuss this point further in
Sec.~\ref{sub:visualisation_of_large_mass_expansion_results_for_gg_to_zz}
and see that the method introduced in the next section provides a way
to circumvent this issue.


\subsubsection{Conformal Mapping and Pad\'e Approximants} 
\label{ssub:conformal_mapping_and_pade_approximants}

Having sufficiently many terms in the $1/m$ expansion at hand allows for a
more powerful resummation method, the \emph{Pad\'e approximation}~\cite{Fleischer:1994ef,Fleischer:1996ju,Harlander:2001sa,Smirnov:2002pj,Press:2007:NRE:1403886}. The \emph{univariate Pad\'e approximant} $[n/m]$ to a given \emph{Maclaurin series} with a non-zero radius of convergence $z_0$
\begin{equation}
	\label{eq:general_series_expansion}
	f(z) = \sum\limits_{n=0}^{\infty}\; a_n z^n
\end{equation}
is defined via the rational function
\begin{equation}
	f_{[n/m]}(z) = \frac{b_0+b_1 z+b_2 z^2+\ldots + b_n z^n}{1+c_1 z+c_2 z^2+\ldots + c_m z^m}
\label{eq:Padedefinition}
\end{equation}
such that its Taylor expansion reproduces the first $n+m$ coefficients
of $f(z)$; the coefficients $b_i$ and $c_i$ are uniquely defined by
this expansion.  The advantage of \emph{Pad\'e approximants} over
other techniques, e.g. \emph{Chebyshev approximation}, lies in the
fact that they can provide genuinely new information about the
underlying function $f(z)$, see~\cite{Press:2007:NRE:1403886} for more
information.

The downside of \emph{Pad\'e approximants} is their
uncontrollability. In general, there is no way to tell how accurate
the approximation is, nor how far the range $z$ can be
extended. Computing the \emph{Pad\'e approximants} $[n/n]$ or
$[n/n\pm1]$ for different orders $n$ allows, at least, checking the
stability of the approximation. We will refer to $[n/n]$ as diagonal
and to $[n/n\pm1]$ as non-diagonal \emph{Pad\'e approximants} in the
following.

Although the \emph{Pad\'e approximation} can be directly applied to
\Eqno{eq:general_series_expansion}, it is advantageous to apply a
\emph{conformal mapping}~\cite{Fleischer:1994ef}
\begin{equation}
	\label{eq:conformal_mapping}
	w(z) = \frac{1-\sqrt{1-z/z_0}}{1+\sqrt{1-z/z_0}}
\end{equation}
first. The amplitudes at hand, $gg(\to H)\to ZZ$, with $z=s/m^2$
develop a branch cut starting from $z_0=4$ and extending to $+\infty$ due to the top
quark pair-production threshold. Applying the mapping,
\Eqno{eq:conformal_mapping}, transforms the entire complex plane
into the unit circle of the $w$-plane, such that the upper (lower)
side of the cut corresponds to the upper (lower) semicircle and the
origin of the original $z$-plane is left unchanged.

The initial power series can now be transformed into a new series in $w$~\cite{Smirnov:2002pj}
\begin{equation}
	f(w) = \sum\limits_{n=0}^\infty\; \Phi_n w^n \,,
\end{equation}
where
\begin{align}
	\Phi_0 = a_0 \quad \text{and}\quad
	\Phi_n = \sum\limits_{k=1}^n\; \frac{(n+k-1)!(-1)^{n-k}}{(2k-1)!(n-k)!}\, (4z_0)^k \,a_k\,, \quad \text{if } n \ge 1\,,
\end{align}
and, subsequently, its \emph{Pad\'e approximants} computed. We will
illustrate these features 
using the example of single Higgs boson production in the next section.


\subsubsection{Comparison of LME with Full Result} 
\label{ssub:ggH_comparison_lme_with_full_result}

Let us briefly compare the results from the large-mass expansions,
\Eqno{eq:ggH_LME1L} and \Eqno{eq:ggH_LME2L}, and their,
previously discussed, improvements to the known LO and NLO QCD result
with full top mass
dependence~\cite{Spira:1995rr,Harlander:2005rq,Anastasiou:2006hc,Aglietti:2006tp}. We
include the subsequent $H\to ZZ$ decay, as given in
\Eqno{decayamplitude}, perform the UV+IR renormalisation and
compute the phase space integral over \Eqno{eq:MEsq_ggH}
including all corresponding phase space factors and coupling
constants. The NLO contribution so obtained is not physical, since we
neglect the real-radiation contribution for now. Considering the
obtained finite parts of the LO and virtual NLO corrections alone, on
the other hand, allow to better verify the validity of our
approximations. To be specific, we set
\begin{equation}
	\label{eq:ggH_sigma_defs_virt}
	\si_\text{H}^\text{LO} \sim \Re\bra{\cF_\cA^{(1)}(m,\mu)}\ket{\cF_\cA^{(1)}(m,\mu)} \quad \text{and} \quad \si_\text{virt,H}^\text{NLO} \sim 2 \, \Re\bra{\cF_\cA^{(1)}(m,\mu)}\ket{\cF_\cA^{(2)}(m,\mu)} \,.
\end{equation}
We utilise the $CT10nlo$ PDF set~\cite{Lai:2010vv} within \texttt{LHAPDF}~\cite{Buckley:2014ana} to determine $\as(\mu_f)$ and use the input parameters
\begin{align*}
	\label{eq:ggH_input_parameters}
	\sqrt{S} &= \unit[13]{TeV}\,, & \mu_f &= \mu_r = \sqrt{s}\,, \\
	m &= \unit[173.5]{GeV}\,, & \mz &= \unit[91.1876]{GeV}\,, \numberthis \\
	\mw &= \unit[80.385]{GeV}\,, & G_F &= \unit[1.16639\cdot 10^{-5}]{GeV^{-2}}\,,
\end{align*}
where $S$ and $s$ denote the hadronic and partonic center-of-mass energy, respectively.

\begin{figure}[h!]
	\centering
	\begin{subfigure}{0.49\textwidth}
		\includegraphics[width=\textwidth]{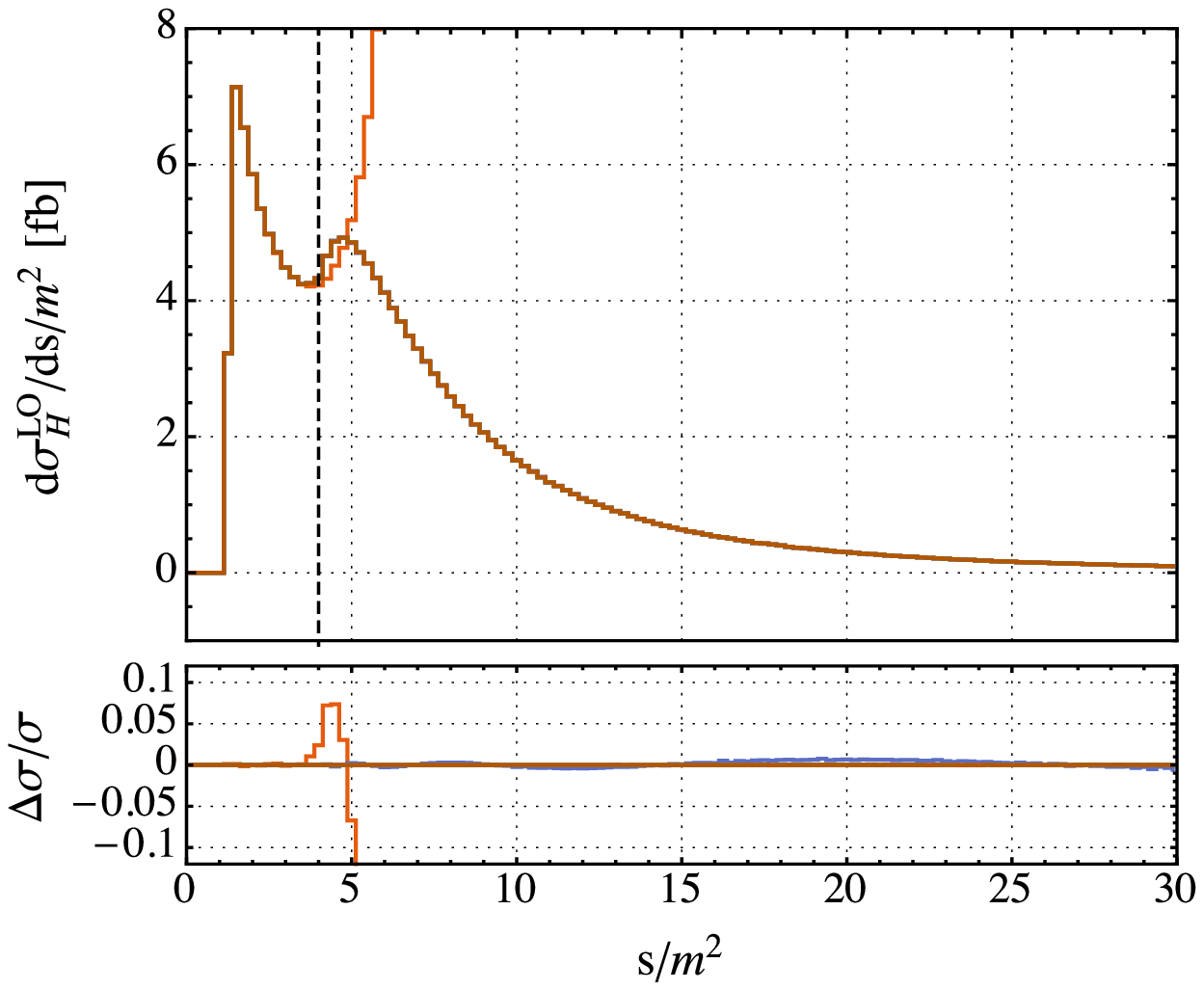}
	\end{subfigure}
	\hfill
	\begin{subfigure}{0.49\textwidth}
		\includegraphics[width=\textwidth]{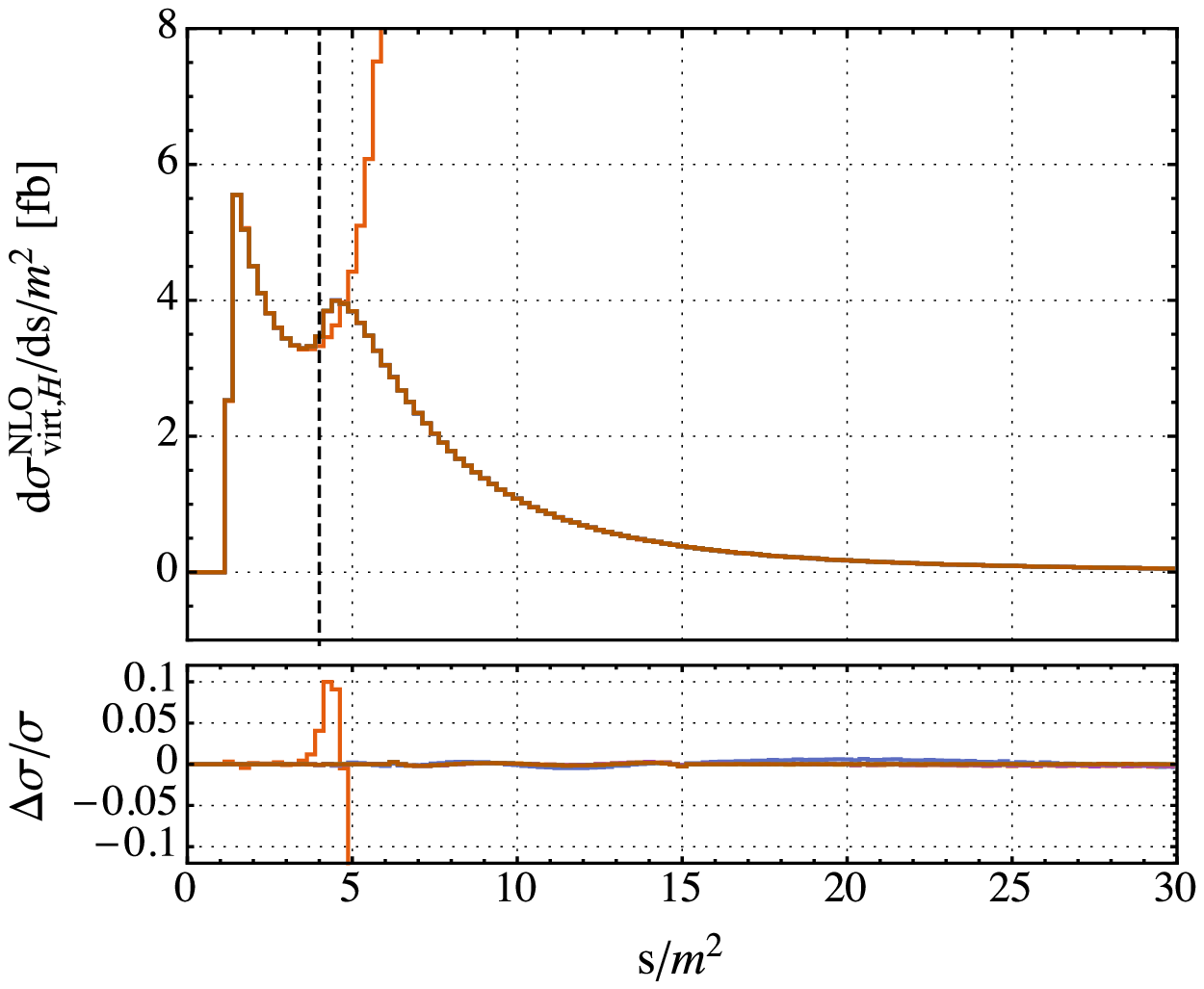}
	\end{subfigure}
	\caption{\textbf{Left panel:} Leading-order $gg\to H\to ZZ$ cross section. 1.) LME up to $1/m^{20}$ (orange). 2.) Exact result (black), LME with conformal mapping (blue) and \emph{Pad\'e approximants} $[4/4],[4/5],[5/4],[5/5]$ (yellow, purple, green, brown) agree perfectly. \textbf{Right panel:} Virtual NLO corrections to $gg\to H\to ZZ$ cross section. See text for details. Color code as in left panel. The bottom plots show the relative deviations with respect to the exact (N)LO results. The vertical dashed line denotes the top quark pair-production threshold.}
	\label{fig:ggHZZ_LO_NLO_LME_Pade}
\end{figure}

The orange curves in Fig.~\ref{fig:ggHZZ_LO_NLO_LME_Pade} depict the
large-mass expansion results of \Eqno{eq:ggH_sigma_defs_virt} for
the LO and the NLO case, where each\footnote{The ambiguity between
expanding the product $\bra{\cF_\cA^{(1)}}\ket{\cF_\cA^{(1,2)}}$ or
expanding each $\ket{\cF_\cA^{(1,2)}}$ separately, consists only of
power corrections which are numerically negligible. We checked that
the difference in $\Delta\si/\si$ at threshold of both approaches is
$\lesssim 1\%$. The same arguments hold for the series expansions
including the conformal mapping.} finite remainder $\cF_\cA^{(1,2)}$ is expanded up to $1/m^{20}$. A minimum cut $\sqrt{s}\ge 2 \mz$ has been imposed and the threshold for top quark pair-production is given by $s/m^2 = 4$. 
The relative deviation
\begin{equation}
	\frac{\Delta \si}{\si} = 1 - \frac{\si^\text{(N)LO}_\text{approx}}{\si^\text{(N)LO}_\text{exact}}
\end{equation}
of the approximated results with respect to the exact result are shown
in the bottom plots.  The large-mass expansion describes the exact LO
and virtual NLO results up to the top threshold very well, with only
$5\%$ deviation at LO and $7\%$ at NLO at $s=4m^2$. As expected
however the large-mass expansion diverges for values above this
threshold. Improvements to this naive approximation by means of the
conformal mapping, \Eqno{eq:conformal_mapping}, are shown in blue. On
top we compute the diagonal, $[5/5]$ (brown) and $[4/4]$ (yellow), and
non-diagonal, $[4/5]$ (purple) and $[5/4]$ (green), \emph{Pad\'e approximants} 
at amplitude level for the mapped series expressions
of each finite remainder, i.e. $\cF_{\cA,[n/m]}^{(i)}$. Both results,
using the \emph{Pad\'e approximants} or the mapped series alone,
excellently reproduce the exact results (black curve) even far above
threshold; with less than $1\%$ deviation from the exact result over
the considered range. As a result the \emph{Pad\'e approximant} $[5/5]$ 
overlays all other curves in
Fig.~\ref{fig:ggHZZ_LO_NLO_LME_Pade}, some of which are scarcely
visible.

\begin{figure}[h!]
	\centering
	\includegraphics[width=.7\textwidth]{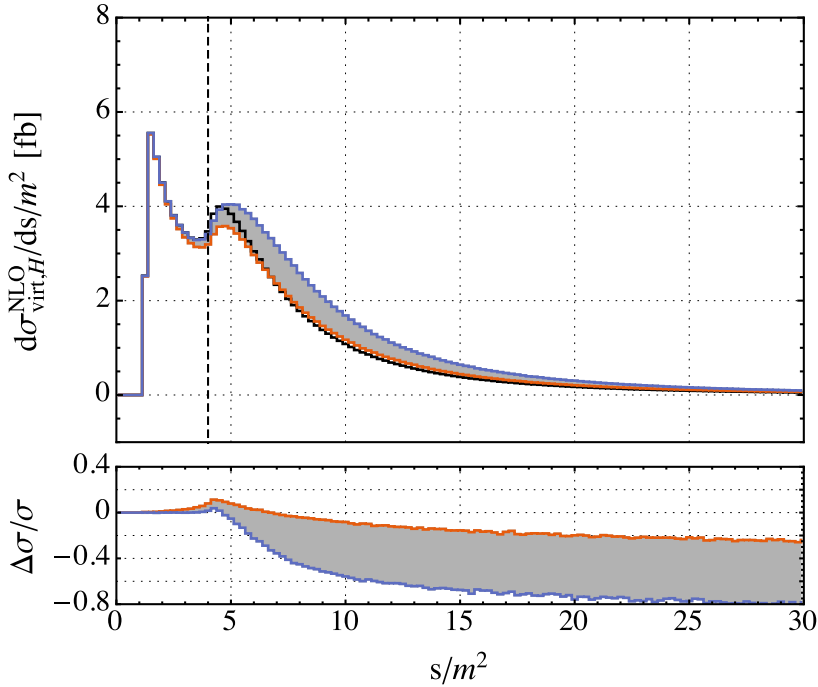}
	\caption{Virtual NLO corrections to $gg\to H\to ZZ$ cross section with rescaling from \Eqno{eq:LME_rescaled}. See text for details. 1.) Exact NLO result (black). 2.) Varying orders of rescaled LMEs are indicated by shaded grey area. Its envelope is given by $\si_{\text{imp},1}^\text{NLO}$ (orange) and $\si_{\text{imp},10}^\text{NLO}$ (blue). The bottom plot shows the relative deviations with respect to the exact NLO results. The vertical dashed line denotes the top quark pair-production threshold.}
	\label{fig:ggHZZ_NLO_melnikov}
\end{figure}

The second choice of improving the naive LME is given by the rescaling
from \Eqno{eq:LME_rescaled}. The results are shown in
Fig.~\ref{fig:ggHZZ_NLO_melnikov}. The exact virtual NLO result is
again shown in black. The rescaled LMEs are indicated by the shaded
grey area and its envelope is given by the expansions
$\si_{\text{imp},1}^\text{NLO}$ (orange) and
$\si_{\text{imp},10}^\text{NLO}$ (blue). Although the heavy-quark
approximation $\si_{\text{imp},1}^\text{NLO}$ gives a reasonable
estimate of the exact result above threshold it fails to describe the
threshold behaviour and peak structure of the exact result. At
threshold the deviation is $10\%$. Taking higher orders in the
expansion into account improves the threshold prescription, with $3\%$
deviation for $\si_{\text{imp},10}^\text{NLO}$ at threshold, but
worsens the trend for higher energies. In both cases we find more than
$20\%$ deviation for $s/m^2>20$.

We end this section by drawing our conclusions from the results
presented. We see that, at least in the single-scale Higgs boson
production and having a sufficient number of terms in the LME at hand,
applying the conformal mapping (and the \emph{Pad\'e approximation})
yields excellent prescriptions of the exact results. The conformal
mapping is imperative, whereas the additional \emph{Pad\'e
  approximants} give only small improvements in terms of uncertainty
reduction and stability of the approximations. We conclude that we
should favour these approximations over the rescaling method.

One important point to notice, however, is that the kinematics change
when moving from the single Higgs boson production to the SM $Z$ boson
pair-production\footnote{Even if the $H\to ZZ$ decay is
included. Effectively, only the $2\to 1$ kinematics of the Higgs boson production matter.}. 
Therefore, the results discussed here may
not necessarily transfer easily. Still, the comparisons within this
chapter should give an idea of the validity of the improved large-mass
expansions. We will discuss analogous considerations for the Z boson
pair-production in
Sec.~\ref{sub:visualisation_of_large_mass_expansion_results_for_gg_to_zz}.


\section{Virtual Corrections to SM $ZZ$ Production via Massive Quark Loops} 
\label{sec:virtual_corrections_to_sm_zz_production_via_massive_quark_loops}

\begin{figure}[ht]
	\centering
	\begin{subfigure}{.3\textwidth}
		\includegraphics[angle=270,scale=.6]{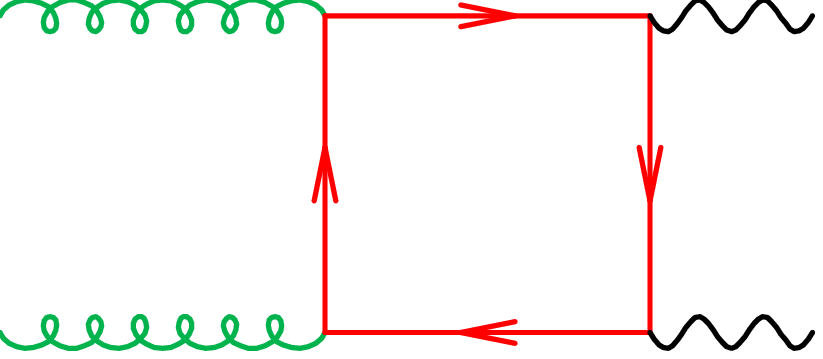}
		\caption{}
		\label{sfig:ggZZ_LO_amp}
	\end{subfigure}
	\hspace{.5cm}
	\begin{subfigure}{.3\textwidth}
		\includegraphics[angle=270,scale=.6]{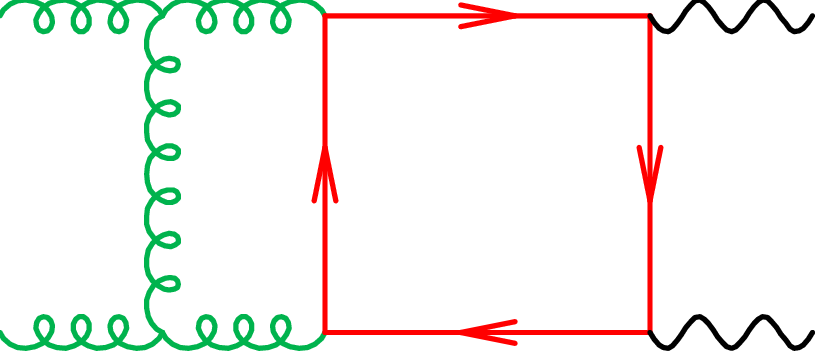}
		\caption{}
	\end{subfigure}
	\hspace{.5cm}
	\begin{subfigure}{.3\textwidth}
		\includegraphics[angle=270,scale=.6]{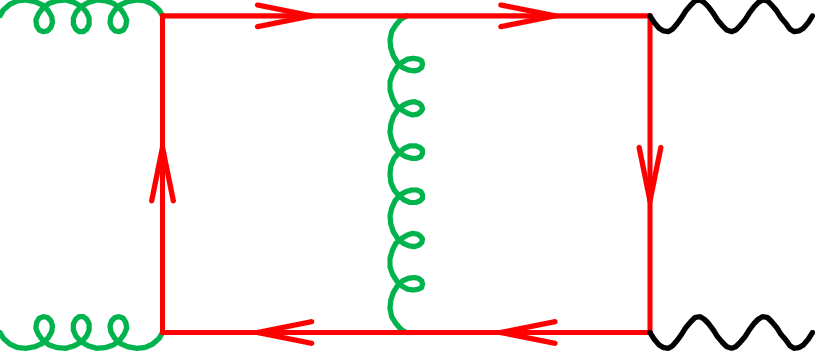}
		\caption{}
	\end{subfigure}
	\begin{subfigure}{.3\textwidth}
		\includegraphics[angle=270,scale=.6]{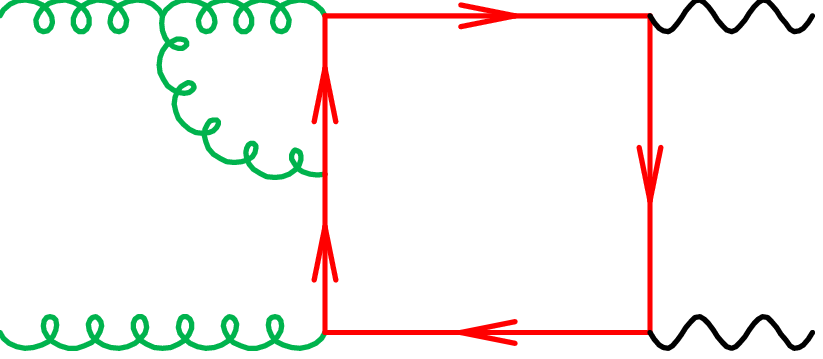}
		\caption{}
	\end{subfigure}
	\hspace{.5cm}
	\begin{subfigure}{.3\textwidth}
		\includegraphics[angle=270,scale=.6]{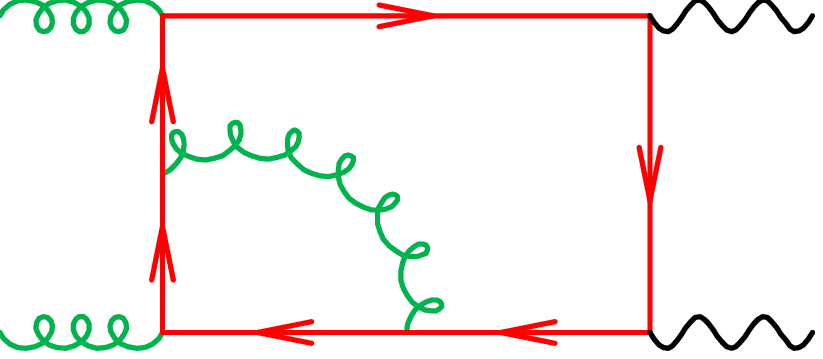}
		\caption{}
	\end{subfigure}
	\hspace{.5cm}
	\begin{subfigure}{.3\textwidth}
		\includegraphics[angle=270,scale=.6]{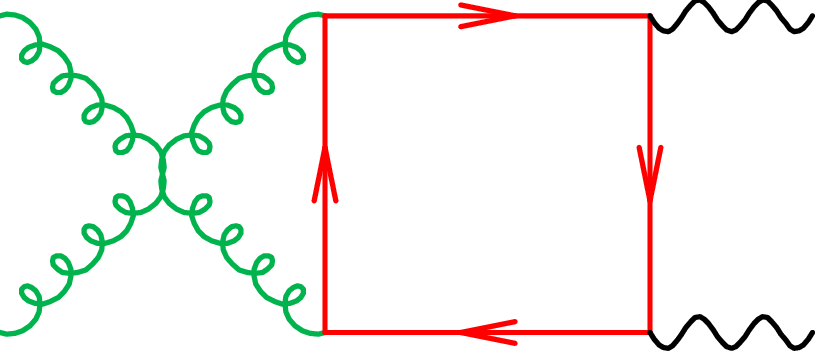}
		\caption{}
	\end{subfigure}
	\caption{Representative diagrams for the LO and virtual NLO $gg\to ZZ$ amplitude.}
	\label{fig:ggZZ_NLO_amps}
\end{figure}

After we set the stage in the previous chapters, including derivation
of known results for the single Higgs amplitudes and extending their
expansion to higher orders, we can now tackle the unknown QCD
corrections to $Z$ boson pair-production via massive quark loops in the
SM. Representative diagrams for the leading-order contribution are
illustrated in Fig.~\ref{sfig:ggZZ_LO_amp} and for the virtual
next-to-leading-order diagrams in Fig.~\ref{fig:ggZZ_NLO_amps}(b)-(f)
and Fig.~\ref{fig:twotriangles}, respectively.

These amplitudes were first studied for on-shell $Z$ bosons in
Ref.~\cite{Glover:1988rg}; more recently, the $Z$ decay and off-shell
effects were also calculated at
leading-order~\cite{Campbell:2013una}. Virtual
two-loop contributions with massless internal quark loops (and
subsequent $Z$ boson decay) became
available only recently\cite{Gehrmann:2014bfa,Cascioli:2014yka,Caola:2014iua,Gehrmann:2015ora,Caola:2015ila,vonManteuffel:2015msa}. Due
to the complexity of the computation and present technical limitations
no full two-loop correction to the amplitudes with \emph{massive}
internal quarks is presently known. The authors
of~\cite{Melnikov:2015laa} made the first attempt in approximating the
virtual NLO corrections with internal top quarks. Their results,
however, includes only the first term of the $1/m$ expansion. At this
order contributions from the vector coupling of the $Z$ bosons to the
quarks are neglected completely. This is not necessarily troubling
since the vector coupling contribution is $a_f/v_f\sim 2.5$ times
smaller than the axial coupling contribution.

However, to fully incorporate the physics of the $Z$ boson
interactions and to give an estimate of power corrections $s/m^2$ we
compute the virtual two-loop corrections up to
$\cO\left(\rt^{-7}\right)$. We keep the $Z$ bosons on-shell, sum over
their polarisations and project onto the tensor structure of the
$gg\to H\to ZZ$ amplitude
(\Eqno{eq:ggHZZ_amplitude_structure}) since we are only
interested in the interference of both.

This chapter is structured as follows: In Sec.~\ref{sub:ggZZ_preliminaries} we give our definitions of the SM $ZZ$ amplitude, as far as the conventions differ from Sec.~\ref{sub:ggH_preliminaries}. The leading-order and next-to-leading-order results are presented in Sec.~\ref{sub:ggZZ_large_mass_expansion_at_one_loop} and Sec.~\ref{sub:ggZZ_large_mass_expansion_at_two_loop}, respectively. The latter is divided into two parts; the first consists of diagrams where both $Z$ bosons couple to one fermion line and the second handles \emph{anomaly style} diagrams where a single $Z$ boson is connected to one fermion string.

\subsection{Preliminaries} 
\label{sub:ggZZ_preliminaries}

The on-shell $Z$ boson pair-production in gluon-gluon fusion
\begin{equation}
	g(p_1,\mu,A)+g(p_2,\nu,B) \to Z(p_3,\mz,\al)+Z(p_4,\mz,\be) \, ,
\end{equation}
via the heavy top quark loop can be completely expressed in terms of kinematical invariants
\begin{equation}
	p_3^2=\mz^2=p_4^2\,,\quad s = (p_1+p_2)^2\,,\quad t=(p_1-p_3)^2\,,\quad u=(p_2-p_3)^2 \text{  and  }\; s+t+u=2\mz^2 \, ,
\end{equation}
or equivalently, using the on-shellness condition, by the rescaled variables
\begin{align}
	r_t &= \frac{m^2}{s}\,,\quad r_Z = \frac{\mz^2}{s}\,,\quad
	x = \frac{\mz^2-t}{s} = \frac{p_1p_3}{p_1p_2}\;\;\text{and}\;\; \tilde{x} = (1-x)x\,.
	\label{eq:massiveggZZ_x_definition}
\end{align}
The SM continuum amplitudes
$\ket{\cB^{0,AB}_{\mu\nu\al\be}(\as^0,m^0,\mu,\ep)}$ admit the same
perturbative expansion as given in \Eqno{eq:ggH_amp_bare_expansion}
for the Higgs-mediated process. The bare amplitudes are renormalized
in accordance with Eqs.~(\ref{eq:ggH_renormalisation_definition}-\ref{eq:Zop_as}),
omitting the superfluous Higgs vertex renormalisation. As mentioned
earlier we project onto the tensor and color structure of the
Higgs-mediated amplitude (\Eqno{eq:ggHZZ_amplitude_structure})
with
\begin{equation}
	\label{eq:massiveggZZ_projected_amplitude}
	\ket{\cB^0(\as^0,m^0,\mu,\ep)} = \frac{\de^{AB}}{N_A} (g^{\mu\nu}\, p_1p_2-p_2^\mu p_1^\nu)\; P_Z^{\al\rho'}(p_3) P_{Z,\rho'}^{\be}(p_4) \ket{\cB^{0,AB}_{\mu\nu\al\be}(\as^0,m^0,\mu,\ep)}\,,
\end{equation}
where $N_A=N_c^2-1=8$ and $P_{Z,\al\be}(p)$ from \Eqno{Zpolsum}.\\
We shall consider a single quark of flavor $f$ to be circulating in the quark loop. The Standard Model coupling of this fermion to a $Z$ boson is given by,
\beq
-i \frac{\gW}{2 \cosW} \gamma^\mu \left(v_f - a_f \gamma_5\right),\;\;v_{f} = \tau_f -2 Q_f \sin^2 \theta_W,\;\;a_{f} = \tau_f,\;\;\tau_f =\pm \frac{1}{2}\, .
\eeq
The superposition of vector and axial coupling allows to write the scattering amplitude as
\begin{equation}
	\label{eq:massiveggZZ_normalisation}
	\ket{\cB^0(\as^0,m^0,\mu,\ep)} = \cN \; \left( v_f^2 \ket{\TW{\cB}_{VV}^0(\as^0,m^0,\mu,\ep)} + a_f^2 \ket{\TW{\cB}_{AA}^0(\as^0,m^0,\mu,\ep)} \right)\,,
\end{equation}
where we factored out the normalisation factor from \Eqno{eq:Ndef}. The mixed coupling structure $v_f a_f$ vanishes due to charge parity conservation. With the amplitudes outlined above it is straightforward to compute the interference.
\begin{align*}
	\label{eq:general_H_ZZ_interference}
	\cB_{gg} &= 2\Re\Bigl\{ \bra{\cA_{\al\be\ro'\si'}^{AB}(\as^{(n_l)},m,\mu,\ep)}\ket{\cB^{AB,\al\be\ro\si}(\as^{(n_l)},m,\mu,\ep)}\; P^{\ro'}_{Z,\ro}(p_3) P^{\si'}_{Z,\si}(p_4) \Bigl\} \\
	&= 2\Re\Bigl\{ \cN^* \frac{8}{3} \frac{s\, N_A}{s-\mh^2} \bra{\cA(\as^{(n_l)},m,\mu,\ep)}\ket{\cB(\as^{(n_l)},m,\mu,\ep)} \Bigl\} \numberthis \\
	&= |\cN|^2 \frac{16}{3} \frac{s\, N_A}{s-\mh^2} \; \Re\Bigl\{ \bra{\cA(\as^{(n_l)},m,\mu,\ep)} \left[v_f^2 \ket{\TW{\cB}_{VV}(\as^{(n_l)},m,\mu,\ep)} + a_f^2 \ket{\TW{\cB}_{AA}(\as^{(n_l)},m,\mu,\ep)} \right] \Bigl\}\,.
\end{align*}
Writing \Eqno{eq:general_H_ZZ_interference} in this way establishes that $\cA(\as^{(n_l)},m,\mu,\ep)$ and $\cB(\as^{(n_l)},m,\mu,\ep)$ are dimensionless quantities, i.e. we compute $\cB(\as^{(n_l)},m,\mu,\ep)$ for $s=1$ in the following.


\subsection{Projected Exact Result at One Loop} 
\label{sub:projected_exact_result_at_one_loop}

The leading-order amplitude for the SM continuum production of two $Z$
bosons is known exactly in $d=4-2\ep$ dimensions.  The usual
normalisation factor \Eqno{eq:loop_measure_factors} is chosen. We
split the result, according to \Eqno{eq:massiveggZZ_normalisation},
into vector-vector ($VV$) and axial-axial ($AA$) contribution.
\begin{align*}
	\label{eq:ggZZ_LO_exact_VV}
	\ket{\TW{\cB}_{VV}^{(1)}(r_t,\mu,\ep)} &= S_\ep\, c_\Gamma \cdot 2\Big\{
	4 \ep(1-\ep) B_{\{1,2\}} + 2\ep \left(B_{\{1,3\}} + B_{\{2,3\}} - 2 B_{\{3\}}\right) \numberthis \\
	&\hspace{-1cm}+ s C_{\{1,2\}} \left[ 8 r_t + 2\ep (1-4 r_t)- 2 \ep^2 \right]
	+ s C_{\{23,1\}} \left[2 (1-4 r_t-2 r_Z) (1-x) - 4\ep (1-r_Z) (1-x) \right] \\
	&\hspace{-1cm}+ s C_{\{12,3\}} \left[ \ep \left(2 (1-4 r_t-2 r_Z) -2\ep (1-2 r_Z) \right)  \right]
	+ s C_{\{1,3\}} \left[ 2 (1-4 r_t-2 r_Z) x - 4 \ep (1-r_Z) x \right] \\
	&\hspace{-1cm}+ s^2 D_{\{1,2,3\}} \left[ 4 r_t (1-2 r_t-r_Z) + \ep \left((1-4 r_t) (1-r_Z)-x \right) + \ep^2 \left(-1+r_Z+x \right) \right] \\
	&\hspace{-1cm}+ s^2 D_{\{2,1,3\}} \left[4 r_t (1-2 r_t-r_Z) + \ep \left( 4 r_t (-1+r_Z)-r_Z+x \right) + \ep^2 \left( r_Z-x \right) \right] \\
	&\hspace{-1cm}+ s^2 D_{\{1,3,2\}} \left[(1-4 r_t-2 r_Z) \left(2 r_t-r_Z+x-x^2\right) +\ep \left( 4 r_t (-1+r_Z)+(1-2 r_Z) (r_Z-(1-x) x) \right)\right. \\
	&\hspace{-1cm}+\left. \ep^2 \left(r_Z-(1-x) x \right) \right]
	\Big\}\,.
\end{align*}

\begin{align*}
	\label{eq:ggZZ_LO_exact_AA}
	\ket{\TW{\cB}_{AA}^{(1)}(r_t,\mu,\ep)} &= \ket{\TW{\cB}_{VV}^{(1)}(r_t,\mu,\ep)} + S_\ep\, c_\Gamma \cdot 2r_t \Big\{  
	s C_{\{1,2\}} \left[(2-4 r_Z)/r_Z^2 \right] \numberthis \\
	&\hspace{-1cm}+ s C_{\{23,1\}} \left[ 4(1-6 r_Z)(-1+x)/r_Z - 16 \ep (1-x) \right]
	+ s C_{\{12,3\}} \left[ \ep \left(24+(2-8 r_Z)/r_Z^2 -16 \ep \right) \right] \\
	&\hspace{-1cm}+ s C_{\{1,3\}} \left[4 \left(6-1/r_Z\right) x -16 \ep x  \right] \\
	&\hspace{-1cm}+ s^2 D_{\{1,2,3\}} \left[ -4+24 r_t+ 2 r_t/r_Z^2- 8 r_t/r_Z + \ep \left( 10-16 r_t+(1-x)/r_Z^2 - (3-2 x)/r_Z \right) -4\ep^2 \right] \\
	&\hspace{-1cm}+ s^2 D_{\{2,1,3\}} \left[-4+24 r_t+2 r_t/r_Z^2 - 8 r_t/r_Z + \ep \left(10-16 r_t - (1+2 x)/r_Z + x/r_Z^2 \right) - 4\ep^2 \right] \\
	&\hspace{-1cm}+ s^2 D_{\{1,3,2\}} \left[ 2 r_t/r_Z^2 - 12 r_Z - (8 r_t+2(1-x) x)/r_Z+2 \
\left(1+12 r_t+6 x-6 x^2\right) \right. \\
	&\hspace{-1cm}+\left. \ep \left( 8 r_Z - 2 \left(8 r_t-(1-2 x)^2\right)+ (1-x)x/r_Z^2 - (1-2 x+2 x^2)/r_Z \right) -4 \ep^2 \right]
	\Big\}\,.
\end{align*}
The notation for the scalar integrals $B,C$ and $D$ is given in Table~\ref{tab:ggZZ_int_def}.
\begin{table}
\begin{center}
\begin{tabular}{|l|l||l|l||l|l|}
\hline
$D_{\{1,2,3\}}$    &$D_0(q_1,q_2,q_3;m,m,m,m)$ &$C_{\{1,2\}}$	&$C_0(q_1,q_2;m,m,m)$ 		& $B_{\{1,2\}}$   &$B_0(q_{12};m,m)$      \\
$D_{\{1,3,2\}}$    &$D_0(q_1,q_3,q_2;m,m,m,m)$ &$C_{\{1,3\}}$	&$C_0(q_1,q_3;m,m,m)$ 		& $B_{\{1,3\}}$   &$B_0(q_{13};m,m)$      \\
$D_{\{2,1,3\}}$    &$D_0(q_2,q_1,q_3;m,m,m,m)$ &$C_{\{12,3\}}$	&$C_0(q_{12},q_3;m,m,m)$   	& $B_{\{2,3\}}$   &$B_0(q_{23};m,m)$\\
  				&                              &$C_{\{23,1\}}$  &$C_0(q_{23},q_1;m,m,m)$   & $B_{\{3\}}$		&$B_0(q_3;m,m)$\\
\hline
\end{tabular}
\end{center}
\caption{Scalar integrals occurring in full LO SM continuum $ZZ$ production. The momenta are defined as $q_1=p_1, q_2=p_2, q_3=-p_3$ and $q_{ij} = q_i+q_j$. The scalar integrals are defined in appendix~\ref{Intdef}.}
\label{tab:ggZZ_int_def}
\end{table}
We re-introduced factors of $s$ in \Eqno{eq:ggZZ_LO_exact_VV} and \Eqno{eq:ggZZ_LO_exact_AA} to indicate the correct dimensionality of the expressions.
We note that, in contrast to the case where the $Z$ bosons are off-shell and their decays included,
these formulae for the interference take a very simple form.  Eq.~(\ref{eq:ggZZ_LO_exact_VV},\ref{eq:ggZZ_LO_exact_AA})
extend the results of Ref~\cite{Campbell:2014gua} to include the terms of order $\epsilon^1$ and $\epsilon^2$.


\subsection{Large-Mass Expansion at One Loop} 
\label{sub:ggZZ_large_mass_expansion_at_one_loop}

Equivalently, \Eqno{eq:ggZZ_LO_exact_VV} and \Eqno{eq:ggZZ_LO_exact_AA} can be expressed by means of the large-mass expansion. The result for the vector-vector part yields
\begin{align*}
	\label{eq:LME_ggZZ_1L_VV}
	\ket{\TW{\cB}_{VV}^{(1)}(r_t,\mu,\ep)} &= S_\ep\, c_\Gamma\,\left(\frac{\mu^2}{m^2}\right)^\ep \Bigg\{ 
	\frac{1}{r_t^2} \left[ \frac{1}{10}-\frac{r_Z}{5}+\ep^2 \left(-\frac{1}{10}+\frac{4 r_Z}{15}-\frac{\tilde{x}}{15}\right)+\ep \left(\frac{1}{15}-\frac{17 r_Z}{45}+\frac{11 \tilde{x}}{45}\right) \right] \\
	&\hspace{-2cm}+\frac{1}{r_t^3} \left[ \frac{2}{315}+\frac{r_Z}{21}-\frac{4 r_Z^2}{35}-\frac{4 \tilde{x}}{315}+\ep^2 \left(-\frac{4}{315}-\frac{149 r_Z}{3780}+\frac{17 r_Z^2}{315}+\left(\frac{143}{3780}+\frac{2 r_Z}{45}\right) \tilde{x}\right) \right. \\
	&\hspace{-2cm}+ \left. \ep \left(-\frac{1}{105}+\frac{61 r_Z}{945}-\frac{68 r_Z^2}{315}+\left(\frac{37}{1890}+\frac{8 r_Z}{63}\right) \tilde{x}\right) \right] 
	+\frac{1}{r_t^4} \left[ \frac{1}{1080}-\frac{r_Z}{1260}+\frac{41 r_Z^2}{1890}-\frac{r_Z^3}{21} \right. \\
	&\hspace{-2cm}+ \left(\frac{1}{945}-\frac{2 r_Z}{315} \right) \tilde{x}+\ep \left(\frac{131}{45360}-\frac{61 r_Z}{4200}+\frac{2171 r_Z^2}{56700}-\frac{59 r_Z^3}{630} + \left(-\frac{47}{28350}+\frac{319 r_Z}{18900}+\frac{16 r_Z^2}{315}\right) \tilde{x} \right. \\
	&\hspace{-2cm}- \left.\left. \frac{13 \tilde{x}^2}{2700} \right) + \ep^2 \left(\frac{1}{1008}-\frac{31 r_Z}{3240}-\frac{7 r_Z^2}{810}-\frac{2 r_Z^3}{945}+\left(-\frac{7}{1620}+\frac{659 r_Z}{22680}+\frac{37 r_Z^2}{945}\right) \tilde{x}-\frac{43 \tilde{x}^2}{22680}\right) \right] \\
	&\hspace{-2cm}+ \frac{1}{r_t^5} \left[ \frac{4}{17325}+\frac{r_Z}{2475}-\frac{43 r_Z^2}{20790}+\frac{r_Z^3}{110}-\frac{4 r_Z^4}{231}+\left(-\frac{1}{2079}+\frac{13 r_Z}{20790}-\frac{r_Z^2}{693}\right) \tilde{x}+\frac{ \tilde{x}^2}{2310} \right. \numberthis \\
	&\hspace{-2cm}+\ep \left(-\frac{1}{4725}+\frac{r_Z}{330}-\frac{2671 r_Z^2}{249480}+\frac{2533 r_Z^3}{138600}-\frac{53 r_Z^4}{1485}+\left( \frac{349}{311850}-\frac{733 r_Z}{178200}+\frac{67 r_Z^2}{5940}+\frac{188 r_Z^3}{10395}\right) \tilde{x} \right. \\
	&\hspace{-2cm}+ \left. \left(\frac{37}{59400}-\frac{4 r_Z}{825}\right) \tilde{x}^2 \right)+\ep^2 \left(-\frac{67}{103950}+\frac{73 r_Z}{32400}-\frac{16871 r_Z^2}{2494800}+\frac{323 r_Z^3}{831600}-\frac{67 r_Z^4}{8910} \right. \\
	&\hspace{-2cm}+\left. \left. \left( \frac{5939}{2494800}-\frac{451 r_Z}{75600}+\frac{839 r_Z^2}{51975}+\frac{611 r_Z^3}{31185}\right) \tilde{x}+\left(-\frac{2083}{2494800}-\frac{61 r_Z}{17325}\right) \tilde{x}^2\right) \right] \\
	&\hspace{-2cm}+ \frac{1}{r_t^6} \left[ \frac{1}{108108}+\frac{5 r_Z}{54054}+\frac{569 r_Z^2}{1801800}-\frac{163 r_Z^3}{108108}+\frac{7 r_Z^4}{1980}-\frac{5 r_Z^5}{858}+\left(\frac{1}{9009}-\frac{19 r_Z}{42900}-\frac{r_Z^2}{90090}+\frac{4 r_Z^3}{19305}\right) \tilde{x} \right. \\
	&\hspace{-2cm}+ \left(-\frac{97}{600600}+\frac{29 r_Z}{60060}\right) \tilde{x}^2 +\ep^2 \left( \frac{191}{1853280}-\frac{28507 r_Z}{32432400}+\frac{444149 r_Z^2}{216216000}-\frac{263839 r_Z^3}{64864800}+\frac{301471 r_Z^4}{194594400} \right.\\
	&\hspace{-2cm}-\frac{5743 r_Z^5}{1297296} + \left(-\frac{8269}{16216200}+\frac{82241 r_Z}{24948000}-\frac{23113 r_Z^2}{4633200}+\frac{11041 r_Z^3}{1389960}+\frac{5185 r_Z^4}{648648}\right) \tilde{x} +\frac{46 \tilde{x}^3}{405405} \\ 
	&\hspace{-2cm}+ \left. \left(\frac{98009}{216216000}-\frac{58703 r_Z}{64864800}-\frac{9979 r_Z^2}{3243240}\right) \tilde{x}^2 \right) + \ep \left( \frac{617}{6486480}-\frac{13037 r_Z}{22702680}+\frac{1785391 r_Z^2}{756756000}-\frac{274301 r_Z^3}{45405360} \right.\\
	&\hspace{-2cm}+ \frac{19199 r_Z^4}{2494800}-\frac{19 r_Z^5}{1512} + \left(-\frac{53}{291060}+\frac{60449 r_Z}{34398000}-\frac{153919 r_Z^2}{37837800}+\frac{137 r_Z^3}{21450}+ \frac{25 r_Z^4}{4158}\right) \tilde{x} \\ 
	&\hspace{-2cm}+ \left. \left. \left(-\frac{1091}{19404000}+\frac{47153 r_Z}{75675600}-\frac{167 r_Z^2}{54054}\right) \tilde{x}^2+\frac{29 \tilde{x}^3}{189189}\right) \right] + \cO\left(1/r_t^7,\ep^3\right) \Bigg\}\,.
\end{align*}
The result for the axial-axial part is
\begin{align*}
	\label{eq:LME_ggZZ_1L_AA}
	\ket{\TW{\cB}_{AA}^{(1)}(r_t,\mu,\ep)} &= S_\ep\,c_\Gamma\, \left( \frac{\mu^2}{m^2} \right)^\ep \Bigg\{
	\frac{1}{r_t} \left[ -2-\frac{1}{6 r_Z^2}+\frac{2}{3 r_Z}+ \ep^2 \left(\frac{4}{3}+\frac{1}{6 r_Z^2}-\frac{1}{r_Z}+\left(\frac{1}{3 r_Z^2}+\frac{2}{3 r_Z}\right) \tilde{x}\right) \right.\\
	&\hspace{-2cm}+\left. \ep \left(2-\frac{1}{3 r_Z}+\left(\frac{1}{3 r_Z^2}+\frac{2}{3 r_Z}\right) \tilde{x}\right)\right]
	+ \frac{1}{r_t^2} \left[ \frac{7}{30}+\frac{1}{90 r_Z^2}-\frac{7}{90 r_Z}-\frac{3 r_Z}{5}+\left(-\frac{1}{30 r_Z^2}+\frac{1}{15 r_Z}\right) \tilde{x} \right.\\
	&\hspace{-2cm}+ \ep^2 \left(-\frac{31}{90}-\frac{1}{90 r_Z^2}+\frac{17}{180 r_Z}+\frac{7 r_Z}{9}+\left(\frac{2}{9}+\frac{1}{60 r_Z^2}\right) \tilde{x}\right)+\ep \left( \frac{1}{180 r_Z^2}-\frac{1}{36 r_Z}+\frac{4 r_Z}{45} \right. \numberthis \\ 
	&\hspace{-2cm}+ \left.\left. \left(\frac{17}{45}-\frac{1}{36 r_Z^2}+\frac{7}{90 r_Z} \right) \tilde{x} \right)\right]
	+ \frac{1}{r_t^3} \left[ -\frac{13}{210}-\frac{1}{280 r_Z^2}+\frac{13}{630 r_Z}+\frac{2 r_Z}{21}-\frac{6 r_Z^2}{35}+\left( \frac{4}{63}+\frac{1}{126 r_Z^2} \right.\right. \\
	&\hspace{-2cm}-\left. \frac{13}{315 r_Z} \right) \tilde{x} + \ep^2 \left( \frac{149}{1890}+\frac{1}{336 r_Z^2}-\frac{403}{15120 r_Z}-\frac{2 r_Z}{15}+\frac{23 r_Z^2}{105}+\left( -\frac{71}{1260}+\frac{43}{15120 r_Z^2}+\frac{337}{7560 r_Z} \right.\right.\\
	&\hspace{-2cm}+ \left.\left.\frac{32 r_Z}{315} \right) \tilde{x} + \left(-\frac{11}{840 r_Z^2}-\frac{11}{420 r_Z}\right) \tilde{x}^2 \right) +\ep \left( -\frac{101}{3780}-\frac{1}{336 r_Z^2}+\frac{13}{1080 r_Z}+\frac{4 r_Z}{135}-\frac{17 r_Z^2}{315} \right. \\
	&\hspace{-2cm}+  \left.\left. \left(\frac{73}{1890}+\frac{89}{7560 r_Z^2}-\frac{109}{3780 r_Z} + \frac{8 r_Z}{63} \right) \tilde{x}+\left(-\frac{1}{140 r_Z^2}-\frac{1}{70 r_Z}\right) \tilde{x}^2 \right) \right]
	+\frac{1}{r_t^4} \left[ \frac{517}{37800}+\frac{2}{4725 r_Z^2} \right. \\
	&\hspace{-2cm}- \frac{2}{525 r_Z}-\frac{13 r_Z}{420}+\frac{13 r_Z^2}{378}-\frac{r_Z^3}{21} + \left(-\frac{32}{945}-\frac{2}{945 r_Z^2}+\frac{4}{315 r_Z}+\frac{4 r_Z}{105}\right) \tilde{x}+\left(\frac{1}{945 r_Z^2}-\frac{2}{945 r_Z}\right) \tilde{x}^2 \\
	&\hspace{-2cm}+ \ep \left( \frac{127}{11340}+\frac{13}{28350 r_Z^2}-\frac{1}{270 r_Z}-\frac{23 r_Z}{1050} + \frac{1163 r_Z^2}{56700}-\frac{19 r_Z^3}{630}+\left(-\frac{257}{11340}-\frac{11}{4050 r_Z^2}+\frac{19}{1350 r_Z} \right.\right.\\
	&\hspace{-2cm}+ \left.\left. \frac{71 r_Z}{2700}+\frac{4 r_Z^2}{105} \right) \tilde{x}+\left(-\frac{283}{18900}+\frac{167}{56700 r_Z^2}-\frac{17}{4050 r_Z}\right) \tilde{x}^2 \right) + \ep^2 \left( -\frac{6829}{453600}-\frac{1}{3780 r_Z^2}+\frac{37}{11340 r_Z} \right. \\
	&\hspace{-2cm}+ \frac{863 r_Z}{25200}-\frac{811 r_Z^2}{18900}+\frac{17 r_Z^3}{315}+\left( \frac{568}{14175}+\frac{1}{2268 r_Z^2}-\frac{11}{1512 r_Z}-\frac{4157 r_Z}{113400} + \frac{4 r_Z^2}{105} \right) \tilde{x}  \\
	&\hspace{-2cm}+ \left.\left. \left( -\frac{401}{16200} + \frac{1}{324 r_Z^2} - \frac{1}{378 r_Z}\right) \tilde{x}^2 \right) \right] 
	+\frac{1}{r_t^5} \left[ -\frac{767}{207900}-\frac{1}{8316 r_Z^2}+\frac{13}{13860 r_Z}+\frac{1699 r_Z}{207900}-\frac{31 r_Z^2}{2310} \right. \\ 
	&\hspace{-2cm}+ \frac{8 r_Z^3}{693}-\frac{r_Z^4}{77}+\left(\frac{83}{6930}+\frac{1}{1980 r_Z^2}-\frac{13}{3465 r_Z}-\frac{437 r_Z}{20790}+\frac{r_Z^2}{55} \right) \tilde{x}+\left(-\frac{1}{315}-\frac{1}{1980 r_Z^2}+\frac{1}{420 r_Z}\right) \tilde{x}^2 \\
	&\hspace{-2cm}+ \ep \left( -\frac{9299}{2494800}-\frac{1}{6480 r_Z^2}+\frac{871}{831600 r_Z}+\frac{19651 r_Z}{2494800}-\frac{5209 r_Z^2}{415800}+\frac{167 r_Z^3}{17325}-\frac{43 r_Z^4}{3780}+\left( \frac{5323}{415800} + \frac{223}{277200 r_Z^2} \right.\right. \\
	&\hspace{-2cm}-\left. \frac{244}{51975 r_Z}-\frac{21121 r_Z}{1247400}+\frac{3271 r_Z^2}{207900}+\frac{113 r_Z^3}{10395} \right) \tilde{x}+\left(-\frac{383}{207900}-\frac{313}{277200 r_Z^2}+\frac{349}{92400 r_Z}-\frac{53 r_Z}{5775}\right) \tilde{x}^2 \\
	&\hspace{-2cm}+\left. \left(\frac{1}{4158 r_Z^2}+\frac{1}{2079 r_Z}\right) \tilde{x}^3 \right)+\ep^2 \left( \frac{16273}{4989600}+\frac{7}{142560 r_Z^2}-\frac{923}{1247400 r_Z}-\frac{3497 r_Z}{453600}+\frac{30463 r_Z^2}{2494800}\right. \\
	&\hspace{-2cm}-\frac{305 r_Z^3}{24948} + \frac{3089 r_Z^4}{249480}+\left(-\frac{1493}{166320}+\frac{47}{311850 r_Z^2}+\frac{4507}{2494800 r_Z}+\frac{5191 r_Z}{226800}-\frac{19627 r_Z^2}{1247400}+\frac{227 r_Z^3}{17820}\right) \tilde{x} \\
	&\hspace{-2cm}+ \left.\left. \left(\frac{5161}{1247400}-\frac{4451}{4989600 r_Z^2}+\frac{1123}{1663200 r_Z}-\frac{3187 r_Z}{207900}\right) \tilde{x}^2+\left(\frac{137}{249480 r_Z^2}+\frac{137}{124740 r_Z} \right) \tilde{x}^3 \right) \right] \\
	&\hspace{-2cm}+ \frac{1}{r_t^6} \left[ \frac{2603}{2910600}+\frac{1}{56056 r_Z^2}-\frac{1}{5096 r_Z}-\frac{1223 r_Z}{491400}+\frac{22381 r_Z^2}{5405400}-\frac{34 r_Z^3}{6435}+\frac{19 r_Z^4}{5148}-\frac{r_Z^5}{286}+\left( -\frac{1019}{257400} \right.\right. \\
	&\hspace{-2cm}-\left. \frac{1}{8008 r_Z^2} + \frac{941}{900900 r_Z}+\frac{11447 r_Z}{1351350}-\frac{5897 r_Z^2}{540540}+\frac{293 r_Z^3}{38610} \right) \tilde{x}+\left(\frac{5167}{1801800}+\frac{167}{900900 r_Z^2}-\frac{697}{600600 r_Z} \right.\\
	&\hspace{-2cm}- \left. \frac{493 r_Z}{180180} \right) \tilde{x}^2 + \left(-\frac{1}{24024 r_Z^2}+\frac{1}{12012 r_Z}\right) \tilde{x}^3 + \ep^2 \left( -\frac{50693}{87318000}-\frac{1}{288288 r_Z^2} + \frac{59}{720720 r_Z} \right. \\
	&\hspace{-2cm}+ \frac{2049041 r_Z}{1135134000}-\frac{3861083 r_Z^2}{1238328000} + \frac{5078077 r_Z^3}{1362160800}-\frac{616361 r_Z^4}{194594400}+\frac{17341 r_Z^5}{6486480}+ \left( \frac{863221}{378378000}-\frac{1}{51480 r_Z^2} \right.\\
	&\hspace{-2cm}- \left. \frac{2159}{11583000 r_Z}-\frac{41229697 r_Z}{6810804000}+\frac{6865009 r_Z^2}{681080400} - \frac{26227 r_Z^3}{4864860}+\frac{12911 r_Z^4}{3243240} \right) \tilde{x}+\left( -\frac{651821}{412776000}+\frac{2917}{11583000 r_Z^2} \right.\\
	&\hspace{-2cm}- \left.\left. \frac{1741}{3861000 r_Z}+\frac{162983 r_Z}{34927200}-\frac{24883 r_Z^2}{3243240} \right) \tilde{x}^2 + \left(\frac{36731}{22702680}-\frac{43}{154440 r_Z^2}+\frac{1}{5616 r_Z}\right) \tilde{x}^3 \right) \\
	&\hspace{-2cm}+ \ep \left( \frac{2523253}{2270268000}+\frac{29}{1121120 r_Z^2}-\frac{23}{86240 r_Z}-\frac{932231 r_Z}{324324000} + \frac{503059 r_Z^2}{108108000}-\frac{674147 r_Z^3}{113513400}+\frac{3541 r_Z^4}{926640} \right. \\
	&\hspace{-2cm}- \frac{4051 r_Z^5}{1081080}+\left( -\frac{3717937}{756756000}-\frac{223}{1121120 r_Z^2}+\frac{63961}{42042000 r_Z}+\frac{908203 r_Z}{94594500} - \frac{802811 r_Z^2}{75675600}+\frac{3889 r_Z^3}{491400} \right. \\
	&\hspace{-2cm}+ \left. \frac{163 r_Z^4}{54054} \right) \tilde{x} + \left(\frac{862991}{252252000}+\frac{48721}{126126000 r_Z^2} -\frac{164711}{84084000 r_Z}-\frac{7159 r_Z}{5821200}-\frac{239 r_Z^2}{54054}\right) \tilde{x}^2 \\
	&\hspace{-2cm}+\left.\left. \left(\frac{283}{378378}-\frac{1789}{10090080 r_Z^2}+\frac{1009}{5045040 r_Z}\right) \tilde{x}^3 \right) \right]
	+ \cO\left(1/r_t^7,\ep^3\right) \Bigg\} \,.
\end{align*}
The leading term in the vector-vector expansion is sub-dominant with respect to the axial-axial part. The reason for this difference has been given in~\cite{Melnikov:2015laa}.

\subsection{Large-Mass Expansion at Two Loops} 
\label{sub:ggZZ_large_mass_expansion_at_two_loop}

The two-loop SM continuum amplitude consists in total of $93+16$
non-zero diagrams. $93$ diagrams belong to topologies where both $Z$
bosons couple to the same fermion string, as illustrated in
Fig.~\ref{fig:ggZZ_NLO_amps}. Due to momentum conservation and
assuming an anti-commuting $\ga_5$ in $d$-dimensions, no $\ga_5$
contribution arises in the fermion traces of the respective
diagrams. The large-mass expansion results for the vector-vector and
axial-axial part of these diagrams are shown in
Sec.~\ref{ssub:non_anomalous_diagrams}.

The remaining $16$ \emph{anomaly style} diagrams belong to the
topology shown in Fig.~\ref{fig:twotriangles}, where the $Z$ bosons
couple to distinct fermion lines. These diagrams must, in principle,
be handled with care when using dimensional regularisation due to the
non-conservation of the axial-current. Furthermore, contributions from
each quark-doublet have to be considered simultaneously. Only the sum
over one quark-doublet leads to a gauge anomaly free theory. In case
of massless quark doublets all contributions vanish and we only have
to consider the third-generation quark doublet, i.e. top and bottom
quarks. Results for these diagrams are presented in
Sec.~\ref{ssub:anomalous_diagrams}.

\subsubsection{Non-Anomalous Diagrams} 
\label{ssub:non_anomalous_diagrams}

In this section we give explicit formulae for the large-mass expansions for the sum of the $93$ anomaly free diagrams. Including again only mass renormalisation, setting $\ket{\bar{\cB}_{XX}^{(1)}(r_t,\mu,\ep)} = \left(S_\ep c_\Gamma \,(\mu^2/m^2)^\ep \,\right)^{-1} \ket{\TW{\cB}_{XX}^{(1)}(r_t,\mu,\ep)} $ and $\log(-r_t)=\log(m^2/(-s-i\ep))$, we can write the divergent two-loop $VV$ part as
\begin{align*}
	\label{eq:LME_ggZZ_2L_VV}
	\ket{\TW{\cB}_{VV}^{(2)}(r_t,\mu,\ep)} &= \left( S_\ep\, c_\Gamma\,\left(\frac{\mu^2}{m^2}\right)^\ep \right)^2 \Bigg\{ C_A \Bigg[ \left(-\frac{2}{\ep^2}\left(\frac{m^2}{-s-i\ep}\right)^\ep + \frac{\pi^2}{3} \right)\ket{\bar{\cB}_{VV}^{(1)}(r_t,\mu,\ep)} \\
	&\hspace{-2cm}+ \frac{1}{r_t^2} \left[\frac{251}{540}-\frac{317 r_Z}{270}-\frac{11 \tilde{x}}{135} \right] 
	+\frac{1}{r_t^3} \left[ -\frac{158129}{1587600}+\frac{127817 r_Z}{396900}-\frac{5563 r_Z^2}{9450}+\left(\frac{22558}{99225}-\frac{8 r_Z}{189}\right) \tilde{x} \right.\\
	&\hspace{-2cm}+ \left. \log(-r_t) \left(-\frac{4}{315}-\frac{8 r_Z}{315}+\frac{8 \tilde{x}}{105} \right) \right]
	+\frac{1}{r_t^4} \left[ \frac{132779}{9525600}-\frac{10421 r_Z}{119070}+\frac{4411999 r_Z^2}{23814000}-\frac{14521 r_Z^3}{66150} \right. \\
	&\hspace{-2cm}+ \left.\left(-\frac{252937}{5953500}+\frac{260483 r_Z}{1701000}-\frac{16 r_Z^2}{945}\right) \tilde{x}+\frac{13 \tilde{x}^2}{20250}+\log(-r_t) \left(\frac{1}{945}-\frac{4 r_Z}{945}-\frac{4 r_Z^2}{315}+\left(-\frac{2}{315}+\frac{4 r_Z}{105}\right) \tilde{x}\right) \right] \\
	&\hspace{-2cm}+\frac{1}{r_t^5} \left[ -\frac{19803283}{5762988000}+\frac{93293203 r_Z}{5762988000}-\frac{61920091 r_Z^2}{1152597600}+\frac{259936363 r_Z^3}{2881494000}-\frac{236332 r_Z^4}{3274425}+\left( \frac{3048977}{209563200} \right.\right. \\
	&\hspace{-2cm}- \left. \frac{55307339 r_Z}{1280664000}+\frac{4889447 r_Z^2}{68607000}-\frac{188 r_Z^3}{31185} \right) \tilde{x}+\left(-\frac{32556823}{2881494000}+\frac{8 r_Z}{12375}\right) \tilde{x}^2+\log(-r_t) \left( -\frac{17}{41580} \right. \\
	&\hspace{-2cm}- \left.\left. \frac{r_Z}{2970}+\frac{r_Z^2}{4158}-\frac{2 r_Z^3}{693}+\left(\frac{1}{315}-\frac{2 r_Z}{693}+\frac{2 r_Z^2}{231}\right) \tilde{x}-\frac{5 \tilde{x}^2}{1386} \right) \right]
	+\frac{1}{r_t^6} \left[ \frac{132076261729}{204528444120000} \right. \numberthis \\
	&\hspace{-2cm}- \frac{857498948879 r_Z}{204528444120000}+\frac{16366567901 r_Z^2}{1377295920000}-\frac{1148974648051 r_Z^3}{40905688824000}+\frac{22637379733 r_Z^4}{584366983200}-\frac{165500519 r_Z^5}{7491884400} \\
	&\hspace{-2cm}+ \left( -\frac{101592736891}{25566055515000} + \frac{32900707079 r_Z}{1826146822500}-\frac{57230507981 r_Z^2}{2065943880000}+\frac{2366153189 r_Z^3}{83480997600}-\frac{25 r_Z^4}{12474}\right) \tilde{x} \\
	&\hspace{-2cm}+ \left(\frac{341063392349}{68176148040000}-\frac{1401690458627 r_Z}{102264222060000}+\frac{167 r_Z^2}{405405}\right) \tilde{x}^2-\frac{29 \tilde{x}^3}{2648646}+\log(-r_t) \left( \frac{101}{1201200}-\frac{23 r_Z}{163800} \right.\\
	&\hspace{-2cm}-\left.\left. \frac{1907 r_Z^2}{1801800}+\frac{361 r_Z^3}{540540}+\frac{8 r_Z^4}{19305}+\left(-\frac{697}{900900}+\frac{2393 r_Z}{900900}+\frac{31 r_Z^2}{30030}-\frac{8 r_Z^3}{6435}\right) \tilde{x}+\left(\frac{97}{72072}-\frac{145 r_Z}{36036}\right) \tilde{x}^2 \right) \right] \Bigg] \\
	&\hspace{-2cm}+ C_F \Bigg[ 
	\frac{1}{r_t^2} \left[ \frac{83}{54}-\frac{83 r_Z}{27} \right]
	+\frac{1}{r_t^3} \left[ \frac{1271}{5400}+\frac{2297 r_Z}{2700}-\frac{1627 r_Z^2}{675}-\frac{841 \tilde{x}}{2025} \right] 
	+\frac{1}{r_t^4} \left[ \frac{18997}{1587600}+\frac{46231 r_Z}{793800} \right. \\
	&\hspace{-2cm}+ \left. \frac{1859807 r_Z^2}{3969000}-\frac{1516 r_Z^3}{1225}+\left(\frac{42367}{567000}-\frac{11663 r_Z}{36750}\right) \tilde{x} \right]
	+\frac{1}{r_t^5} \left[ \frac{533671}{47628000}-\frac{41471 r_Z}{5103000}-\frac{28573 r_Z^2}{1984500} \right. \\
	&\hspace{-2cm}+ \left. \frac{16657657 r_Z^3}{71442000}-\frac{312829 r_Z^4}{595350}+\left(-\frac{2963}{105840}+\frac{278843 r_Z}{3572100}-\frac{833081 r_Z^2}{5953500}\right) \tilde{x}+\frac{229471 \tilde{x}^2}{11907000} \right] \\
	&\hspace{-2cm}+ \frac{1}{r_t^6} \left[ -\frac{100790441}{302556870000}+\frac{10415465807 r_Z}{1210227480000}-\frac{2539049 r_Z^2}{488980800}-\frac{3259592869 r_Z^3}{121022748000}+\frac{904486537 r_Z^4}{8644482000} \right. \\
	&\hspace{-2cm}- \frac{1175777 r_Z^5}{5880600}+\left(\frac{3025021753}{403409160000}-\frac{513331337 r_Z}{14407470000}+\frac{1216539691 r_Z^2}{27505170000}-\frac{10216127 r_Z^3}{246985200}\right) \tilde{x} \\ 
	&\hspace{-2cm}+ \left. \left(-\frac{10969455173}{1210227480000}+\frac{3151546457 r_Z}{121022748000}\right) \tilde{x}^2 \right] \Bigg] 
	+ \cO\left(1/r_t^7,\ep\right) \Bigg\}\,.
\end{align*}

And the $AA$ part as
\begin{align*}
	\label{eq:LME_ggZZ_2L_AA}
	\ket{\TW{\cB}_{AA}^{(2)}(r_t,\mu,\ep)} &= \left( S_\ep\, c_\Gamma\,\left(\frac{\mu^2}{m^2}\right)^\ep \right)^2 \Bigg\{ C_A \Bigg[ \left(-\frac{2}{\ep^2}\left(\frac{m^2}{-s-i\ep}\right)^\ep + \frac{\pi^2}{3} \right)\ket{\bar{\cB}_{AA}^{(1)}(r_t,\mu,\ep)} \\
	&\hspace{-2cm}+ \frac{1}{r_t} \left[ -6-\frac{5}{9 r_Z^2}+\frac{25}{9 r_Z}+\left(-\frac{1}{9 r_Z^2}-\frac{2}{9 r_Z}\right) \tilde{x} \right]
	+ \frac{1}{r_t^2} \left[ \frac{3563}{2700}-\frac{17}{450 r_Z^2}-\frac{761}{2700 r_Z}-\frac{367 r_Z}{270}+\left( -\frac{17}{135} \right.\right. \\
	&\hspace{-2cm}+ \left.\left. \frac{47}{108 r_Z^2}-\frac{1127}{1350 r_Z} \right) \tilde{x}+\log(-r_t) \left(\frac{2}{15}-\frac{1}{30 r_Z^2}+\left(\frac{1}{5 r_Z^2}-\frac{2}{5 r_Z}\right) \tilde{x}\right) \right]
	+\frac{1}{r_t^3} \left[ -\frac{27023}{176400}+\frac{5}{294 r_Z^2} \right. \\
	&\hspace{-2cm}- \frac{23623}{793800 r_Z}+\frac{8062 r_Z}{14175}-\frac{37 r_Z^2}{135}+\left( -\frac{142847}{198450}-\frac{2419}{22050 r_Z^2}+\frac{28793}{56700 r_Z}-\frac{8 r_Z}{189} \right) \tilde{x}+\left( \frac{1}{1050 r_Z^2} \right. \\
	&\hspace{-2cm}+ \left.\left. \frac{1}{525 r_Z} \right) \tilde{x}^2+\log(-r_t) \left(-\frac{2}{105}+\frac{1}{126 r_Z^2}-\frac{8}{315 r_Z}+\frac{8 r_Z}{63}+\left(-\frac{8}{21}-\frac{1}{21 r_Z^2}+\frac{26}{105 r_Z}\right) \tilde{x}\right) \right] \\
	&\hspace{-2cm}+ \frac{1}{r_t^4} \left[ -\frac{60007}{4762800}-\frac{1322471}{285768000 r_Z^2}+\frac{2690033}{142884000 r_Z}-\frac{100241 r_Z}{1190700}+\frac{5018071 r_Z^2}{23814000}-\frac{101 r_Z^3}{2100} \right. \numberthis \\
	&\hspace{-2cm}+ \left(\frac{312127}{793800}+\frac{321799}{10206000 r_Z^2}-\frac{736874}{4465125 r_Z}-\frac{1195489 r_Z}{2976750}-\frac{4 r_Z^2}{315}\right) \tilde{x}+\left( \frac{283}{141750}-\frac{609137}{28576800 r_Z^2} \right.\\
	&\hspace{-2cm}+ \left. \frac{2876437}{71442000 r_Z} \right) \tilde{x}^2+\log(-r_t) \left( -\frac{1}{135}-\frac{11}{5670 r_Z^2}+\frac{23}{2835 r_Z}-\frac{2 r_Z}{63}+\frac{8 r_Z^2}{105}+\left( \frac{188}{945}+\frac{38}{2835 r_Z^2} \right.\right.\\
	&\hspace{-2cm}- \left.\left.\left. \frac{214}{2835 r_Z}-\frac{8 r_Z}{35} \right) \tilde{x}+\left(-\frac{5}{567 r_Z^2}+\frac{10}{567 r_Z}\right) \tilde{x}^2 \right) \right]
	+\frac{1}{r_t^5} \left[ \frac{37805989}{2881494000}+\frac{163591}{137214000 r_Z^2}-\frac{22106653}{3293136000 r_Z} \right. \\
	&\hspace{-2cm}- \frac{48911 r_Z}{426888000}-\frac{236326427 r_Z^2}{5762988000}+\frac{12436082 r_Z^3}{180093375}-\frac{32867 r_Z^4}{5239080}+\left(-\frac{857358413}{5762988000}-\frac{19562657}{2305195200 r_Z^2} \right. \\
	&\hspace{-2cm}+ \left.\frac{620361529}{11525976000 r_Z}+\frac{80429929 r_Z}{349272000}-\frac{74350621 r_Z^2}{411642000}-\frac{113 r_Z^3}{31185} \right) \tilde{x}+\left( \frac{10032541}{174636000}+\frac{121092001}{11525976000 r_Z^2} \right. \\
	&\hspace{-2cm}- \left. \frac{532057187}{11525976000 r_Z}+\frac{106 r_Z}{86625} \right) \tilde{x}^2+\left(-\frac{1}{58212 r_Z^2}-\frac{1}{29106 r_Z}\right) \tilde{x}^3+\log(-r_t) \left( \frac{13}{3780}+\frac{1}{2376 r_Z^2}-\frac{13}{5544 r_Z} \right. \\
	&\hspace{-2cm}+ \frac{221 r_Z}{41580}-\frac{281 r_Z^2}{10395}+\frac{2 r_Z^3}{55}+\left(-\frac{1439}{20790}-\frac{1}{297 r_Z^2}+\frac{16}{693 r_Z}+\frac{83 r_Z}{693}-\frac{6 r_Z^2}{55}\right) \tilde{x}+\left( \frac{5}{189}+\frac{5}{1188 r_Z^2} \right. \\
	&\hspace{-2cm}- \left.\left.\left. \frac{5}{252 r_Z} \right) \tilde{x}^2 \right) \right] 
	+\frac{1}{r_t^6} \left[ -\frac{2192630176559}{409056888240000}-\frac{940653073}{3408807402000 r_Z^2}+\frac{794421072481}{409056888240000 r_Z} \right. \\
	&\hspace{-2cm}+ \frac{213480554017 r_Z}{37186989840000}+\frac{1605227229157 r_Z^2}{409056888240000}-\frac{140834719651 r_Z^3}{8181137764800}+\frac{568078963 r_Z^4}{27826999200}-\frac{2423 r_Z^5}{152895600} \\
	&\hspace{-2cm}+ \left(\frac{11162940545851}{204528444120000}+\frac{4147271311}{1818030614400 r_Z^2}-\frac{6823304481131}{409056888240000 r_Z}-\frac{2930170878121 r_Z}{29218349160000} \right. \\
	&\hspace{-2cm}+ \left.\frac{2108194453399 r_Z^2}{18593494920000}-\frac{209636149507 r_Z^3}{2921834916000}-\frac{163 r_Z^4}{162162} \right) \tilde{x}+\left( -\frac{11003872588483}{204528444120000}-\frac{819450691}{202905202500 r_Z^2} \right. \\
	&\hspace{-2cm}+ \left. \frac{1913822650681}{81811377648000 r_Z}+\frac{972425439991 r_Z}{20452844412000}+\frac{239 r_Z^2}{405405} \right) \tilde{x}^2+\left(-\frac{283}{5297292}+\frac{15282853319}{14609174580000 r_Z^2} \right. \\
	&\hspace{-2cm}- \left.\frac{11256827563}{5681345670000 r_Z} \right) \tilde{x}^3+\log(-r_t) \left( -\frac{13043}{10810800}-\frac{1031}{10810800 r_Z^2}+\frac{311}{514800 r_Z}-\frac{983 r_Z}{900900}+\frac{48407 r_Z^2}{5405400} \right. \\
	&\hspace{-2cm}- \frac{1019 r_Z^3}{60060}+\frac{293 r_Z^4}{19305}+\left(\frac{9689}{415800}+\frac{9391}{10810800 r_Z^2}-\frac{4001}{600600 r_Z}-\frac{2434 r_Z}{51975}+\frac{386 r_Z^2}{6435}-\frac{293 r_Z^3}{6435}\right) \tilde{x} \\
	&\hspace{-2cm}+ \left.\left.\left(-\frac{733}{30888}-\frac{1697}{1081080 r_Z^2}+\frac{3473}{360360 r_Z}+\frac{2465 r_Z}{108108}\right) \tilde{x}^2+\left(\frac{7}{17160 r_Z^2}-\frac{7}{8580 r_Z}\right) \tilde{x}^3 \right) \right] \Bigg] \\
	&\hspace{-2cm}+ C_F \Bigg[ 
	\frac{1}{r_t} \left[-20-\frac{5}{3 r_Z^2}+\frac{20}{3 r_Z} \right]
	+\frac{1}{r_t^2} \left[\frac{289}{90}+\frac{3}{20 r_Z^2}-\frac{148}{135 r_Z}-\frac{1241 r_Z}{135}+\left(-\frac{74}{135 r_Z^2}+\frac{28}{27 r_Z}\right) \tilde{x} \right] \\
	&\hspace{-2cm}+\frac{1}{r_t^3} \left[ -\frac{49789}{37800}-\frac{2377}{28350 r_Z^2}+\frac{52057}{113400 r_Z}+\frac{104413 r_Z}{56700}-\frac{15919 r_Z^2}{4725}+\left(\frac{38069}{28350}+\frac{347}{1890 r_Z^2}-\frac{7463}{8100 r_Z}\right) \tilde{x} \right] \\
	&\hspace{-2cm}+ \frac{1}{r_t^4} \left[ \frac{1401083}{3969000}+\frac{179797}{15876000 r_Z^2}-\frac{115447}{1134000 r_Z}-\frac{618371 r_Z}{793800}+\frac{3248243 r_Z^2}{3969000}-\frac{36853 r_Z^3}{33075} \right. \\
	&\hspace{-2cm}+ \left.\left( -\frac{904243}{992250} - \frac{469939}{7938000 r_Z^2}+\frac{2780101}{7938000 r_Z}+\frac{643859 r_Z}{661500} \right) \tilde{x}+\left(\frac{11717}{396900 r_Z^2}-\frac{1907}{33075 r_Z}\right) \tilde{x}^2  \right] \\
	&\hspace{-2cm}+ \frac{1}{r_t^5} \left[ -\frac{30333497}{261954000}-\frac{154673}{38808000 r_Z^2}+\frac{18997889}{628689600 r_Z}+\frac{385440967 r_Z}{1571724000}-\frac{100907407 r_Z^2}{261954000}+\frac{250618673 r_Z^3}{785862000} \right.\\
	&\hspace{-2cm}- \frac{4549253 r_Z^4}{13097700}+\left(\frac{600883}{1587600}+\frac{138091}{8316000 r_Z^2}-\frac{2467709}{20412000 r_Z}-\frac{127455007 r_Z}{196465500}+\frac{6359273 r_Z^2}{11907000}\right) \tilde{x} \\
	&\hspace{-2cm}+ \left. \left(-\frac{7134431}{71442000}-\frac{415409}{24948000 r_Z^2}+\frac{120692261}{1571724000 r_Z}\right) \tilde{x}^2 \right]
	+\frac{1}{r_t^6} \left[ \frac{611068021}{19209960000}+\frac{10264144487}{15732957240000 r_Z^2} \right. \\
	&\hspace{-2cm}- \frac{5329718861}{749188440000 r_Z}-\frac{75765552307 r_Z}{874053180000}+\frac{2191285707617 r_Z^2}{15732957240000}-\frac{88529188963 r_Z^3}{524431908000}+\frac{12872473853 r_Z^4}{112378266000} \\
	&\hspace{-2cm}- \frac{390253649 r_Z^5}{3745942200}+\left( -\frac{1119558376063}{7866478620000}-\frac{582826007}{125863657920 r_Z^2}+\frac{300649440137}{7866478620000 r_Z}+\frac{1495708237 r_Z}{5016886875} \right. \\
	&\hspace{-2cm}- \left.\frac{983884905989 r_Z^2}{2622159540000}+\frac{27984051973 r_Z^3}{112378266000} \right) \tilde{x}+\left( \frac{60247899487}{582702120000}+\frac{7750560019}{1123782660000 r_Z^2}-\frac{335249449013}{7866478620000 r_Z} \right. \\
	&\hspace{-2cm}- \left.\left. \frac{756269740169 r_Z}{7866478620000} \right) \tilde{x}^2+\left(-\frac{1732023911}{1123782660000 r_Z^2}+\frac{70960381}{23412138750 r_Z}\right) \tilde{x}^3 \right]
	\Bigg]
	+ \cO\left(1/r_t^7,\ep\right) \Bigg\}\,.
\end{align*}

The leading term in \Eqno{eq:LME_ggZZ_2L_AA} can be compared to the projected results of~\cite{Melnikov:2015laa}. We find agreement with 
their formula\footnote{Both Eq.(5) and Eq.(7) of~\cite{Melnikov:2015laa} contain typographical errors.}. We also performed a consistency check of the renormalisation scale dependence of the presented two-loop expansions by means of the technique given in Sec.~\ref{sec:scale_dependence_of_the_finite_remainder}.


\subsubsection{Anomalous Diagrams}
\label{ssub:anomalous_diagrams}

\begin{figure}[hb]
	\centering
	\begin{subfigure}{.4\textwidth}
		\centering
	\includegraphics[angle=270,scale=.5]{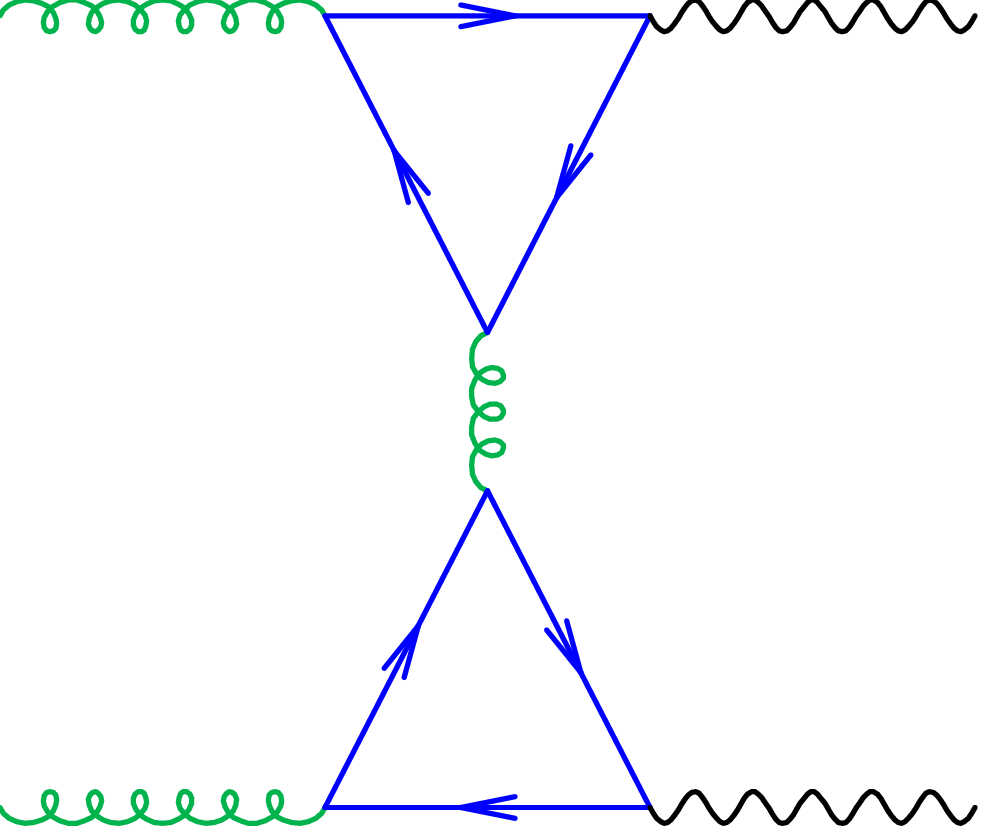}
		\caption{}
	\end{subfigure}
	\hspace{1cm}
	\begin{subfigure}{.4\textwidth}
		\centering
	\includegraphics[angle=270,scale=.5]{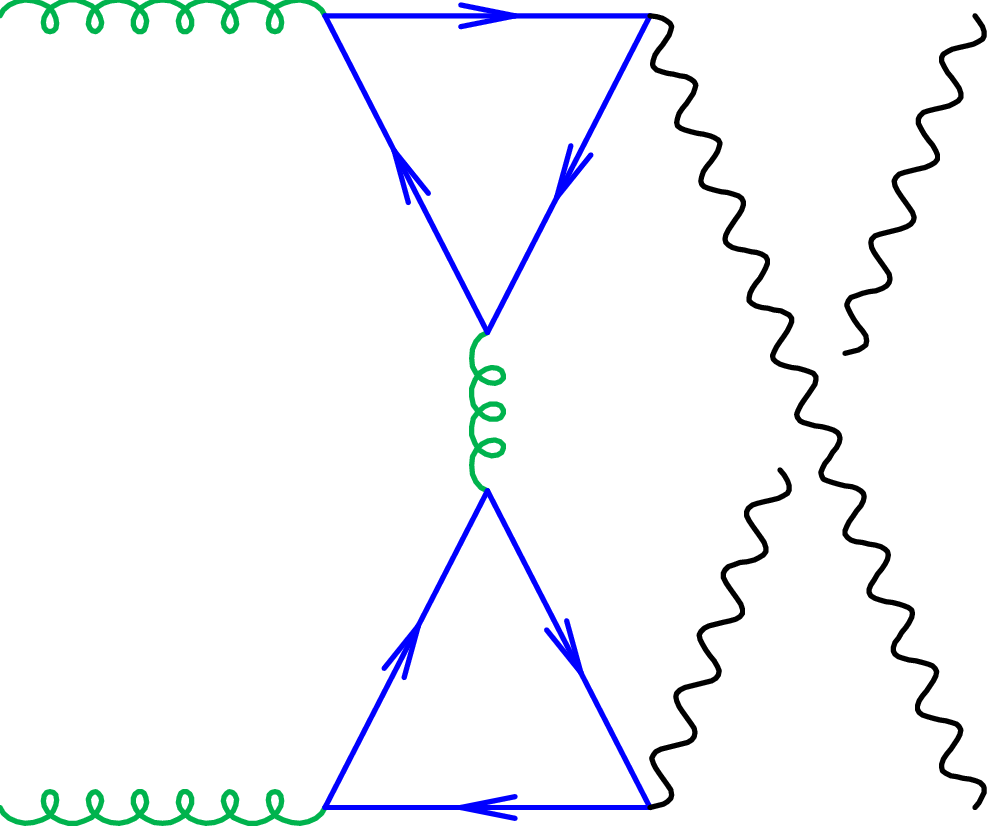}
		\caption{}
	\end{subfigure}
	\caption{Two-loop \emph{anomaly style} diagrams for the production of $Z$ boson pairs.}
	\label{fig:twotriangles}
\end{figure}

The two-loop $gg\to ZZ$ amplitude contains, in addition, two
topologies which consist of products of one-loop sub-diagrams. On the
one hand diagrams containing gluon self-energy contributions vanish
due to color conservation. The diagrams in
Fig.~\ref{fig:twotriangles}, on the other hand, give a finite mass
dependent contribution as long as both $Z$ bosons couple to distinct
fermion loops. These diagrams are proportional only to the axial
coupling of the $Z$ bosons to fermions; the vector component vanishes
due to $C$ invariance (\emph{Furry's theorem}). The diagrams have been
omitted in the previous section since they can be computed with their
full top mass dependence and, therefore, need no large-mass
expansion~\cite{Kniehl:1989qu,Campbell:2007ev}.
\begin{figure}[t]
  \centering \includegraphics[angle=270,width=12cm]{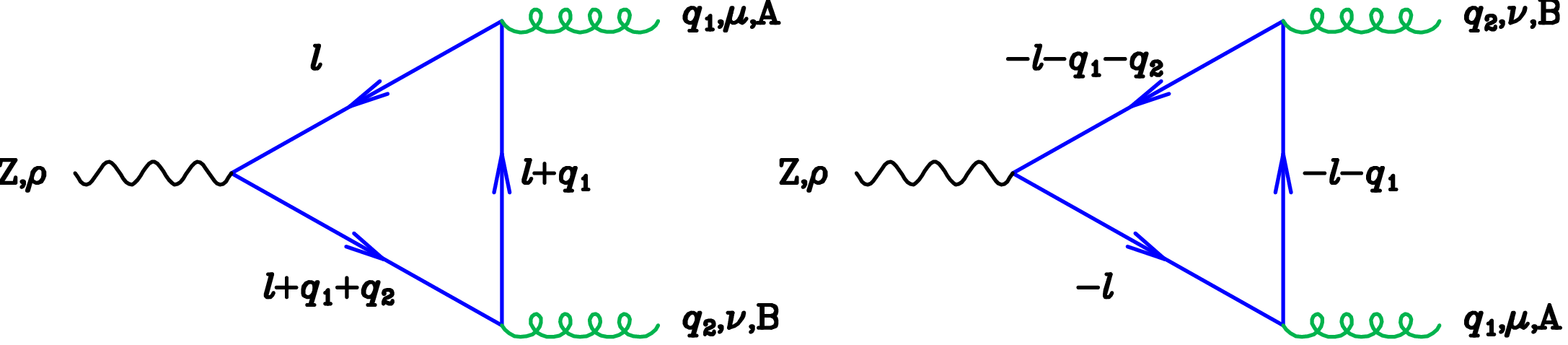}
\caption{Triangle diagrams representing the $Zgg$ form factor at lowest order.}
\label{fig:anomaly}
\end{figure}

In brevity we repeat the results from~\cite{Campbell:2007ev} and give the result in terms of our conventions. Let us denote the amplitude for a $Z$ coupling to two gluons by $T^{\mu \nu \ro}_{AB}$. We calculate the triangle shown in Fig.~\ref{fig:anomaly}, where all
momenta are outgoing $q_1+q_2+q_3=0$ and to begin with $q_1^2 \neq 0,
q_2^2 \neq 0$. The result for the two triangle diagrams (including
the minus sign for a fermion loop) is,
\begin{equation} \label{EqT}
T^{\mu \nu \rho}_{AB}(q_1,q_2) = i \frac{\gs^2}{16 \pi^2} \frac{1}{2} 
\delta_{AB} 
\Big(\frac{\gW}{2 \cosW} \Big) \tau_f\;  \Gamma^{\mu \nu \rho}\,,
\end{equation}
where $\tau_f=\pm 1/2$ and,
\begin{equation}
\Gamma^{\mu \nu \rho}(q_1,q_2,m) = \frac{2}{i \pi^2} \int \; d^d l
\; \mbox{Tr}\Big\{ \gamma^\rho \gamma_5 \frac{1}{\slsh{l}-m}
\gamma^\mu \frac{1}{\slsh{l}+\slsh{q}_1-m}
\gamma^\nu \frac{1}{\slsh{l}+\slsh{q}_1+\slsh{q}_2-m} \Big\}\,.
\end{equation}
The most general form of $\Gamma$ consistent with QCD gauge invariance,
\begin{equation}
q_{1}^{\mu} \Gamma_{\mu \nu \rho}=
q_{2}^{\nu} \Gamma_{\mu \nu \rho} =0 \; ,
\end{equation}
can be written as,
\begin{eqnarray}
\Gamma^{\mu \nu \rho}&=& F_1(q_1,q_2,m) \;
      \Big\{ \mbox{Tr}[\gamma^\rho \gamma^\nu \slsh{q_1} \slsh{q_2}\gamma_5 ] q_1^\mu
            +\mbox{Tr}[\gamma^\rho \gamma^\mu \gamma^\nu \slsh{q_2}\gamma_5 ] q_1^2 \Big\}\nonumber \\
  &+& F_2(q_1,q_2,m) \;
        \Big\{\mbox{Tr}[\gamma^\rho\gamma^\mu\slsh{q_1}\slsh{q_2}\gamma_5 ] q_2^\nu
        +\mbox{Tr}[\gamma^\rho\gamma^\mu\gamma^\nu\slsh{q_1}\gamma_5 ] q_2^2 \Big\}\nonumber \\
  &+& F_3(q_1,q_2,m)\; (q_1^\rho+q_2^\rho)
     \Big\{ \mbox{Tr}[ \gamma^\mu\gamma^\nu\slsh{q_1}\slsh{q_2}\gamma_5 ] \Big\}\nonumber \\
  &+& F_4(q_1,q_2,m)\; (q_1^\rho-q_2^\rho)
      \Big\{ \mbox{Tr}[\gamma^\mu\gamma^\nu\slsh{q_1}\slsh{q_2}\gamma_5 ]  \Big\}\,. 
\end{eqnarray}
By direct calculation it is found that $F_4=0$. 

Contracting with the momentum of the
$Z$ boson we find that, $q_3=-q_1-q_2$
\beq
(q_3)_\rho \, \Gamma^{\mu \nu \rho}= \Big[ -q_1^2 \, F_1(q_1,q_2,m) 
  +q_2^2 \,F_2(q_1,q_2,m) -q_3^2 \, F_3(q_1,q_2,m) \Big] 
  \mbox{Tr}[ \gamma^\mu\gamma^\nu\slsh{q_1}\slsh{q_2}\gamma_5 ]\,.
\eeq
The divergence of the axial current is found by direct calculation to be,
\beq
(q_3)_\rho \, \Gamma^{\mu \nu \rho}=\Big[ 4 m^2 C_0(q_1,q_2;m,m,m) +2 \Big] 
\mbox{Tr}[ \gamma^\mu\gamma^\nu\slsh{q_1}\slsh{q_2}\gamma_5 ] 
\eeq
showing the contribution of the pseudoscalar current proportional to $m^2$ and the anomalous piece. Summation over one complete quark doublet ($\tau_f = \pm 1/2$) cancels the anomaly term and solely the piece proportional to the top mass remains.

For the particular case at hand we are
interested in on-shell $Z$'s and in 
$q_2^2=\varepsilon_2 \cdot q_2=0,\, \varepsilon_3 \cdot q_3=0,\, q_3=-q_1-q_2$, so we get a contribution only from $F_1$. The result for $F_1$ is
\begin{eqnarray}
F_1(q_1,q_2,m) &=&\frac{1}{2 q_1 \cdot q_2}
\Bigg[2+4 m^2 C_0(q_1,q_2;m,m,m) \nonumber \\
 &+&\Big(2 + \frac{q_1^2}{q_1 \cdot q_2}\Big)
 \Big[B_0(q_1+q_2;m,m)-B_0(q_1;m,m)\Big]\Bigg]\,, \\
F_1(q_1,q_2,0) &=&\frac{2}{(q_3^2-q_1^2)}
\Bigg[1+\frac{q_3^2}{(q_3^2-q_1^2)} \log(\frac{q_1^2}{q_3^2})\Bigg]\,.
\end{eqnarray}

We further define a subtracted $F_1$ to take into account the contribution of the top and the bottom quarks,
\beq
{\cF_1}(q_1,q_2,m)=\Big[ F_1(q_1,q_2,m)-F_1(q_1,q_2,0) \Big] \,.
\eeq
Analogous to \Eqno{eq:massiveggZZ_projected_amplitude} we define the projected matrix element for the anomaly piece
\begin{equation}
	\label{eq:anomaly_2L_definition}
	\ket{\cB^0_\text{anom}(\as^0,m^0,\mu,\ep)} = \frac{\de^{AB}}{N_A} \left(g^{\mu\nu}\,p_1p_2-p_2^\mu p_1^\nu \right) P_Z^{\al\rho'}(p_3)P_{Z,\rho'}^\be(p_4) \ket{\cB_{\text{anom},\mu\nu\al\be}^{0,AB}(\as^0,m^0,\mu,\ep)}\,.
\end{equation}
The amplitude defined in \Eqno{eq:anomaly_2L_definition} is UV and IR finite and requires no renormalisation. Including the effect of both the $b$ quark (taken to be massless) and the $t$ quark we obtain (No statistical factor for identical $Z$ bosons is included).
\begin{align*}
	\label{eq:anomaly_result_full}
	\ket{\cB_\text{anom}(\as^{(n_l)},m,\mu)} &= a_t^2 s^2 \mathcal{N} \cdot \left(\frac{\as^{(n_l)}}{4\pi}\right)^2 \numberthis \\
	&\hspace{-2.5cm}\times \Bigg\{
	           \Big[(\rZ-x) \big(1+(\rZ-x) (1/\rZ-1/(2 \rZ^2))\big)\Big] 
	            \cF_1(p_1-p_3,-p_1,m) \cF_1(p_3-p_1,-p_{2},m) \\
	           &\hspace{-2.5cm}+ \Big[(\rZ-1+x)\big(1+(\rZ-1+x) (1/\rZ-1/(2 \rZ^2))\big)\Big]
	            \cF_1(p_1-p_4,-p_1,m) \cF_1(p_4-p_1,-p_2,m)) \Bigg\}\,,
\end{align*}
where $\cN$ is given in \Eqno{eq:Ndef}. Again we include the factors
$s^2$ to indicate the correct dimensionality of
$\cF_1(p_1,p_2,m)$. For completeness we also give the mass expansion
of \Eqno{eq:anomaly_result_full} in case only the top quark
contribution is considered, i.e. $ F_1(p_1,p_2,0)\to 0$. As expected
the expansion starts at $1/m^4$.
\begin{align*}
	\label{eq:anomaly_result_top_exp}
	\ket{\cB_\text{anom,t}(\as^{(n_l)},m,\mu)} &= a_t^2 \mathcal{N} \cdot \left(\frac{\as^{(n_l)}}{4\pi}\right)^2 \;\Bigg\{
	\frac{1}{r_t^2} \left[-\frac{1}{9}+\frac{r_Z}{9}+\frac{1-\tilde{x}}{18 r_Z}+\frac{-1+2 \tilde{x}}{72 r_Z^2}\right]
+ \frac{1}{r_t^3} \left[ -\frac{8 r_Z}{135}+\frac{2 r_Z^2}{45} \right.\\
&\hspace{-2cm}+ \left. \frac{(13-18 \tilde{x})}{270} +  \frac{1-3 \tilde{x}}{270 r_Z^2}+\frac{-11+26 \tilde{x}}{540 r_Z} \right]
+ \frac{1}{r_t^4} \left[ -\frac{22 r_Z^2}{945}+\frac{22 r_Z^3}{1575}+\frac{r_Z (1511-2362 \tilde{x})}{56700} \right.\\
&\hspace{-2cm}+ \left. \frac{-3845+9892 \tilde{x}}{226800} -  \frac{191 \left(1-4 \tilde{x}+2 \tilde{x}^2\right)}{226800 r_Z^2}+\frac{646-2129 \tilde{x}+382 \tilde{x}^2}{113400 r_Z} \right]
+ \frac{1}{r_t^5} \left[ -\frac{38 r_Z^3}{4725}+\frac{19 r_Z^4}{4725} \right. \\
&\hspace{-2cm}+ \frac{r_Z^2 (113-188 \tilde{x})}{9450} + \frac{r_Z (-783+2104 \tilde{x})}{75600}+\frac{-111+472 \tilde{x}-306 \tilde{x}^2}{75600 r_Z} + \frac{1-5 \tilde{x}+5 \tilde{x}^2}{5400 r_Z^2} \\
&\hspace{-2cm}+ \left. \frac{197-688 \tilde{x}+194 \tilde{x}^2}{37800} \right]
+ \frac{1}{r_t^6} \left[-\frac{1613 r_Z^4}{623700}+\frac{1613 r_Z^5}{1455300}+\frac{r_Z^3 (41432-71573 \tilde{x})}{8731800} \right. \numberthis \\
&\hspace{-2cm}+\frac{r_Z^2 (-457682+1261401 \tilde{x})}{87318000} + \frac{-1049213+4652126 \tilde{x}-3464248 \tilde{x}^2}{698544000} \\
&\hspace{-2cm}+ \frac{r_Z \left(622783-2250826 \tilde{x}+764954 \tilde{x}^2\right)}{174636000} +  \frac{42658-222727 \tilde{x}+251038 \tilde{x}^2-18874 \tilde{x}^3}{116424000 r_Z} \\
&\hspace{-2cm}+ \left. \frac{9437 \left(-1+6 \tilde{x}-9 \tilde{x}^2+2 \tilde{x}^3\right)}{232848000 r_Z^2}\right] + \cO\left(1/r_t^7,\ep\right)\Bigg\} \,.
\end{align*}

\subsection{Visualisation of Large-Mass Expansion Results for $gg\to ZZ$} 
\label{sub:visualisation_of_large_mass_expansion_results_for_gg_to_zz}

Let us turn towards the graphical representations of the large-mass
expansion results for the SM continuum, Eqs.~(\ref{eq:LME_ggZZ_1L_VV}-\ref{eq:LME_ggZZ_2L_AA}), and their
improvements. We proceed analogously to
Sec.~\ref{ssub:ggH_comparison_lme_with_full_result} and compute the
UV+IR renormalised version of \Eqno{eq:general_H_ZZ_interference} and
again integrate over the $ZZ$ phase space. The setup from
\Eqno{eq:ggH_input_parameters} is utilised. Since we focus our
discussion in this section mainly on the different improvements of the
large-mass expansions we, again, do not take into account the full NLO
correction. We merely focus on the unknown virtual massive two-loop
contribution of the SM continuum interfered with the Higgs-mediated
process. That is, we set
\begin{equation}
	\label{eq:ggZZ_sigma_defs_virt}
	\si_\text{int}^\text{LO} \sim 2\,\Re\bra{\cF_\cA^{(1)}(m,\mu)}\ket{\cF_\cB^{(1)}(m,\mu)} \quad \text{and} \quad \si_\text{virt,int}^\text{NLO} \sim 2\,\Re\bra{\cF_\cA^{(1)}(m,\mu)}\ket{\cF_\cB^{(2)}(m,\mu)} \,,
\end{equation}
which also excludes the \emph{anomaly style} contribution from eq~\eqref{eq:anomaly_result_full} since this part can be computed without the necessity of any approximation.

It is important to notice the following conventions for our
approximations using \emph{Pad\'e approximants} below. As in
Sec.~\ref{ssub:ggH_comparison_lme_with_full_result} the \emph{Pad\'e approximants} 
are computed at amplitude level for each finite
remainder $\cF_{\cA,\cB}$, including the conformal
mapping\footnote{Computing the \emph{Pad\'e approximants} for the
expanded product $\bra{\cF_\cA^{(1)}}\ket{\cF_\cB^{(1,2)}}$ yield no
reasonable result above threshold. We have checked this by
explicitly computing the \emph{homogeneous bivariate Pad\'e Approximants} 
$[2/2]$-$[3/3]$~\cite{Cuyt1979,Guillaume2000197} for
the LO interference $\Re\bra{\cF_\cA^{(1)}}\ket{\cF_\cB^{(1)}}$,
where we treated the mapped variable $w$,
\Eqno{eq:conformal_mapping}, and its complex conjugated $\bar{w}$ as
independent variables.}. We know from our previous discussion that
the best approximation of the LO as well as the virtual NLO
contribution of the Higgs-mediated process is given by
$\cF_{\cA,[5/5]}^{(1,2)}$. It is understood that we will always use
this approximant in the following considerations. In principle, we can
also substitute the approximated Higgs-mediated amplitude
$\cF_{\cA,[5/5]}^{(1)}$ with its exact LO result. Doing so would
remove any uncertainties from the Higgs-mediated contribution. On the
other hand the numerical difference between both approaches is
negligible as discussed in
Sec.~\ref{ssub:ggH_comparison_lme_with_full_result}.

The vector-vector part of the SM continuum gives only a minor contribution to the total cross section, $\si_{VV}/\si_{AA}\sim 10^{-3}$. This relies on the fact that the mass expansion of the $VV$ part starts only at $1/m^4$ whereas the $AA$ part starts at $1/m^2$ and additionally $a_t^2/v_t^2 \sim 7$.

\begin{figure}[h!]
	\centering
	\begin{subfigure}{0.495\textwidth}
		\includegraphics[width=\textwidth]{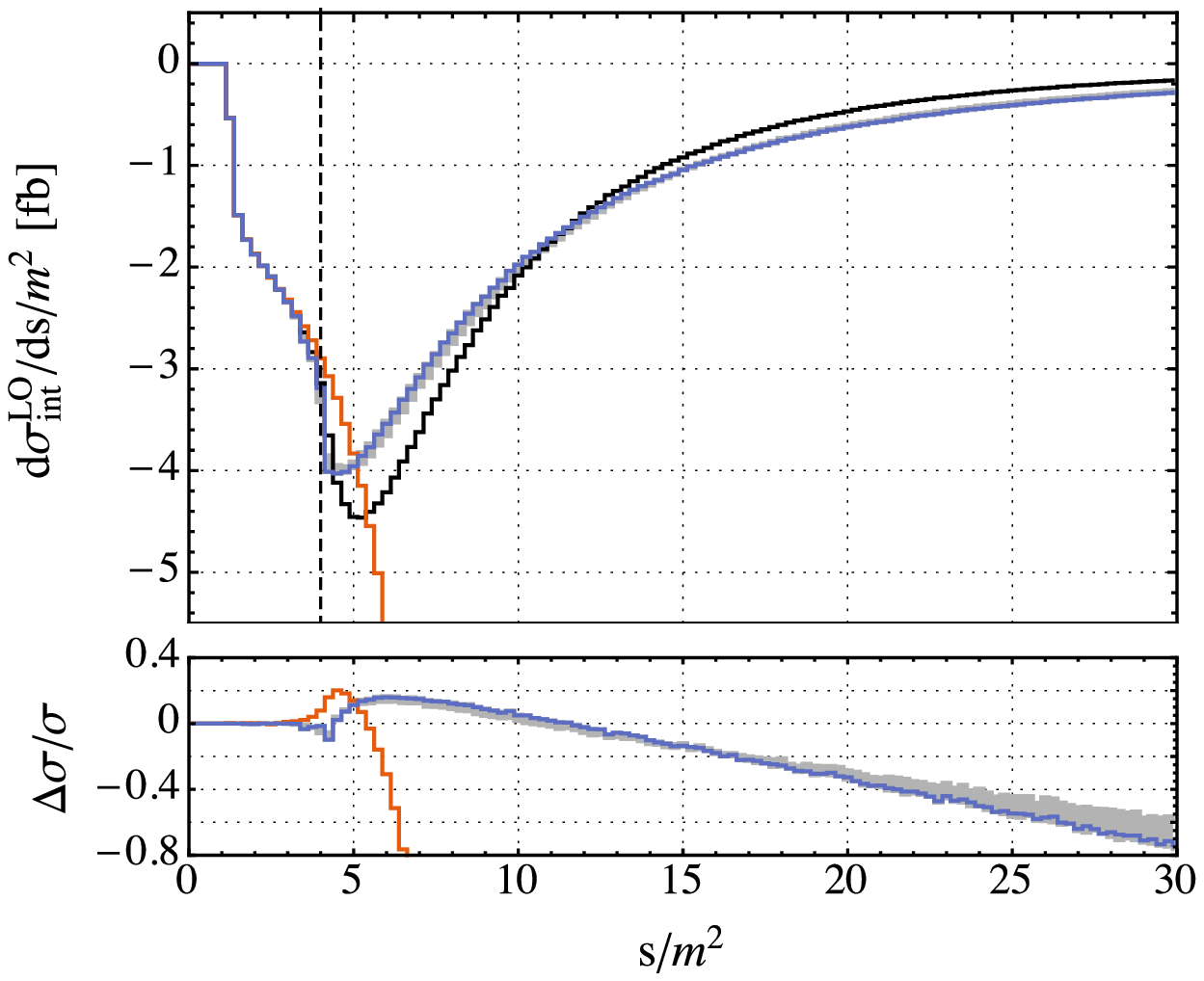}
	\end{subfigure}
	\hfill
	\begin{subfigure}{0.495\textwidth}
		\includegraphics[width=\textwidth]{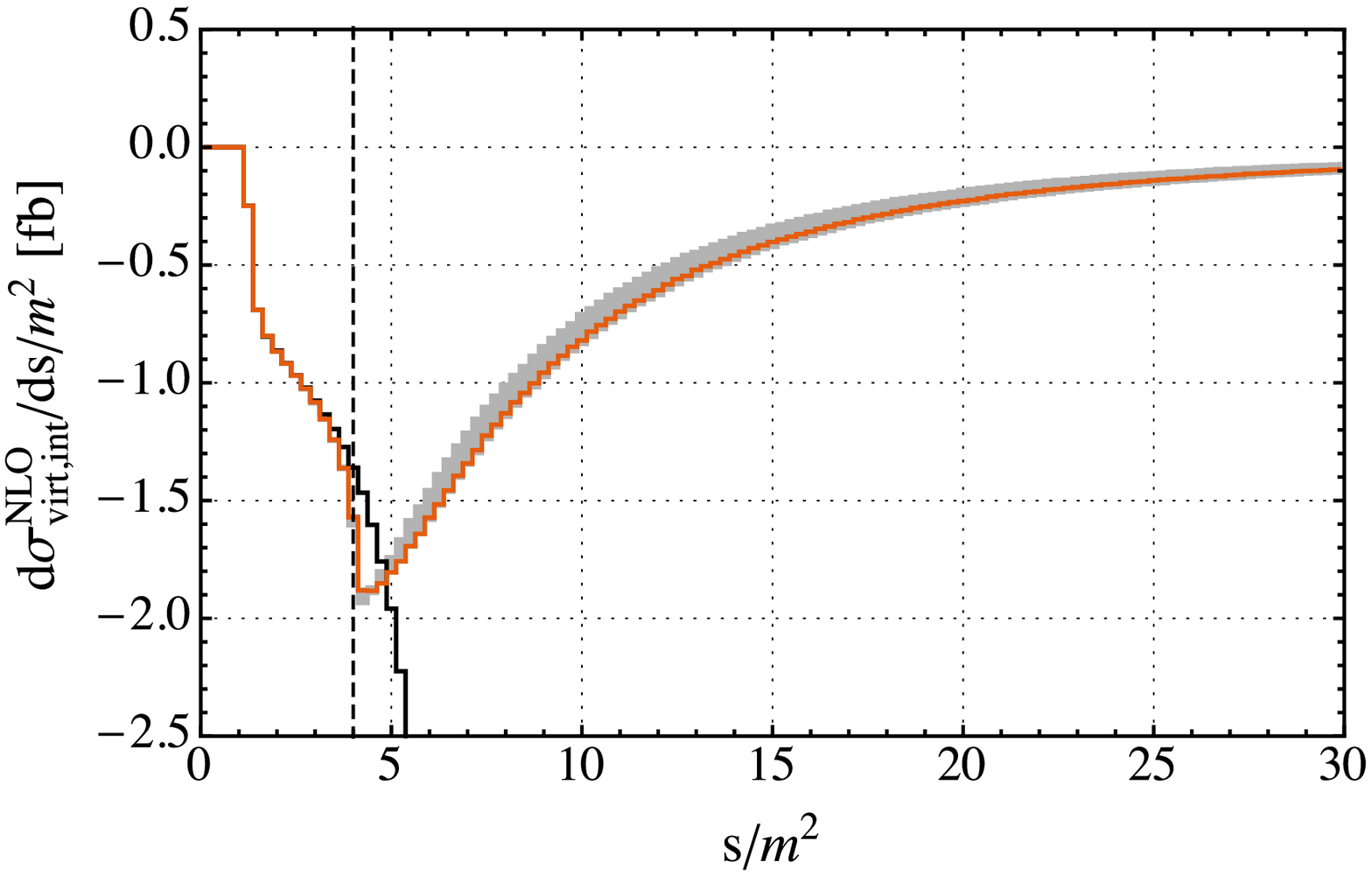}
	\end{subfigure}
	\caption{\textbf{Left panel:} Leading-order interference $\Re\bra{\cF_\cA^{(1)}(m,\mu)}\ket{\cF_\cB^{(1)}(m,\mu)}$. Exact result (black), LME up to $1/m^{12}$ (orange) and envelope of \emph{Pad\'e approximants} $[2/2],[2/3],[3/2]$ and $[3/3]$ (blue) as grey area. Bottom plot shows the relative deviation from the exact result. \textbf{Right panel:} Next-to-leading-order interference $\Re\bra{\cF_\cA^{(1)}(m,\mu)}\ket{\cF_\cB^{(2)}(m,\mu)}$. LME up to $1/m^{12}$ (black) and envelope of \emph{Pad\'e approximants} $[2/2],[2/3],[3/2]$ and $[3/3]$ (orange) as grey area. The vertical dashed line denotes the top quark pair-production threshold. See text for details.}
	\label{fig:ggZZ_LO_NLO_LME_Pade}
\end{figure}

The interference including the exact top mass dependence is only known
at leading-order, which is shown in the left panel of
Fig.~\ref{fig:ggZZ_LO_NLO_LME_Pade}. Comparing the exact result
(black) and its naive large-mass approximation up to $1/m^{12}$
(orange) shows excellent agreement up to $s\sim 3 m^2$, with
approximately $1\%$ deviation from the exact result. At threshold the
deviation rises to $12\%$. In contrast the \emph{Pad\'e approximant}
$\cF_{\cB,[3/3]}^{(1)}$ (blue) deviates from the exact result by $6\%$
at threshold. The shaded grey area indicates the variation from
computing the \emph{Pad\'e approximants} $[2/2],[2/3],[3/2],[3/3]$
with $3-8\%$ deviation at threshold.  Due to the change of sign of
their derivatives we get a better approximation closely above
threshold, as can be seen in the bottom plot of
Fig.~\ref{fig:ggZZ_LO_NLO_LME_Pade}.  Nevertheless, the peak of the
exact LO result at $s\sim 5.2 m^2$ is with $10-11\%$ deviation quite
poorly approximated. Ineptly this is the region of interest for our
later analysis of the Higgs boson width. Going to large values of $s$
the deviations inevitably become larger, but the contribution to the
cross section is small due to the suppression by the flux.

This situation seems to continue in case of the next-to-leading-order
large-mass expansion as shown in the right panel of
Fig.~\ref{fig:ggZZ_LO_NLO_LME_Pade}. Evidently no exact result is
available and we have to rely on the approximate results. All
\emph{Pad\'e approximants} $[2/2],[2/3],[3/2],[3/3]$ for
$\cF_\cB^{(2)}$ show a stable trend over the entire $s/m^2$ range. The
deviations between the diagonal and non-diagonal \emph{Pad\'e
  approximants} are again indicated by the shaded grey area and the
approximant $\cF_{\cB,[3/3]}^{(2)}$ is shown in orange. The steeper
rise near the top threshold suggests a better description of the
actual threshold properties of the NLO result with exact top mass
dependence in contrast to the naive large-mass expansion
(black). Comparing the trend above threshold with its analogous LO
situation we can only guess that we have to expect comparable
deviations from our \emph{Pad\'e approximations} with respect to the
unknown exact NLO result.

\begin{figure}[h!]
	\centering
	\begin{subfigure}{0.495\textwidth}
		\includegraphics[width=\textwidth]{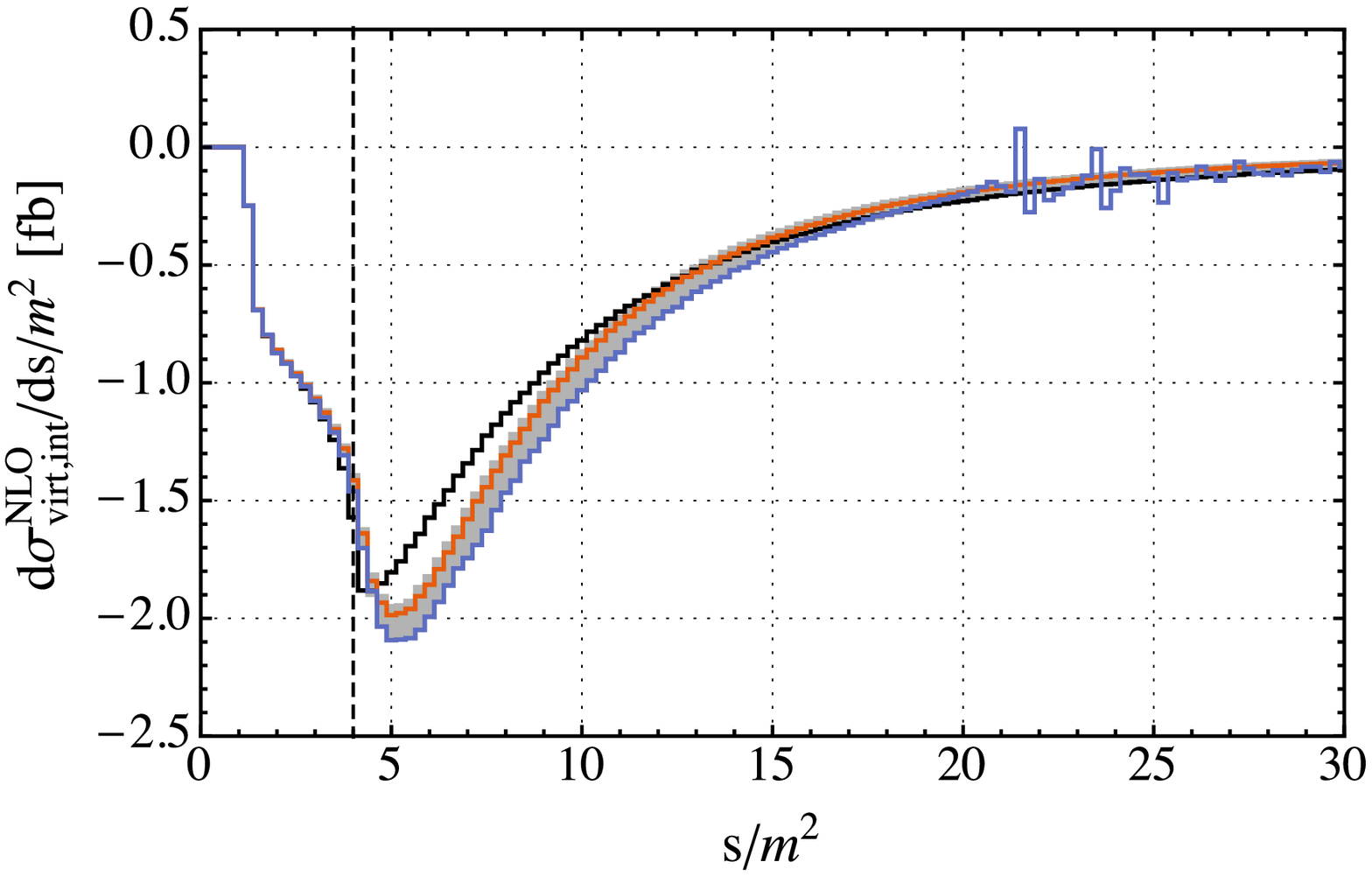}
	\end{subfigure}
	\hfill
	\begin{subfigure}{0.495\textwidth}
		\includegraphics[width=\textwidth]{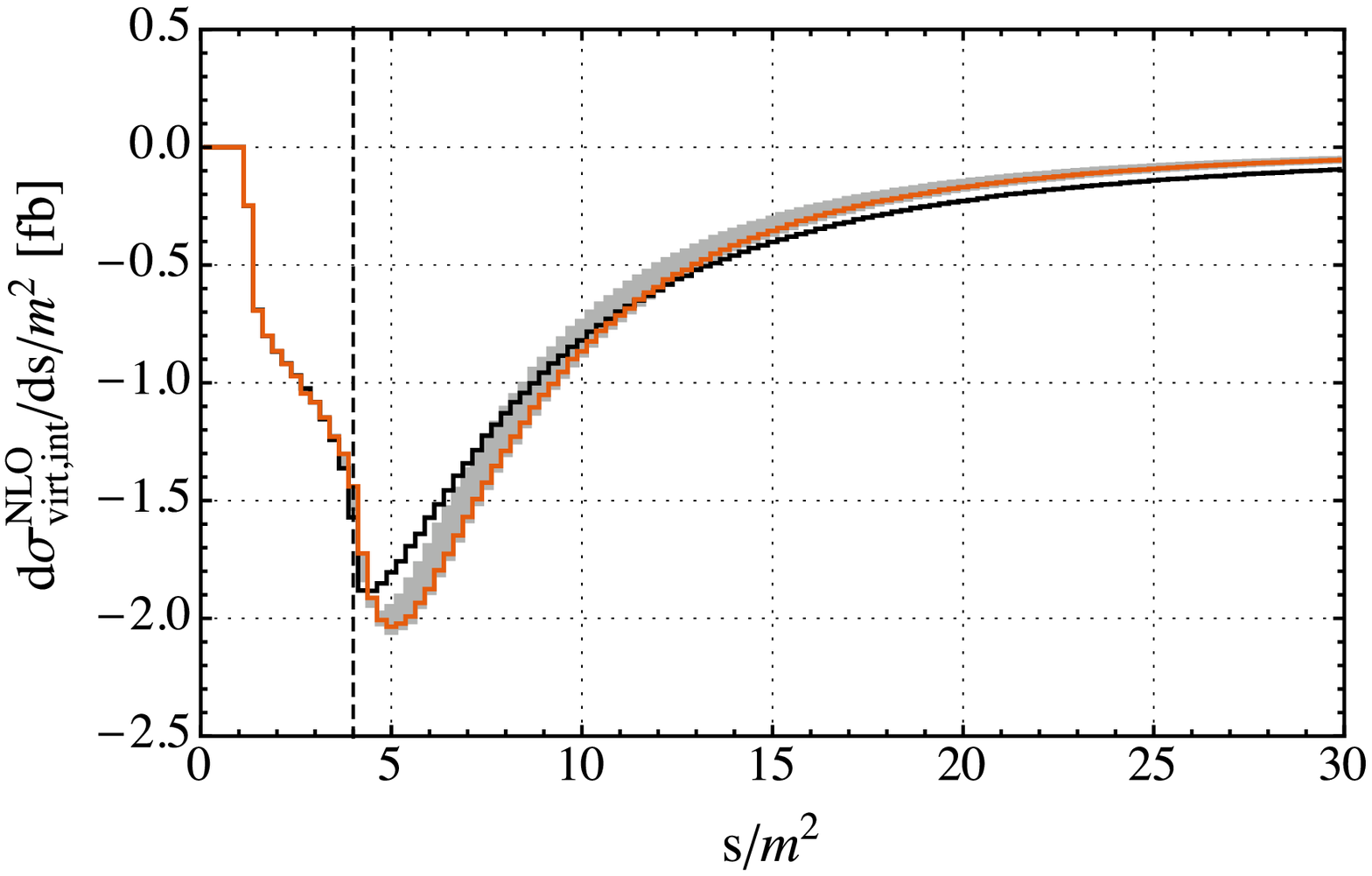}
	\end{subfigure}
	\caption{Next-to-leading-order interference 
	$\Re\bra{\cF_\cA^{(1)}(m,\mu)}\ket{\cF_\cB^{(2)}(m,\mu)}$. 
	\textbf{Left panel:} Interference by rescaling, \Eqno{eq:LME_rescaled}. 
	\emph{Pad\'e approximant} $[3/3]$ as comparison 
	(black). Envelope of $\si_{\text{imp},n}^\text{NLO}$ for $n=\{1,\ldots,6\}$ 
	as grey area; $n=1$ (orange) and $n=6$ (blue) shown explicitly. 
	\textbf{Right panel:} Interference by alternative rescaling, 
	\Eqno{eq:LME_rescaled_Pade}. \emph{Pad\'e approximant} 
	$[3/3]$ as comparison 
	(black). Grey area given by envelope of $\si_{\text{imp},[n/m]}^\text{NLO}$ 
	with $n,m=\{2,3\}$; $[3/3]$ shown explicitly (orange). The vertical dashed 
	line denotes the top quark pair-production threshold. See text for details.}
	\label{fig:ggZZ_NLO_melnikov}
\end{figure}

We can also consider rescaling the NLO large-mass expansion as
described in \Eqno{eq:LME_rescaled}. The resulting curves are
shown in the left panel of Fig.~\ref{fig:ggZZ_NLO_melnikov}. To guide
the eye we also include $\cF_{\cB,[3/3]}^{(2)}$ (black). The envelope
of the different orders $n$ in the expansion
$\si_{\text{imp},n}^\text{NLO}$ is shown as grey area. For $s\le
20m^2$ the envelope is determined from $n=\{1,\ldots,6\}$, whereas for
$s>20m^2$ we only use $n=\{1,\ldots,5\}$ due to the instabilities for
$n=6$ in the high energy regime. The most interesting curves, namely
the heavy-quark approximation $n=1$ and the highest order in the
expansion $n=6$, are shown in orange and blue, respectively. Factoring
out the exact LO result seems to give a more natural description of
the threshold behaviour and peak structure in comparison to the plain
use of the \emph{Pad\'e approximation}.

The origin of the numerical instabilities of the $n=6$ expansion is
probably due to delicate numerical cancellations in the $(s/m^2)^6$
coefficients. One could try to cure this problem by switching to a
higher numerical precision or by a proper \emph{economisation}~\cite{Press:2007:NRE:1403886} of the
power series. With the \emph{Pad\'e approximation} we already have an
excellent method at hand and we adopt the idea of factoring out the
exact LO interference,
\begin{equation}
	\label{eq:LME_rescaled_Pade}
	\si_{\text{imp},[n/m]}^\text{NLO} = \si_\text{exact}^\text{LO} \cdot \frac{\si_{[n/m]}^\text{NLO}}{\si_{[n/m]}^\text{LO}}\,.
\end{equation}
Keeping our usual definition in mind
$\si_{\text{imp},[n/m]}^\text{(N)LO}$ denotes the (virtual N)LO
contribution using $\cF_{\cA,[5/5]}^{(1)}$ and
$\cF_{\cB,[n/m]}^{(1,2)}$. The result is shown in
Fig.~\ref{fig:ggZZ_NLO_melnikov}, right panel. We immediately see the
advantages of this approach. Firstly we also get a similar, more
natural behaviour at threshold and of the peak structure above
threshold. Secondly we get a stable result across the entire range of
$s/m^2$. The grey area is again given by the envelopes due to the
variation between the (non-)diagonal \emph{Pad\'e approximants}
$[2/2],[2/3],[3/2]$ and $[3/3]$(orange). Ultimately by using the
\emph{Pad\'e approximants} in contrast to \Eqno{eq:LME_rescaled}
we could entirely remove the uncertainty of having to use an
approximation for the involved Higgs-mediated amplitude and fall back
to using the exactly known result for $\cF_{\cA}^{(1)}$.

Some concluding remarks. In contrast to the purely Higgs-mediated case, Sec.~\ref{ssub:ggH_comparison_lme_with_full_result}, it turns out that we require the \emph{Pad\'e approximation} in the interference case. Using the conformal mapping alone without an additional \emph{Pad\'e approximant} on top gives no reasonable approximation for the quantities discussed above. On the other hand we have seen that we hugely benefit by using \emph{Pad\'e approximations} due to their stability and the possibility of removing any uncertainty besides the approximated virtual massive two-loop $gg\to ZZ$ amplitude.


\section{Real Corrections to SM $ZZ$ Production} 
\label{sec:real_corrections_to_sm_zz_production}

\begin{figure}[ht]
\begin{center}
\includegraphics[angle=270,scale=0.8]{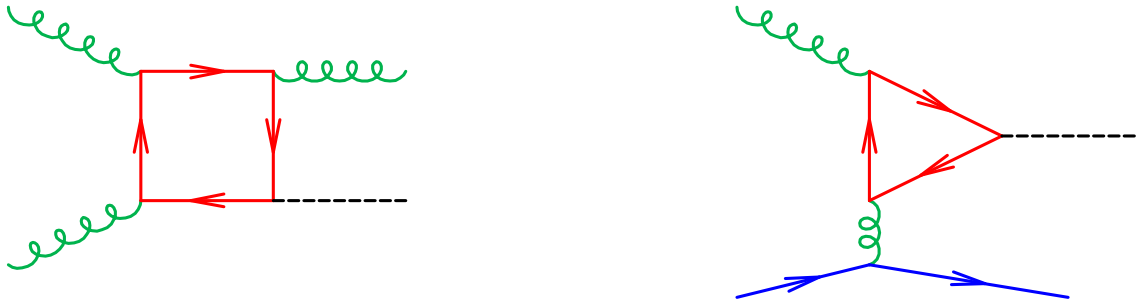}
\caption{Representative diagrams for the $0 \to ggHg$  and the $0 \to gHq \bar{q}$ amplitudes.}
\label{fig:ggHg_amps}
\end{center}
\end{figure}
\begin{figure}[ht]
\begin{center}
\includegraphics[angle=270,scale=0.6]{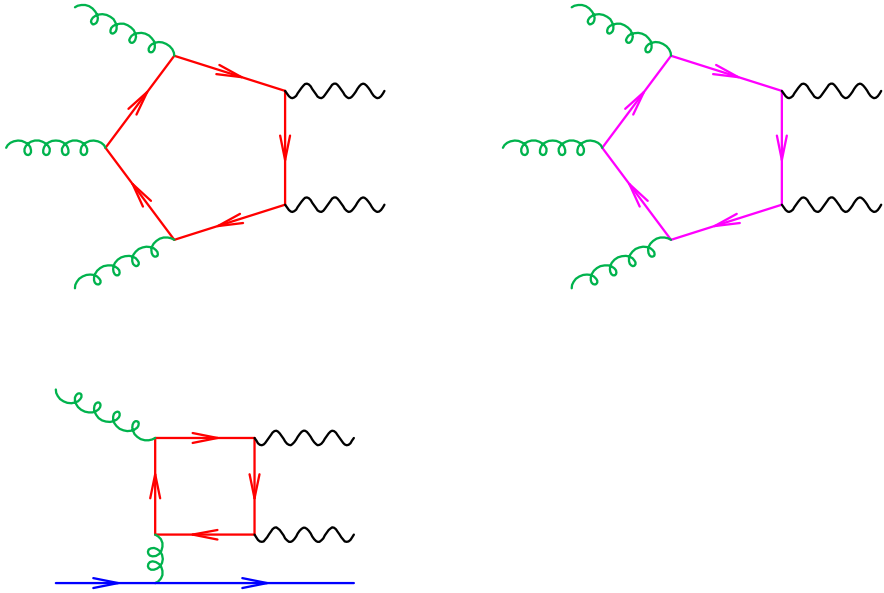}
\caption{Representative diagrams for the $0 \to ggZZg$  and the $0 \to gZZq \bar{q}$ amplitudes.}
\label{fig:ggZZg_amps}
\end{center}
\end{figure}
Representative diagrams for the real radiation contributions to this process are shown in
Figs.~\ref{fig:ggHg_amps} and \ref{fig:ggZZg_amps}. 
The Higgs-mediated diagrams have previously been computed in~\cite{Ellis:1987xu}.  They
can easily be adapted to our calculation by combining those results with the decay amplitude given in
\Eqno{decayamplitude} and $\cN$ from \Eqno{eq:Ndef}.  This procedure, together with
the strategy for handling the amplitudes for diagrams without a Higgs boson,  is described
in detail in~\cite{Campbell:2014gua}.    We adopt this implementation here.
Our calculation of the pure-Higgs contribution involves the computation of
the square of the diagrams shown in Fig.~\ref{fig:ggHg_amps}, together with
all crossings of the quarks in Fig.~\ref{fig:ggHg_amps}~(right) into the
initial state.  Similarly, the interference contribution includes all
crossings of the diagrams shown in Fig.~\ref{fig:ggZZg_amps}.
In principle another contribution to the interference occurs
at this order, between tree-level amplitudes for the process $qg \to ZZ q$ and the $qg$-initiated diagrams
shown in Fig.~\ref{fig:ggHg_amps}~(right) and \ref{fig:ggZZg_amps}~(bottom-left).  However this contribution
is subleading~\cite{Campbell:2014gua}, particularly for high invariant masses of the $ZZ$ system, so we do not
consider it here.

The real radiation diagrams contain infrared singularities, of soft and collinear origin, that
must be isolated and combined with the corresponding poles in the two-loop amplitudes.  This
is handled using the dipole subtraction procedure~\cite{Catani:1996vz}.

\section{Results}
\label{sec:results}

The individual components of the calculation that have been extensively
discussed above have been included in the parton-level Monte Carlo
code MCFM~\cite{Campbell:1999ah,Campbell:2011bn,Campbell:2015qma}.
The bulk of the calculation is performed in a
straightforward manner using the normal operation of MCFM at NLO.
The exception is the finite contribution to the two-loop amplitude containing
a closed loop of massless quarks.  Since these contributions are computationally
expensive to evaluate, we choose to include their effects by reweighting
an unweighted sample of LO events.

For the two-loop amplitudes containing
massive loops of quarks the approximations used are as follows.  The Higgs
amplitude is evaluated using the $[5/5]$~\emph{Pad\'e approximant} to the LME
after conformal mapping.  As demonstrated in
Sec.~\ref{sec:higgs_production_via_gluon_fusion}, this is virtually identical
to the exact result.  The massive quark box contributions are computed
by factoring out the exact LO amplitude according to
\Eqno{eq:LME_rescaled_Pade}, with the \emph{Pad\'e approximant} corresponding
to $n = m = 3$ in the definition given in \Eqno{eq:Padedefinition}.
The anomalous diagrams of Sec.~\ref{ssub:anomalous_diagrams} are not
included in the discussion of the massive quark loops below, but instead are
accounted for only when the sum of all loops is considered.

For massless quarks circulating in the loop the calculation is simplified by
the fact that the entire amplitude is proportional to the combination of
couplings $(v_f^2 + a_f^2)$, i.e. in the decomposition given in 
\Eqno{eq:massiveggZZ_normalisation} the quantities
$\ket{\TW{\cB}_{VV}^0}$ and $\ket{\TW{\cB}_{AA}^0}$ are equal.
The calculation requires the one-loop master integrals up to $\ep^2$,
for which all orders results are given in ref.~\cite{Smirnov:2006ry} for bubble integrals
and refs.~\cite{Bern:1994zx,Bern:1993kr,Brandhuber:2004yw,Brandhuber:2005kd,Cachazo:2004zb} for the \emph{easy box}
(two opposite off-shell legs). The necessary results for the \emph{three-mass} triangle with massless propagators
and the \emph{hard box} (two adjacent off-shell legs) can be taken from refs.~\cite{Chavez:2012kn}
and~\cite{Anastasiou:2014nha} respectively. We use the \emph{coproduct} formalism~\cite{Duhr:2012fh,Duhr:2014woa}
to \emph{analytical continue} the results to the physical phase space regions. All master integrals have been numerically
cross-checked with \texttt{SecDec}~\cite{Borowka:2015mxa}.  The two-loop master integrals for $gg\to ZZ$ are taken
from ref.~\cite{Gehrmann:2014bfa} and \texttt{GiNaC} is used to evaluate the polylogarithms.
Our results for this contribution agree with the earlier calculation of ref.~\cite{vonManteuffel:2015msa}.

\begin{figure}[ht!]
	\centering
	\begin{subfigure}{0.5\textwidth}
		\includegraphics[width=\textwidth]{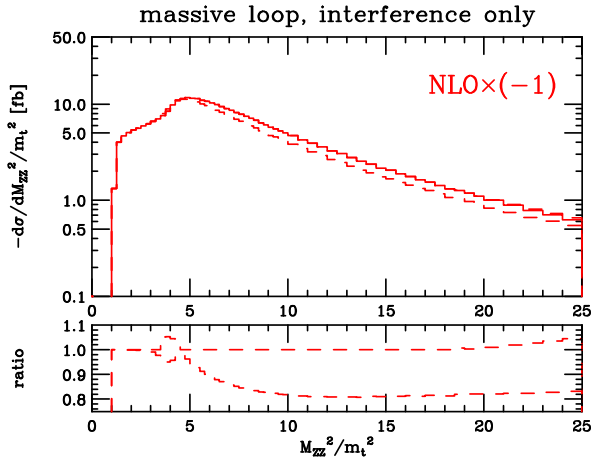}
	\end{subfigure}
	\caption{The uncertainty on the calculation of the massive loop interference contribution
        stemming from the use of the LME expansion and \emph{Pad\'e approximants}.   The central result is
        shown as a solid histogram, with the dashed lines indicating deviations that correspond
        to the grey area in Fig.~\ref{fig:ggZZ_NLO_melnikov}, multiplied by a factor of
        two.  All curves are computed for the central scale choice, $\mu_r = \mu_f = M_{ZZ}/2$.}
	\label{fig:NLO_massless_LMEuncertainty}
\end{figure}

The parameters for the following results have already been specified
in Sec.~\ref{ssub:ggH_comparison_lme_with_full_result}.  Here we make only one
change:  our central scale corresponds to the choice $\mu_r = \mu_f = M_{ZZ}/2$, where $M_{ZZ}$ is the
invariant mass of the $ZZ$ pair.  As an estimate of the theoretical uncertainty we consider variations
by a factor of two about this value.  We also introduce an uncertainty that is based on our combination of LME
and \emph{Pad\'e approximants} in the calculation of the massive quark loops, that has already been explored in
Fig.~\ref{fig:ggZZ_NLO_melnikov} (right).  In order to obtain a more conservative error estimate we multiply the
deviations of the extremal values in the grey area with respect to $\si_{\text{imp},[3/3]}^\text{NLO}$ by a factor of two.
The impact of this variation on the complete NLO prediction for the massive loop is shown in Fig.~\ref{fig:NLO_massless_LMEuncertainty}.
Even for this choice, the impact of the approximation is estimated to be less than $20$\% throughout the distribution.
For the remaining plots in this section we no longer show the impact of this uncertainty, but it will be explicitly
included in Tables~\ref{tab:xsecs} and~\ref{tab:xsecsabove300} later on.

\begin{figure}[ht!]
	\centering
	\begin{subfigure}{0.49\textwidth}
		\includegraphics[width=\textwidth]{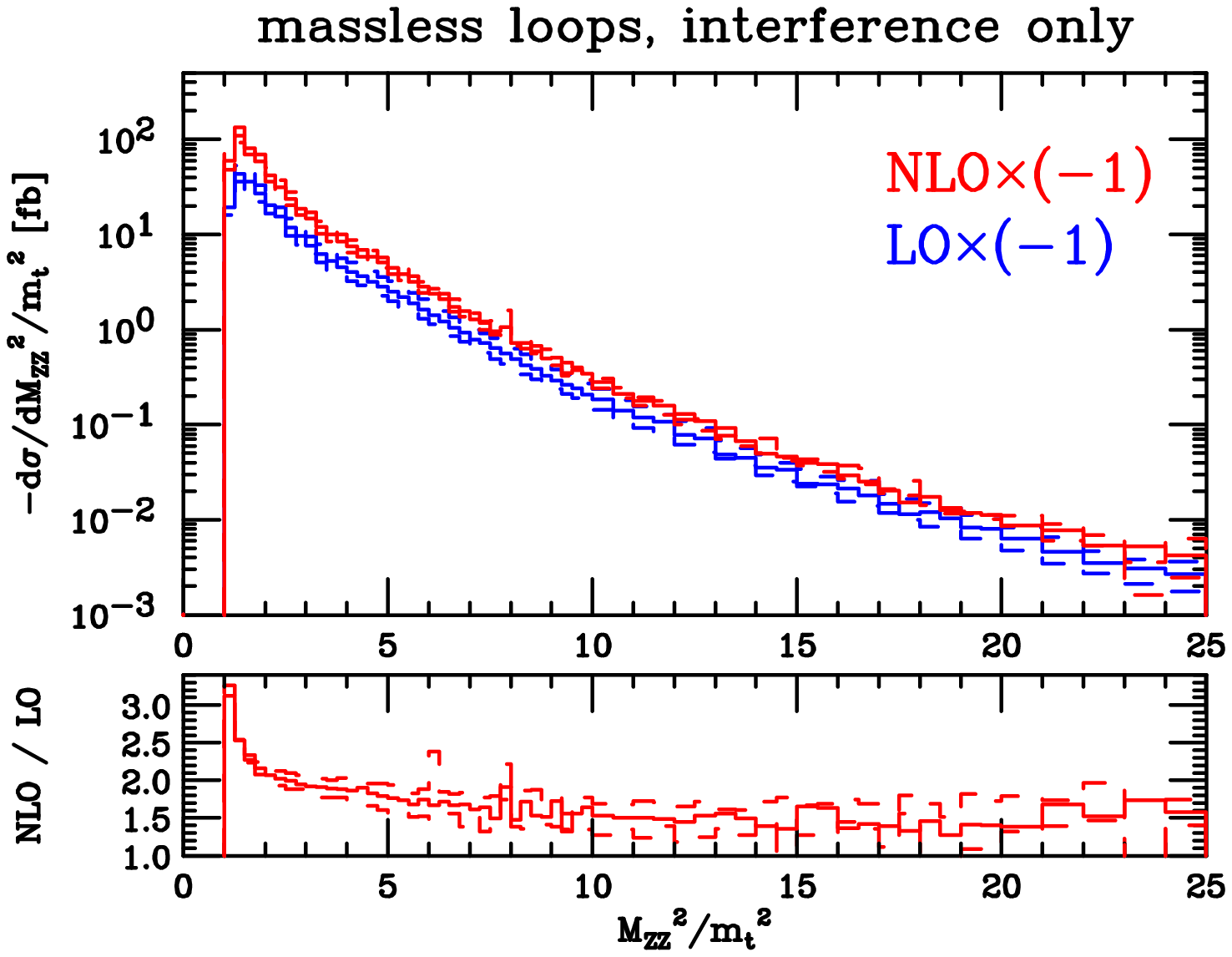}
	\end{subfigure}
	\hfill
	\begin{subfigure}{0.49\textwidth}
		\includegraphics[width=\textwidth]{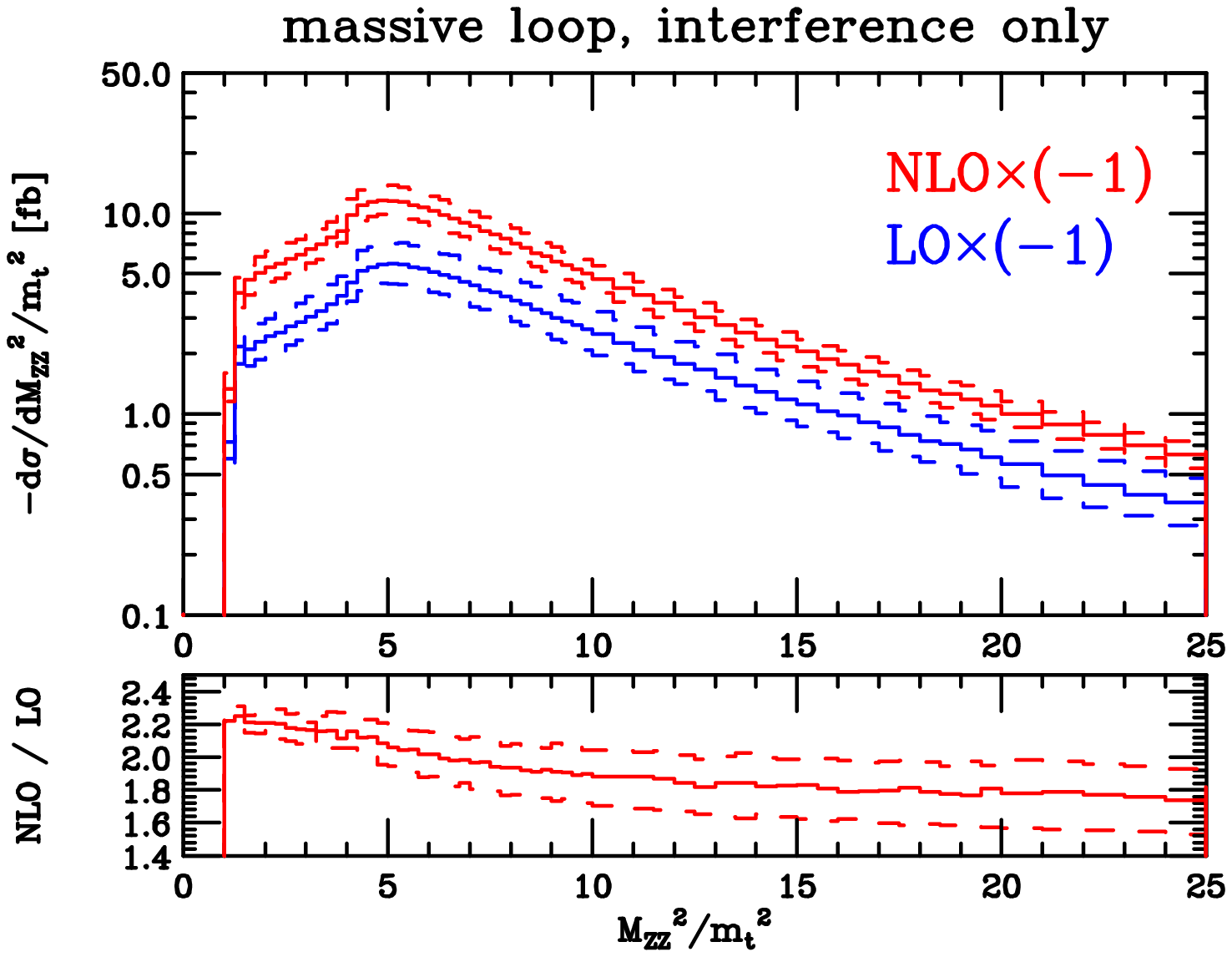}
	\end{subfigure}
	\caption{\textbf{Left panel:} Interference of the Higgs amplitude and massless quark loops at LO and NLO,
	with the scale uncertainty indicated by the dashed histograms.  The ratio of the
        NLO and LO results is shown in the lower panel.
                 \textbf{Right panel:} The equivalent results for the interference of the Higgs amplitude and the top quark loops.}
	\label{fig:NLO_massless_massive}
\end{figure}

Results for both the massless and massive quark contributions to the interference, including the effects of scale variation,
are shown in Fig.~\ref{fig:NLO_massless_massive}. The interference is negative for both
the massless and massive quark contributions and is shown in Fig.~\ref{fig:NLO_massless_massive}
reversed in sign. In both cases the $K$-factor
decreases as the invariant mass of the $Z$-boson pair increases.   The $K$-factor at small invariant
masses is larger for the massless loops;  as the invariant mass increases, the NLO corrections are more
important for the massive loop.  The NLO corrections are larger for the top quark loops
and exhibit a stronger dependence on $M_{ZZ}$.  In both cases the NLO result lies outside the
estimated LO uncertainty bands and the scale uncertainty is not significantly reduced at NLO.

\begin{figure}[ht!]
	\centering
	\begin{subfigure}{0.6\textwidth}
		\includegraphics[width=\textwidth]{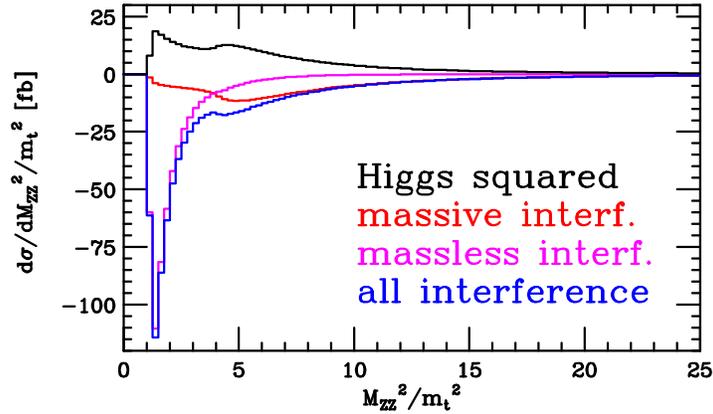}
	\end{subfigure}
	\caption{Comparison of the effect of the massless (magenta) and massive (red) loops in the NLO interference.
        Also shown is the sum (blue) and the corresponding result for the Higgs amplitude squared (black).
        All curves are computed for the central scale choice, $\mu_r = \mu_f = M_{ZZ}/2$.}
	\label{fig:compareNLO}
\end{figure}

\begin{figure}[ht!]
	\centering
	\begin{subfigure}{0.49\textwidth}
		\includegraphics[width=\textwidth]{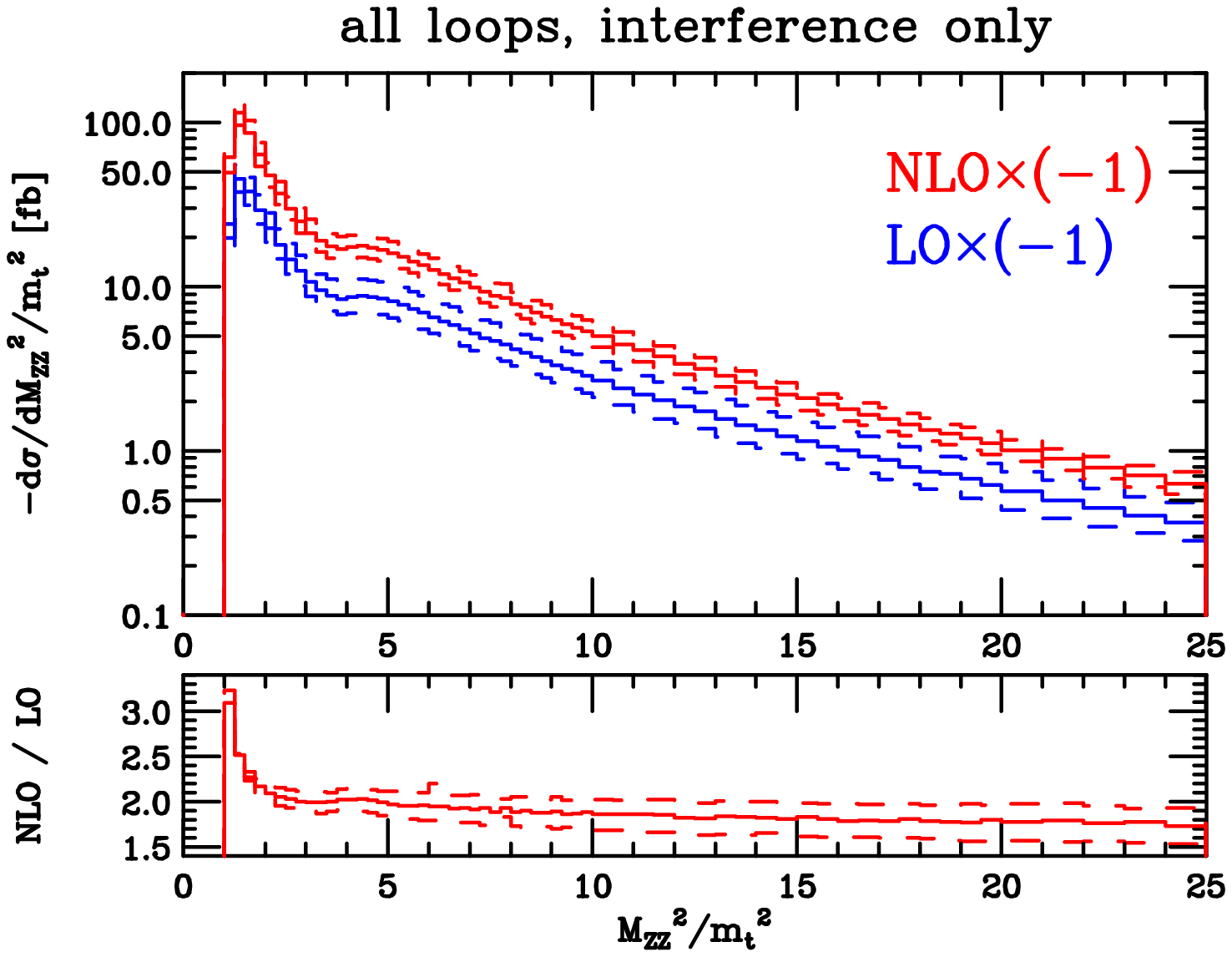}
	\end{subfigure}
	\hfill
	\begin{subfigure}{0.49\textwidth}
		\includegraphics[width=\textwidth]{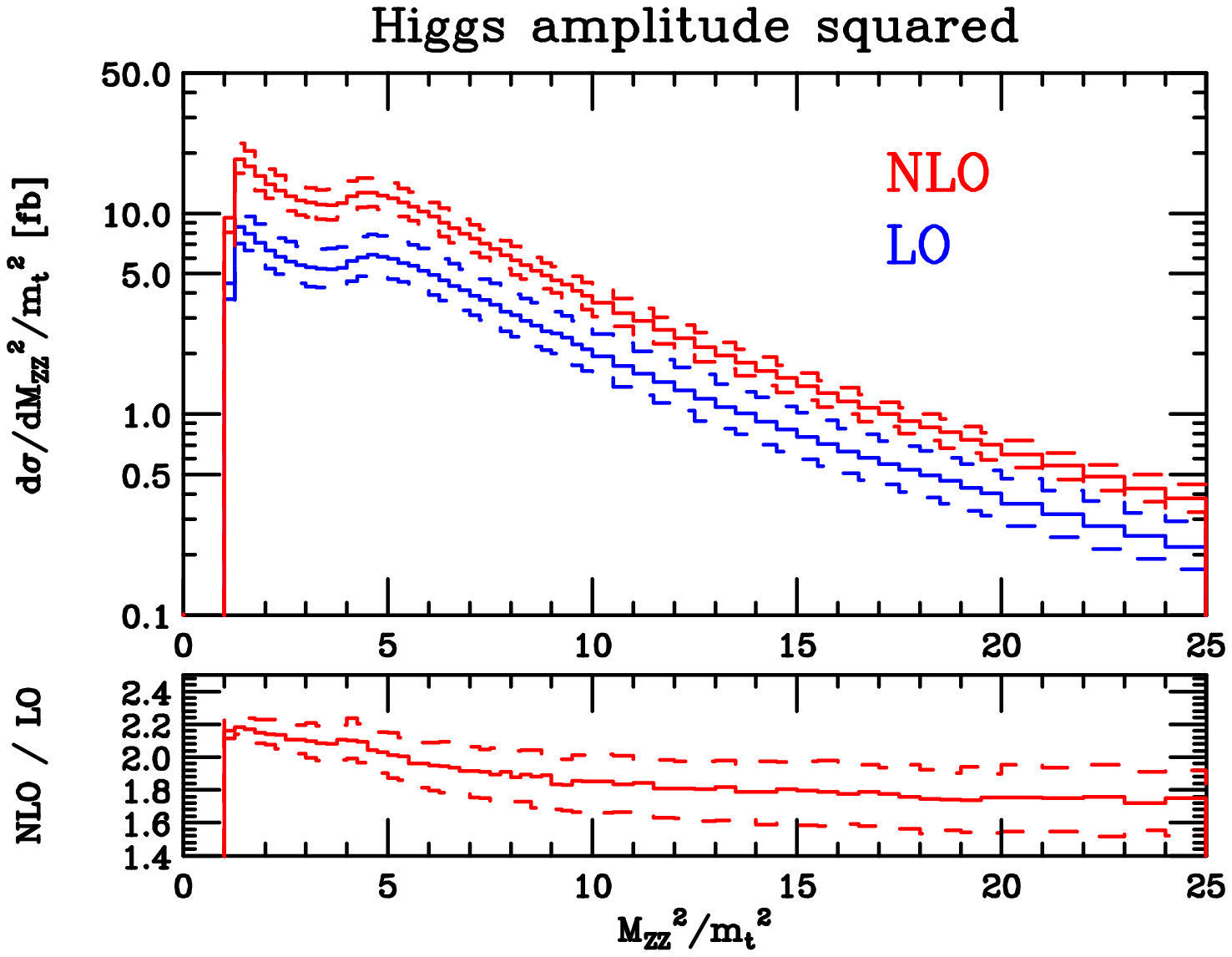}
	\end{subfigure}
	\caption{\textbf{Left panel:} Interference of the Higgs amplitude and quark loops at LO and NLO,
	with the scale uncertainty indicated by the dashed histograms.  The ratio of the
        NLO and LO results is shown in the lower panel.
                 \textbf{Right panel:} The equivalent results for the Higgs amplitude squared.}
	\label{fig:NLO_full_higgs}
\end{figure}

\begin{figure}[ht!]
	\centering
	\begin{subfigure}{0.5\textwidth}
		\includegraphics[angle=-90,width=\textwidth]{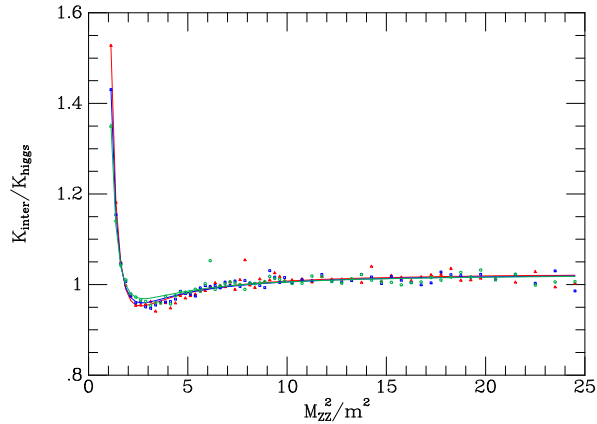}
	\end{subfigure}
	\caption{The ratio of the $K$-factors for the square of the Higgs diagrams alone ($K_{\mbox{higgs}}$)
        and the one for the interference ($K_{\mbox{inter}}$).  The lines are fits to the individual histogram
        bins that are good to the level of a few percent and are shown for the central scale (blue) as well as
        the scale variations (red, green).}
	\label{fig:Krat}
\end{figure}

The relative importance of the massive and massless loops can be
better-assessed from the NLO predictions shown in
Fig.~\ref{fig:compareNLO}.  At smaller invariant masses, below the
top-pair threshold, the massless loops are most important.  Around the
top-pair threshold the two are of a similar size, but at high energies
the massless loops are insignificant.  In contrast, the top quark loop
quickly becomes the dominant contribution beyond this threshold and
exhibits a long tail out to invariant masses of around one TeV.
The full prediction for the interference that is obtained by summing over
both massless and top quark loops, as well as the numerically-small
anomalous contribution discussed in Sec.~\ref{ssub:anomalous_diagrams},
is shown in Fig.~\ref{fig:NLO_full_higgs}.  The
relative size of the massless and top quark loops discussed above
means that the behaviour of the $K$-factor for the sum of both
contributions interpolates between the massless-loop $K$-factor for
small $M_{ZZ}$ and the massive loop one for high $M_{ZZ}$.  It
therefore decreases from around $3$ at the peak of the distribution to
approximately $1.8$ in the tail.  This is to be contrasted with the
$K$-factor distribution for the pure Higgs amplitudes alone, shown in
the right panel of Fig.~\ref{fig:NLO_full_higgs}.  In that case the
$K$-factor decreases slowly from around $2.2$ at small invariant
masses to around $1.8$ in the far tail.  We note that the $K$-factor
for the Higgs amplitudes alone, and the one for the interference with
the top quark loops, is almost identical. 
In the high-energy limit this is guaranteed to be the case, due to the cancellation between
these two processes.  This behaviour is shown explicitly in Fig.~\ref{fig:Krat}.

The integrated cross-sections for the interference contributions and the Higgs amplitude squared
are shown in Table~\ref{tab:xsecs}.  Note that, in this table, the total interference differs from
the sum of the massive and massless loops by a small amount that is due to the anomalous contribution. 
At this level the differences between the effects of the NLO
corrections on the various contributions is quite small, with all corresponding to a NLO enhancement
by close to a factor of two.  The $K$-factor for the massless loops is slightly larger, which
is also reflected in the result for the total interference.  In addition to the scale uncertainty, we have
also indicated our estimate of the residual uncertainty related to the LME expansion that is indicated
in Fig.~\ref{fig:NLO_massless_LMEuncertainty}.  The impact of this uncertainty is relatively small, at the
level of around $5$\%, due to the fact that the integrated cross-section is dominated by the region
$M_{ZZ}^2 \lesssim 5 m^2$ where the LME is expected to work well.
\begin{table}
\begin{center}
\begin{tabular}{|l|l|l|l|}
\hline
Contribution            & $\sigma_{LO}$ [fb]      & $\sigma_{NLO}$ [fb]          & $\sigma_{NLO}/\sigma_{LO}$\\
\hline
Higgs mediated diagrams  & ~~~$56.3^{+15.3}_{-11.4}$  & ~~$111.0^{+20.1}_{-16.6}$ & 1.97 \\
interference (total)    & $-113.5^{+22.2}_{-29.5}$ & $-237.8^{+36.4}_{-45.4}$(scale)${}^{+5.4}_{-0.4}$(LME)    & 2.09 \\
\hline
interference (massless loops) & $-60.2^{+11.0}_{-14.2}$ & $-132.7^{+20.5}_{-26.3}$     & 2.20 \\
interference (massive loop)  & $-53.3^{+11.2}_{-15.3}$ & $-104.2^{+15.8}_{-18.7}$(scale)${}^{+5.4}_{-0.4}$(LME)     & 1.95 \\
\hline
\end{tabular}
\end{center}
\caption{Integrated cross-sections at $\sqrt{S}=13$~TeV, using the input parameters of
Sec.~\ref{ssub:ggH_comparison_lme_with_full_result} and $\mu=M_{ZZ}/2$. Uncertainties correspond to scale variation
as described in the text and, for NLO results that include massive quarks, an estimate of the limitations
of the LME. The $K$-factor is computed using only the central result.}
\label{tab:xsecs}
\end{table}

For obtaining a bound on the width of the Higgs boson it is useful to focus on a high-mass region where 
backgrounds from the continuum processes, represented at tree-level by $q\bar q \to ZZ$, are small but
the effect of the interference is still significant~\cite{Caola:2013yja,Campbell:2013una}. To that
end, in Table~\ref{tab:xsecsabove300} we show the cross-sections after the application of the
cut $M_{ZZ} > 300$~GeV.  We see that, as expected, the impact of the massive top loop on the interference
is much greater, compared to the massless loops.  This also has the effect of ensuring that the $K$-factors
for the Higgs amplitude squared and the total interference are almost equal.  To estimate the cross-section
after the decays of the $Z$-bosons into electrons and muons we can simply take these results and multiply by a factor
of $4 \times BR(Z \to e^- e^+)^2$, where $BR(Z \to e^- e^+) = 3.363 \times 10^{-2}$.  Assuming that the on-shell
Higgs cross-section takes its Standard Model value and that the Higgs boson couplings and width are related accordingly,
we can write the predictions for the off-shell region as,
\begin{eqnarray}
\sigma_{4\ell}^{LO} (m_{4\ell} > 300~\mbox{GeV}) = 
 \left(0.190^{+0.055}_{-0.040}\right) \times \left( \frac{\Gamma_H}{\Gamma_H^{SM} }\right)
-\left(0.275^{+0.079}_{-0.058}\right) \times \sqrt{ \frac{\Gamma_H}{\Gamma_H^{SM} }} \; \mbox{fb} \,, \label{eq:offshellLO} \\
\sigma_{4\ell}^{NLO} (m_{4\ell} > 300~\mbox{GeV}) = 
 \left(0.365^{+0.064}_{-0.054}\right) \times \left( \frac{\Gamma_H}{\Gamma_H^{SM} }\right)
-\left(0.526^{+0.092}_{-0.103}\right) \times \sqrt{ \frac{\Gamma_H}{\Gamma_H^{SM} }} \; \mbox{fb} \,. \label{eq:offshellNLO}
\end{eqnarray}
The linear terms derive from the Higgs cross-sections in Table~\ref{tab:xsecsabove300} while the terms that
scale as the square-root of the modified width reflect the total interference contributions.
The uncertainties reflect those shown in Table~\ref{tab:xsecsabove300}, with the scale and LME uncertainties
added linearly. It is interesting to compare these results with the corresponding on-shell Higgs cross-sections.
These are given by,
\begin{eqnarray}
\sigma_{4\ell}^{LO} (m_{4\ell} < 130~\mbox{GeV}) = 1.654^{+0.249}_{-0.220} \; \mbox{fb} \;, \qquad 
\sigma_{4\ell}^{NLO} (m_{4\ell} < 130~\mbox{GeV}) = 3.898^{+0.770}_{-0.560} \; \mbox{fb} \,, \label{eq:onshellLOandNLO}
\end{eqnarray}
where the uncertainties correspond to our usual scale variation procedure.
From the results in Eqs.~(\ref{eq:offshellLO}) and~(\ref{eq:offshellNLO}) it is clear that the absolute rate of off-shell events varies
considerably between LO and NLO.  On the other hand, the cross-sections in Eq.~(\ref{eq:onshellLOandNLO}) imply that the ratio of the
number of events in the off-shell region compared to the peak region is much better predicted,
\begin{eqnarray}
\frac{\sigma_{4\ell}^{LO} (m_{4\ell} > 300~\mbox{GeV})}{\sigma_{4\ell}^{LO} (m_{4\ell} < 130~\mbox{GeV})} = 
 \left(0.115^{+0.014}_{-0.010}\right) \times \left( \frac{\Gamma_H}{\Gamma_H^{SM} }\right)
-\left(0.166^{+0.020}_{-0.015}\right) \times \sqrt{ \frac{\Gamma_H}{\Gamma_H^{SM} }}  \,, \nonumber \\
\frac{\sigma_{4\ell}^{NLO} (m_{4\ell} > 300~\mbox{GeV})}{\sigma_{4\ell}^{NLO} (m_{4\ell} < 130~\mbox{GeV})} = 
 \left(0.094^{+0.000}_{-0.002}\right) \times \left( \frac{\Gamma_H}{\Gamma_H^{SM} }\right)
-\left(0.135^{+0.000}_{-0.008}\right) \times \sqrt{ \frac{\Gamma_H}{\Gamma_H^{SM} }}  \,. 
\end{eqnarray}
The uncertainties in this equation are obtained by using both the LME uncertainty estimate and the scale
variation, but ensuring that the cross-sections that appear in the numerator and denominator are evaluated
at the same scale.
\begin{table}
\begin{center}
\begin{tabular}{|l|l|l|l|}
\hline
Contribution            & $\sigma_{LO}$ [fb]      & $\sigma_{NLO}$ [fb]          & $\sigma_{NLO}/\sigma_{LO}$\\
\hline
Higgs mediated diagrams & ~~$42.1^{+12.1}_{-8.8}$ & ~~~$80.7^{+14.2}_{-12.0}$ & 1.92 \\
interference (total)    & $-60.7^{+12.8}_{-17.4}$ & $-116.3^{+17.5}_{-19.9}$(scale)${}^{+5.4}_{-0.4}$(LME) & 1.91 \\
\hline
interference (massless loops) & $-12.5^{+2.5}_{-3.4}$   & $-22.5^{+3.2}_{-3.2}$    & 1.80 \\
interference (massive loop)  & $-48.2^{+10.3}_{-14.1}$  & $-93.0^{+14.0}_{-16.4}$(scale)${}^{+5.4}_{-0.4}$(LME)  & 1.93 \\
\hline
\end{tabular}
\end{center}
\caption{Cross-sections at $\sqrt{S}=13$~TeV in the region defined by $M_{ZZ} > 300$~GeV,
using the input parameters of
Sec.~\ref{ssub:ggH_comparison_lme_with_full_result}.   Uncertainties correspond to scale variation
as described in the text and, for NLO results that include massive quarks, an estimate of the limitations
of the LME. The $K$-factor is computed using only the central result.}
\label{tab:xsecsabove300}
\end{table}

\section{Conclusions} 
\label{sec:conclusions}

In this paper we have presented a calculation of on-shell $Z$-boson pair
production via gluon fusion at the two-loop level.  This occurs both
through diagrams that are mediated by a Higgs boson, with $H \to ZZ$,
and by continuum contributions in which the $Z$ bosons couple through loops
of quarks.  We have considered contributions up to the two-loop level, corresponding
to NLO corrections, for the Higgs diagrams alone and also for the interference
between the two sets of diagrams.

In the continuum contribution the two-loop corrections containing loops of
massless quarks are known and we have reproduced results from the
literature.  Our treatment of the massive quark loops is based on a large-mass
expansion up to order $1/m^{12}$, that is extended to the high-mass
region by using a combination of conformal mapping and \emph{Pad\'e approximation}.
This procedure was shown to provide an excellent approximation of the
Higgs contribution alone, where the exact result is known.
Additionally, applying the large-mass expansion in combination 
with the conformal mapping and the \emph{Pad\'e approximation} to the 
$gg\to ZZ$ amplitudes is obviously not limited to the interference calculation
alone. The same procedure can also be applied to the virtual two-loop 
$gg\to ZZ$ amplitude including its full tensor structure. It might be 
desirable to apply the presented procedure also to the Higgs-boson 
pair-production process, because the latter offers identical kinematics. 
Comparing those results with the recently published results including the full 
top mass effects~\cite{Borowka:2016ehy} could lead to interesting insights 
concerning the error estimate of the used approximation. However, this is kept
as future work.

We have used our calculation to provide theoretical predictions for the impact
of the interference contribution on the invariant mass distribution of
$Z$-boson pairs at the $13$~TeV LHC.  In the high-mass region we have shown
that the impact of the NLO corrections to the interference are practically
identical to those for Higgs production alone.   This explicit calculation
justifies using a procedure for estimating the number of off-shell events
due to the interference by rescaling the LO prediction by the on-shell
$K$-factor.



\acknowledgments
S.K. would like to thank Claude Duhr and Robert M. Schabinger for clarifying 
conversations about the \emph{coproduct} formalism, and Robert V. Harlander and 
Paul Fiedler for helpful discussions.
S.K. was supported by the Deutsche Forschungsgemeinschaft through 
Graduiertenkolleg GRK 1675.
Fermilab is operated by Fermi Research Alliance, LLC under Contract No. 
De-AC02-07CH11359 with the United States Department of Energy.


\appendix
\section{Definition of Scalar Integrals} 
\label{Intdef}

We work in the Bjorken-Drell metric so that
$l^2=l_0^2-l_1^2-l_2^2-l_3^2$. 
The definition of the integrals is as follows
\begin{eqnarray}
&& A_0(m)  =
 \frac{\mu^{4-d}}{i \pi^{\frac{d}{2}}\rG}
\int d^d l \;
 \frac{1}
{(l^2-m^2+i\varepsilon)}\,, \\
&& B_0(p_1;m,m)  =
 \frac{\mu^{4-d}}{i \pi^{\frac{d}{2}}\rG}
\int d^d l \;
 \frac{1}
{(l^2-m^2+i\varepsilon)
((l+p_1)^2-m^2+i\varepsilon)}\,, \\
&& C_0(p_1,p_2;m,m,m)  =
 \frac{\mu^{4-d}}{i \pi^{\frac{d}{2}}\rG}  \\
&& \times \int d^d l \;
 \frac{1}
{(l^2-m^2+i\varepsilon)
((l+p_1)^2-m^2+i\varepsilon)
((l+p_1+p_2)^2-m^2+i\varepsilon)}\,,\nn \\
&&
D_0(p_1,p_2,p_3;m,m,m,m) =
\frac{\mu^{4-d}}{i \pi^{\frac{d}{2}}\rG} \\
\nn \\
&&
\times \int d^d l \;
 \frac{1}
{(l^2-m^2+i\varepsilon)
((l+p_1)^2-m^2+i\varepsilon)
((l+p_1+p_2)^2-m^2+i\varepsilon)
((l+p_1+p_2+p_3)^2-m^2+i\varepsilon)}\,, \nn
\end{eqnarray}
We have removed the overall constant which occurs in $d$-dimensional integrals, ($d=4-2\ep$)
\beq
\rG\equiv\Gamma(1+\e)=
1-\e \gamma+\e^2\Big[\frac{\gamma^2}{2}+\frac{\pi^2}{12}\Big]
\eeq
with the Euler-Mascheroni constant $\ga=0.57721\ldots$. The large mass expansion of some of these integrals are
\begin{align*}
	B_0\left((p_1+p_2)^2,m,m\right) &\overset{s=1}{=} \left(\frac{\mu^2}{m^2}\right)^\ep \Bigg\{
	\frac{1}{\ep}
	+ \frac{1}{6} \frac{1}{r_t}
	+ \frac{1+\ep}{60} \frac{1}{r_t^2}
	+ \frac{(1+\ep) (2+\ep)}{840} \frac{1}{r_t^3}
	+ \frac{6+11 \ep+6 \ep^2}{15120} \frac{1}{r_t^4} \numberthis\\
	&\hspace{-2.5cm}+ \frac{24+50 \ep+35 \ep^2}{332640} \frac{1}{r_t^5}
	+ \frac{120+274 \ep+225 \ep^2}{8648640} \frac{1}{r_t^6}
	+ \frac{180+441 \ep+406 \ep^2}{64864800} \frac{1}{r_t^7}
	+ \frac{1260+3267 \ep+3283 \ep^2}{2205403200} \frac{1}{r_t^8} \\
	&\hspace{-2.5cm}+ \frac{10080+27396 \ep+29531 \ep^2}{83805321600} \frac{1}{r_t^9}
	+ \frac{10080+28516 \ep+32575 \ep^2}{391091500800} \frac{1}{r_t^{10}}
	+ \cO\left(1/r_t^{11},\ep^3\right) \Bigg\}
\end{align*}
and
\begin{align*}
	C_0\left(p_1,p_2,m,m,m\right) &\overset{s=1}{=} -\left(\frac{\mu^2}{m^2}\right)^\ep \Bigg\{
	 \frac{1}{2} \frac{1}{r_t}
	+ \frac{1+\ep}{24} \frac{1}{r_t^2}
	+ \frac{(1+\ep) (2+\ep)}{360} \frac{1}{r_t^3}
	+ \frac{6+11 \ep+6 \ep^2}{6720} \frac{1}{r_t^4} \numberthis\\
	&\hspace{-2.5cm}+ \frac{24+50 \ep+35 \ep^2}{151200} \frac{1}{r_t^5}
	+ \frac{120+274 \ep+225 \ep^2}{3991680} \frac{1}{r_t^6}
	+ \frac{180+441 \ep+406 \ep^2}{30270240} \frac{1}{r_t^7}
	+ \frac{1260+3267 \ep+3283 \ep^2}{1037836800} \frac{1}{r_t^8}\\
	&\hspace{-2.5cm}+ \frac{10080+27396 \ep+29531 \ep^2}{39697257600} \frac{1}{r_t^9}
	+ \frac{10080+28516 \ep+32575 \ep^2}{186234048000} \frac{1}{r_t^{10}}
	+\cO\left(1/r_t^{11},\ep^3\right) \Bigg\}\,.
\end{align*}
for $p_1^2=p_2^2=0,(p_1+p_2)^2=s$.


\section{Scale Dependence of the Finite Remainder} 
\label{sec:scale_dependence_of_the_finite_remainder}
In this section we shortly summarise a convenient, and well-known, way to determine the dependence on the renormalisation scale $\mu=\mu_r$ of the one- and two-loop finite remainders used within this work, i.e. processes with a loop-induced leading-order matrix element. This determination is possible by exploiting the \emph{renormalisation group equation} (RGE) properties of the individual building block, e.g. $\as^{(n_f)}(\mu)$, as discussed below. Knowledge of this scale dependence, in return, offers a simple way to compute finite remainder results at arbitrary scales, provided the results at a starting scale $\mu_0$ are known. We mostly recycle our definitions from Sec.~\ref{sub:ggH_preliminaries}. In the following, however, we stick to a slightly more general notation when applicable. To this end we drop the amplitude specifications $\cA$ and $\cB$ from the finite remainder definition in Eq.~\eqref{eq:FinRem_def} and denote our previous amplitudes $\cA$ and $\cB$ simply by $\cM$. We also replace our, to the $gg\to ZZ$ process specialised, IR constant $\Zop_{gg}^{(n_l)}$ from Eq.~\eqref{eq:Zop_as} by a more general IR constant $\Zop_{IR}$ following the notation in~\cite{Ferroglia:2009ii,Baernreuther:2013caa,Czakon:2014oma}.  The finite remainder for $n_f$ quark flavours is thus defined by
\begin{align*}
	\label{eq:scalecheck_FinRem_def}
	\ket{\cF(\as^{(n_f)},m,\mu)} &= \frac{1}{\Zop_{IR}} \,\ket{\cM^r(\as^{(n_f)},m,\mu)}  = \frac{Z_{UV}^{(n_f)}}{\Zop_{IR}} \left(\frac{N^\ep Z^{(n_f)}_{\as} \as^{(n_f)}(\mu)}{4\pi}\right) \Bigg[ \ket{\cM^{(1),0}(m)} \\
	&+ \left(\frac{N^\ep Z^{(n_f)}_{\as} \as^{(n_f)}(\mu)}{4\pi}\right) \ket{\cM^{(2),0}(m)} \Bigg] + \cO\left((\as^{(n_f)})^3\right) \,. \numberthis
\end{align*}
The mass dependence does not play any important role in the subsequent discussion and, hence, all results are valid for arbitrary masses $m$. $Z_{UV}^{(n_f)}$ denotes the process dependent $UV$ renormalisation constants and the mass renormalisation $m^0=Z_m m$ is again kept implicit. The strong coupling constants $\as$ is renormalised according to
\begin{equation}
	\as^0 = N^\ep Z^{(n_f)}_{\as} \as^{(n_f)}(\mu) \quad \text{with} \quad N^\ep = \mu^{2\ep} \frac{e^{\ep\,\ga_E}}{(4\pi)^\ep}\,,
\end{equation}
where the explicit $\mu$ dependence from the loop measure in Eq.~\eqref{eq:loopnorm_ggH} was shifted to $N^\ep$. The renormalisation constant $Z^{(n_f)}_{\as}$ and the coefficient of the beta function $\be_0^{(n_f)}$ are given in Eq.~\eqref{eq:Zas_as}. The explicit scale and flavour dependence of $\as=\as^{(n_f)}(\mu)$ is neglected in the following for simplicity.

Equivalently to Eq.~\eqref{eq:scalecheck_FinRem_def} we define the perturbative expansion of the finite remainder as
\begin{equation}
	\label{eq:scalecheck_FinRem_expansion}
	\ket{\cF(\as,m,\mu)} = \frac{\as}{4\pi} \ket{\cF^{(1)}(m,\mu)} + \left(\frac{\as}{4\pi}\right)^2 \ket{\cF^{(2)}(m,\mu)} + \cO\left(\as^3\right) \,.
\end{equation}
Taking the derivative with respect to $\mu^2$ of Eq.~\eqref{eq:scalecheck_FinRem_def} and Eq.~\eqref{eq:scalecheck_FinRem_expansion} leads to
\begin{align*}
	\label{eq:scalecheck_FinRem_der}
	\mu^2 \frac{d\,}{d\mu^2} \ket{\cF(\as,m,\mu)} &= \left(\mu^2 \frac{d}{d\mu^2}\left(\frac{\as}{4\pi}\right)\right) \ket{\cF^{(1)}(m,\mu)} + \frac{\as}{4\pi} \;\mu^2 \frac{d\,}{d\mu^2} \ket{\cF^{(1)}(m,\mu)} \\
	&+2 \left(\frac{\as}{4\pi}\right) \left(\mu^2 \frac{d\,}{d\mu^2} \left(\frac{\as}{4\pi}\right)\right) \ket{\cF^{(2)}(m,\mu)} + \left(\frac{\as}{4\pi}\right)^2\; \mu^2 \frac{d\,}{d\mu^2} \ket{\cF^{(2)}(m,\mu)} \numberthis\\
	&= \mu^2 \frac{d\,}{d\mu^2} \Bigg\{ \frac{Z_{UV}^{(n_f)}}{\Zop_{IR}} \left(\frac{N^\ep Z^{(n_f)}_{\as} \as^{(n_f)}(\mu)}{4\pi}\right) \Bigg[ \ket{\cM^{(1),0}(m)} \\
	&+ \left(\frac{N^\ep Z^{(n_f)}_{\as} \as^{(n_f)}(\mu)}{4\pi}\right) \ket{\cM^{(2),0}(m)} \Bigg] \Bigg\} \,.
\end{align*}
The derivatives of $Z_{UV}^{(n_f)}$ and $Z_m$ vanish because these renormalisation constants are defined in the on-shell scheme. The explicit $\mu$ dependence within these expressions cancels against the $\as$ scale dependence. The derivative of $\Zop_{IR}$ with respect to $\mu$ is given by its RGE~\cite{Ferroglia:2009ii,Baernreuther:2013caa,Czakon:2014oma} and therefore
\begin{equation}
	\mu^2 \frac{d\,}{d\mu^2} \frac{1}{\Zop_{IR}} = -\frac{1}{\Zop_{IR}^2} \frac{1}{2}\; \underbrace{\frac{d}{d\log\mu}\Zop_{IR}}_{-\hat\Gamma \cdot \Zop_{IR}} = \frac{\as}{4\pi} \frac{\hat\Gamma^{(1)}}{2}\cdot \frac{1}{\Zop_{IR}} + \cO\left(\as^2\right)\,.
\end{equation}
The anomalous dimension operator $\hat\Gamma$ can be taken from~\cite{Czakon:2014oma} and references therein. For our $gg\to ZZ$ processes $\hat{\Gamma}$ simplifies to
\begin{align*}
	\hat\Gamma = \frac{\as}{4\pi}\, \hat\Gamma^{(1)} + \cO\left(\as^2\right) &= \frac{\as}{4\pi} \left(-4C_A\log\left(\frac{\mu^2}{-s-i\ep}\right)  -2\be_0^{(n_f)} \right) + \cO\left(\as^2\right) \numberthis\\
	&= \frac{\as}{4\pi} \left( \hat{K}^{(1)} + \hat{D}^{(1)}\cdot \log\left(\frac{\mu^2}{\mu_0^2} \right)\right)+ \cO\left(\as^2\right)
\end{align*}
with
\begin{equation}
	\label{eq:scalecheck_KD_def}
	\hat{K}^{(1)} = -4C_A \,\log\left(\frac{\mu_0^2}{-s-i\ep}\right) - 2\be_0^{(n_f)} \quad \text{and} \quad \hat{D}^{(1)} = -4C_A\,.
\end{equation}
The remaining derivatives up to $\cO\left(\as^2\right)$
\begin{equation}
	\mu^2 \frac{d}{d\mu^2} \left(\frac{g_s^2}{4\pi}\right) = \as \left(-\ep- \be_0^{(n_f)}\,\frac{\as}{4\pi}\right),\quad \mu^2 \frac{d\,}{d\mu^2}\; N^\ep = \ep\;N^\ep \quad \text{and}\quad \mu^2 \frac{d\,}{d\mu^2}\; Z_{\as}^{(n_f)} = Z_{\as}^{(n_f)}\;\be_0^{(n_f)} \frac{\as}{4\pi}
\end{equation}
combine to
\begin{equation}
	\mu^2 \frac{d\,}{d\mu^2}\; \left(\frac{Z_{UV}^{(n_f)}}{\Zop_{IR}} \;\frac{N^\ep Z_{\as}\as}{4\pi}\right) = \frac{Z_{UV}^{(n_f)}}{\Zop_{IR}}\;\frac{N^\ep Z_{\as}\as}{4\pi} \left[ \frac{\as}{4\pi} \frac{\hat\Gamma^{(1)}}{2} \right]\,.
\end{equation}
Using the shorthand notation $\mu^2 \frac{d\,}{d\mu^2} \ket{\cF} = \frac{d}{d\log\mu^2} \ket{\cF} = \ket{\cF^{'}}$ Equation~\eqref{eq:scalecheck_FinRem_der} becomes
\begin{align*}
	\mu^2 \frac{d\,}{d\mu^2}\; \ket{\cF(\as,m,\mu)} &= \frac{\as}{4\pi} \left(-\ep- \be_0^{(n_f)}\frac{\as}{4\pi} \right) \ket{\cF^{(1)}(m,\mu)} + \frac{\as}{4\pi} \ket{\cF^{(1)'}(m,\mu)} \\
	&+ 2 \left(\frac{\as}{4\pi}\right)^2 \left(-\ep- \be_0^{(n_f)}\frac{\as}{4\pi} \right)\ket{\cF^{(2)}(m,\mu)} + \left(\frac{\as}{4\pi}\right)^2 \ket{\cF^{(2)'}(m,\mu)} \numberthis\\
	&=\left(\frac{\as}{4\pi}\right)^2 \frac{\hat\Gamma^{(1)}}{2} \ket{\cF^{(1)}(m,\mu)} + \cO\left(\as^3\right)\,.
\end{align*}
Comparing each order in $\as$ yields the system of differential equations
\begin{align}
	\label{eq:scalecheck_diffeq_sys}
	\Rightarrow& \left(\frac{\as}{4\pi}\right) \left(\ket{\cF^{(1)'}(m,\mu)}-\ep \ket{\cF^{(1)}(m,\mu)}\right) = 0 \\
	\Rightarrow& \left(\frac{\as}{4\pi}\right)^2 \left(\ket{\cF^{(2)'}(m,\mu)}-2\ep \ket{\cF^{(2)}(m,\mu)}-\left(\be_0^{(n_f)}+\frac{\hat\Gamma^{(1)}}{2}\right)\ket{\cF^{(1)}(m,\mu)}\right) = 0\,.
\end{align}
Solving the homogeneous differential equations for the leading- and next-to-leading-order finite remainder results in
\begin{equation}
	\label{eq:scalecheck_hom_eq}
	\ket{\cF^{(1)}(m,\mu)} = \left(\frac{\mu^2}{\mu_0^2}\right)^\ep \ket{\cF^{(1)}(m,\mu_0)} \quad \text{and} \quad \ket{\cF^{(2)}(m,\mu)}_h = \left(\frac{\mu^2}{\mu_0^2}\right)^{2\ep} \ket{\cF^{(2)}(m,\mu_0)}\,.
\end{equation}
The inhomogeneous equation for the NLO finite remainder can easily be solved by \emph{variation of constants}. We make an ansatz for the solution of the inhomogeneous equation and write the homogeneous solution as
\begin{equation}
	\ket{\cF^{(2)}(m,\mu)} = C(\mu)\,e^{F(\log\mu^2)}\, \quad \text{with} \quad F(\log\mu^2) = \int\limits_{\log\mu_0^2}^{\log\mu^2} \; 2\ep \;d\log\mu^2\,.
\end{equation}
Reinsertion into Eq.~\eqref{eq:scalecheck_diffeq_sys} yields the differential equation for $C(\mu)$
\begin{equation}
	\label{eq:scalecheck_Cdiff}
	C^{'}(\mu) = e^{-F(\log\mu^2)}\cdot \left(\be_0^{(n_f)} + \frac{\hat\Gamma^{(1)}}{2}\right) \ket{\cF^{(1)}(m,\mu)} \overset{\eqref{eq:scalecheck_hom_eq}}{=} \left(\frac{\mu^2}{\mu_0^2}\right)^{-\ep} \left(\be_0^{(n_f)}+\frac{\hat\Gamma^{(1)}}{2}\right) \ket{\cF^{(1)}(m,\mu_0)}\,.
\end{equation}
Solving Eq.~\eqref{eq:scalecheck_Cdiff} by an elementary integration using the decomposition of $\hat\Gamma^{(1)}$ into $\hat{K}^{(1)}$ and $\hat{D}^{(1)}$ from Eq.~\eqref{eq:scalecheck_KD_def} and combining the particular solution with the homogeneous solution from Eq.~\eqref{eq:scalecheck_hom_eq} yields for the scale dependence of the one- and two-loop finite remainders
\begin{align*}
	\ket{\cF^{(1)}(m,\mu)} &\overset{\ep\to 0}{=} \ket{\cF^{(1)}(m,\mu_0)} \qquad \text{and} \numberthis\\
	\ket{\cF^{(2)}(m,\mu)} &\overset{\ep\to 0}{=} \ket{\cF^{(2)}(m,\mu_0)} + \left[\log\left(\frac{\mu^2}{\mu_0^2}\right) \left(\be_0^{(n_f)} +\frac{\hat{K}^{(1)}}{2} \right) + \frac{\hat{D}^{(1)}}{4}\log^2\left(\frac{\mu^2}{\mu_0^2}\right)\right] \ket{\cF^{(1)}(m,\mu_0)} \numberthis \\
	&= \ket{\cF^{(2)}(m,\mu_0)} -2 C_A \log\left(\frac{\mu^2}{\mu_0^2}\right) \left[ \log\left(\frac{\mu_0^2}{-s-i\ep}\right) + \frac{1}{2} \log\left(\frac{\mu^2}{\mu_0^2}\right) \right] \ket{\cF^{1}(m,\mu_0)}\,.
\end{align*}



\bibliography{paper}
\bibliographystyle{JHEP}

\end{document}